\documentclass[useAMS,usenatbib,usegraphicx]{mn2e}
\usepackage{times}
\usepackage{graphicx}
\usepackage{epsfig}
\usepackage{amssymb}
\usepackage{natbib}
\usepackage{array}
\usepackage{color}

\newcommand{\sauron}{\texttt{SAURON}}
\newcommand{\atlas}{ATLAS$^{\rm 3D}$ }
\newcommand{\lreps}{$\lambda_R - \epsilon$}

\title[Simulations of binary galaxy mergers]{The \atlas{} Project - VI. Simulations of binary galaxy mergers and the link with Fast Rotators, Slow Rotators, and Kinematically Distinct Cores.}
\author[Maxime Bois et al.]{Maxime Bois$^{1,2}$\thanks{E-mail: maxime.bois@obspm.fr}, Eric Emsellem$^{2,1}$, Fr\'ed\'eric Bournaud$^{3}$, Katherine Alatalo$^4$, \newauthor Leo Blitz$^4$, Martin Bureau$^5$, Michele Cappellari$^5$, Roger L. Davies$^5$, \newauthor Timothy A. Davis$^5$, P. T. de Zeeuw$^{2,6}$, Pierre-Alain Duc$^{3}$, Sadegh Khochfar$^7$, \newauthor Davor Krajnovi\'c$^2$, Harald Kuntschner$^{8}$, Pierre-Yves Lablanche$^{1,2}$, \newauthor Richard M. McDermid $^{9}$, Raffaella Morganti$^{10,11}$,  Thorsten Naab$^{12}$, \newauthor Tom Oosterloo$^{10,11}$, Marc Sarzi$^{13}$,Nicholas Scott$^5$, Paolo Serra$^{10}$, \newauthor Anne-Marie Weijmans$^{14}$\thanks{Dunlap Fellow} and Lisa M. Young$^{15}$ \\ \\
$^1$Universit\'e Lyon 1, Observatoire de Lyon, Centre de Recherche Astrophysique de Lyon and Ecole Normale Sup\'erieure de Lyon,\\ 9 avenue Charles Andr\'e, F-69230 Saint-Genis Laval, France\\
$^2$European Southern Observatory, Karl-Schwarzschild-Str. 2, 85748 Garching, Germany\\
$^3$Laboratoire AIM Paris-Saclay, CEA/IRFU/SAp – CNRS – Universit\'e Paris Diderot, 91191 Gif-sur-Yvette Cedex, France\\
$^4$Department of Astronomy, Campbell Hall, University of California, Berkeley, CA 94720, USA\\
$^5$Sub-department of Astrophysics, University of Oxford, Denys Wilkinson Building, Keble Road, Oxford OX1 3RH, UK\\
$^6$Sterrewacht Leiden, Leiden University, Postbus 9513, 2300 RA Leiden, the Netherlands\\
$^7$Max-Planck Institut f\"ur extraterrestrische Physik, PO Box 1312, D-85478 Garching, Germany\\
$^8$Space Telescope European Coordinating Facility, European Southern Observatory, Karl-Schwarzschild-Str. 2, 85748 Garching, Germany\\
$^9$Gemini Observatory, Northern Operations Centre, 670 N. A`ohoku Place, Hilo, HI 96720, USA\\
$^{10}$Netherlands Institute for Radio Astronomy (ASTRON), Postbus 2, 7990 AA Dwingeloo, The Netherlands\\
$^{11}$Kapteyn Astronomical Institute, University of Groningen, Postbus 800, 9700 AV Groningen, The Netherlands\\
$^{12}$Max-Planck-Institut f\"ur Astrophysik, Karl-Schwarzschild-Str. 1, 85741 Garching, Germany\\
$^{13}$Centre for Astrophysics Research, University of Hertfordshire, Hatfield, Herts AL1 9AB, UK\\
$^{14}$Dunlap Institute for Astronomy \& Astrophysics, University of Toronto, 50 St. George Street, Toronto, ON M5S 3H4, Canada\\
$^{15}$Physics Department, New Mexico Institute of Mining and Technology, Socorro, NM 87801, USA\\
}

\begin{document}

\date{Accepted ??. Received ??; in original form ??}

\pagerange{\pageref{firstpage}--\pageref{lastpage}} \pubyear{??}

\maketitle

\label{firstpage}

\clearpage

\begin{abstract}
We study the formation of early-type galaxies (ETGs) through mergers with a sample of 70 high-resolution (softening length~$<$~60~pc and 12$\times10^6$ particles) numerical simulations of binary mergers of disc galaxies (with 10 per cent of gas) and 16 simulations of ETG remergers. 
These simulations, designed to accompany observations and models conducted within the \atlas{} project, encompass various mass ratios (from 1:1 to 6:1), initial conditions and orbital parameters. The progenitor disc galaxies are spiral-like with bulge to disc ratios typical of Sb and Sc galaxies and high central baryonic angular momentum.
We find that binary mergers of disc galaxies with mass ratios of 3:1 and 6:1 are nearly always classified as Fast Rotators according to the \atlas{} criterion (based on the $\lambda_R$ parameter, see Emsellem et al. 2011 -- \atlas{} Paper III): they preserve the structure of the input fast rotating spiral progenitors. They have intrinsic ellipticities larger than 0.5, cover intrinsic $\lambda_R$ values between 0.2 and 0.6, within the range of observed Fast Rotators. The distribution of the observed Fastest Rotators does in fact coincide with the distribution of our disc progenitors.
Major disc mergers (mass ratios of 2:1 and 1:1) lead to both Fast and Slow Rotators. Most of the Fast Rotators produced in major mergers have intermediate flattening, with ellipticities $\epsilon$ between 0.4 and 0.6. Most Slow Rotators formed in these binary disc mergers hold a stellar Kinematically Distinct Core (KDC) in their $\sim 1$-3 central kilo-parsec: these KDCs are built from the stellar components of the progenitors. However, these remnants are still very flat with $\epsilon$ often larger than 0.45 and sometimes as high as 0.65. Besides a handful of specific observed systems --~the counter-rotating discs (2-$\sigma$ galaxies, see Krajnovi\'c et al. 2011, \atlas{} Paper II)~-- these therefore cannot reproduce the observed population of Slow Rotators in the nearby Universe.
This sample of simulations supports the notion of Slow and Fast Rotators: these two families of ETGs present distinct characteristics in term of their angular momentum content (at all radii) and intrinsic properties: the Slow Rotators are not simply velocity-scaled down versions of Fast Rotators. The mass ratio of the progenitors is a fundamental parameter for the formation of Slow Rotators in these binary mergers, but it also requires a retrograde spin for the earlier-type (Sb) progenitor galaxy with respect to the orbital angular momentum.
We also study re-mergers of these merger remnants: these produce relatively round Fast Rotators or systems near the threshold for Slow Rotators. In such cases, the orbital angular momentum dominates the central region, and these systems no longer exhibit a KDC, as KDCs are destroyed during the remergers and do not reform in these relatively dry events.
\end{abstract}

\begin{keywords}
galaxies:~formation -- galaxies:~elliptical and lenticulars, cD -- galaxies:~interactions -- galaxies:~kinematics and dynamics -- methods:~N-body simulations
\end{keywords}

\section{Introduction}
Numerical simulations, intensively used for more than two decades, have clearly shown that the global characteristics of the remnants of binary merger between two equal-mass spiral galaxies, called major mergers, resemble those of Early-Type Galaxies (\textit{i.e} ellipticals \& lenticulars), hereafter ETGs \citep{toomre72,HB91,Barnes92, mihos95,sprin00, naab03, bour05}. Remnants with properties similar to ETGs  can also be recovered via multiple minor mergers with the total accreted mass being at least half of the initial mass of the main progenitor \citep{weil94, weil96, bour07}. This picture of the formation of ETGs via accretion and merging of stellar bodies would fit well within the frame of the hierarchical assembly of galaxies provided by $\Lambda$CDM cosmology.

Owing to large statistical samples, modern work tends to quantify in detail the properties of major and minor merger remnants, together with thorough comparisons with observed properties of ETGs \citep{naab03, bour04, bour05, NJB06, coxal06,coxal08,roth06, dimatteo1,burkert-sauron, jesseitlr, hoff, chili10}. It is, however, still difficult to build large samples of simulations with sufficiently high numerical resolution. Many studies have indeed shown that the resolution affects the properties of the simulated objects, either in cosmological simulations \citep{naabresol, navresol} or on simulations of star formation in mergers \citep{coxresol, hopkinsresol,dimatteo2,wuyts10}. \citet{boisresol} claimed that a high-enough resolution is required to resolve properly the fluctuations of the gravitational potential during the merger, \textit{i.e} violent relaxation, which can significantly impact the morphology and kinematics of merger remnants.

In this paper, we further examine the role of the initial conditions (impact parameter, incoming velocities, inclination and spins of the progenitors) on the global characteristics of the remnants of binary galaxy mergers. We then propose to study the morphology and the kinematics of binary galaxy merger remnants simulated with a high numerical resolution. With that purpose in mind, we build two-dimensional momentum (intensity, velocity and velocity dispersion) maps of the merger remnants and analyse their apparent properties, directly linked with their orbital structures \citep{JNB05}. Using two-dimensional maps enables us to compare our merger remnants directly with modern spectroscopic observations of resolved local galaxies: the emergence of integral field spectrographs, such as the \sauron{} spectrograph \citep{bacon}, allowed the mapping of local ETGs up to about one effective radius. Numerical simulations of binary mergers of disc galaxies or binary remergers are not necessarily representative of observed early-type galaxies \citep{roth06,burkert-sauron} but they are a powerful tool to constrain the formation mechanisms of different features observed in 2D maps of local galaxies like \textit{e.g.} the formation of a Kinematically Distinct Cores (KDC) \citep[\textit{e.g.}][]{HB91,jesseitlr,boisresol, hoff}, the presence of gaseous discs \citep[\textit{e.g.}][]{H09,martig09}, the relation between nuclear black holes and the dynamics of the stars \citep[\textit{e.g.}][]{SMH05,johan09}, the formation of plumes, tails, stellar clusters at large radii, or giant gas rings \citep{feld08, B08, leo10}.

The \sauron{} survey \citep{zeeuw} has introduced a new view of ETGs, classifying them in two families --~the fast and the slow rotators~-- according not only to their morphology, but also to their kinematics \citep[][ see also Section~\ref{sec:physparam}]{ems07,cap07}. These two classes are intrinsically different and seem to correspond to different populations of galaxies: while the fast rotators present regular rotation patterns aligned with the photometry, the slow rotators have low angular momentum and show misalignments between the photometry and the velocity axes. Furthermore, slow rotators often exhibit KDCs, usually defined as a central stellar component with a rotation axis distinct from the outer stellar body \citep[see \textit{e.g.}][]{kraj08}. KDCs have been claimed to be a signature of a past interaction \citep[\textit{e.g.}][]{franx88,jedr88,zeeuw91,scorza,davies01}. 

To further constrain the formation scenarios for these two classes, an ambitious program --~\atlas{}~-- is being conducted, combining a multi-wavelength observational survey of a complete volume-limited sample of ETGs with various numerical simulation and modelling efforts. The \atlas{} project\footnote{http://purl.org/atlas3d} (Cappellari et al. 2011, hereafter Paper I) aims to quantify the distribution and kinematics of the stellar and gaseous components of a statistically significant sample of ETGs to relate their detailed properties to their mass assembly, star formation history and evolution. Krajnovi\'c et al. (2011, hereafter Paper II) introduced new sub-classes within the slow rotator class: non-rotators which do not show any apparent sign of rotation, galaxies with a KDC when there is an abrupt change in the radial profile of the kinematical position angle ($> 30^{\circ}$), galaxies with non regular velocity pattern but with no noticeable kinematic feature, and the 2-$\sigma$ galaxies characterised by two off-centred symmetric dispersion peaks along the major-axis. These so-called 2-$\sigma$ galaxies are generally thought to have two stellar counter-rotating components (\textit{i.e.} a KDC at an 180$^{\circ}$ away from the outer body) \citep[see also][for 2D-map examples of these counter-rotating components in simulations]{JNPB07}.

In the context of the \atlas{} project, an extensive set of numerical simulations is being conducted to support the survey: cosmological simulations, semi-analytic modelling and idealized binary galaxy mergers. In the present study, we have simulated a substantial sample ($\sim$ 90) of binary mergers of disc galaxies and binary galaxy remergers at an unprecedented resolution, sufficient to properly follow the merging process and the resulting galaxy remnant \citep{boisresol}. We aim to constrain the formation of the slow and fast rotators revealed by the \sauron{} survey. For this purpose, we look at the morphology and the kinematics of the binary merger and remerger remnants and we link the initial conditions of the mergers to the formation of the fast and slow rotators as well as the formation of the KDCs. In Section~\ref{sec:simus}, we present the initial progenitor galaxies, which are spirals of Hubble type Sb and Sc, and the different initial conditions of merging. We present in Section~\ref{sec:binary} the global properties of the merger remnants and the impact of the initial conditions: we thus show that the mass ratio, the angular momentum content and the Hubble type of the progenitors play an important role in the formation of the slow and fast rotators and the KDCs. In Section~\ref{sec:remergers}, we analyse the remnants of galaxy remergers, these remnants are all classified as fast rotators or near the threshold for slow rotators and do not show any trace of a KDC. The formation mechanism of the KDCs via binary mergers of disc galaxies is presented in Section~\ref{sec:kdc}. We then discuss our results and compare it with previous studies (Section~\ref{sec:discu}) and with the observations of the 260 ETGs of the \atlas{} survey (Section~\ref{sec:compa}). We then conclude and sum-up all our results in Section~\ref{sec:conclu}.

\begin{table*}
 \begin{center}
  \caption{Main physical parameters used to model the initial progenitor galaxies in our merger models. (1) Spiral Hubble type (2) $M_{stars}$ in the bulge / total $M_{stars}$ (3) Fraction of the baryonic mass in gas (4) Baryonic mass in unit of $1.3 \times 10^{11}$~M$_{\odot}$. (5) Total number of particles (stars, gas and dark matter) \label{tab:tabsetup}}
  \begin{tabular}{|c||c|c|c|c|c||c|c|c|c|c|c|}
  \hline
  Name     & Prog 1$^{(1)}$   & $B/T^{(2)}$ & $f_{gas}^{(3)}$    & $m_B^{(4)}$ & $N_{part}^{(5)}$  & Prog 2$^{(1)}$    & $B/T^{(2)}$   & $f_{gas}^{(2)}$ & $m_B^{(3)}$ & $N_{part}^{(5)}$  \\
  \hline
  m11      & Sb               &  0.2        & 10\%               & 1           & 6$\times10^6$     & Sc                &  0.12         & 10\%            & 1    &  6$\times10^6$   \\ 
  m21g10   & Sb               &  0.2        & 10\%               & 1           & 6$\times10^6$     & Sc                &  0.12         & 10\%            & 0.5  &  3$\times10^6$  \\ 
  m21g33   & Sb               &  0.2        & 33\%               & 1           & 6$\times10^6$     & Sc                &  0.12         & 33\%            & 0.5  &  3$\times10^6$  \\ 
  m31      & Sb               &  0.2        & 10\%               & 1           & 6$\times10^6$     & Sc                &  0.12         & 10\%            & 0.33 &  2$\times10^6$  \\ 
  m61      & Sb               &  0.2        & 10\%               & 1           & 6$\times10^6$     & Sc                &  0.12         & 15\%            & 0.17 &  1$\times10^6$  \\ 
  \hline
  \end{tabular}
 \end{center}
\end{table*}

\begin{figure*}
\begin{center}
 \includegraphics[width=0.98\columnwidth]{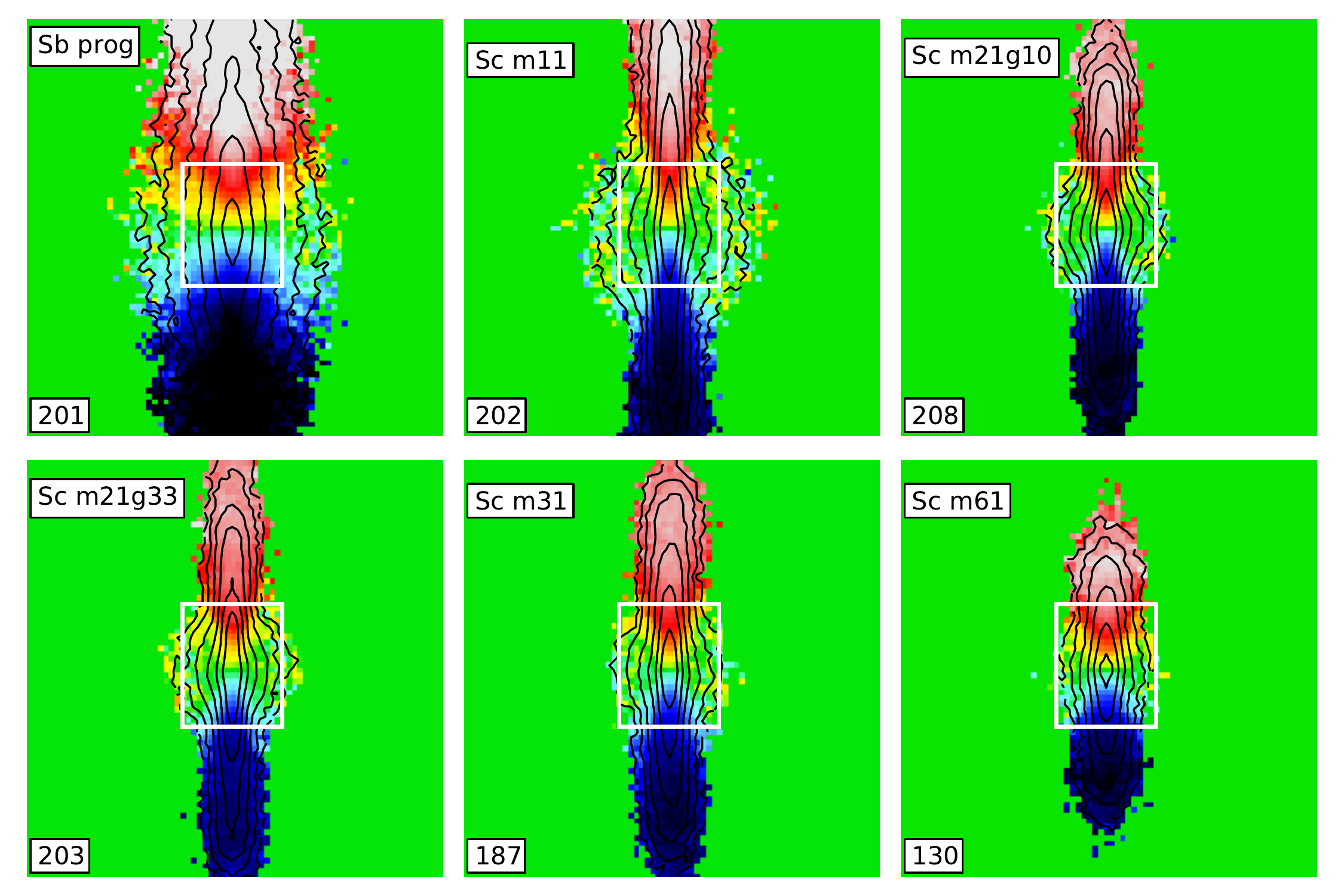}
 \includegraphics[width=0.11\columnwidth]{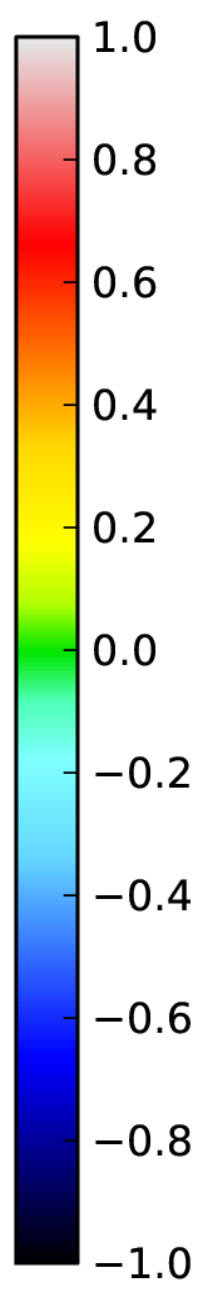}
 \includegraphics[width=0.91\columnwidth]{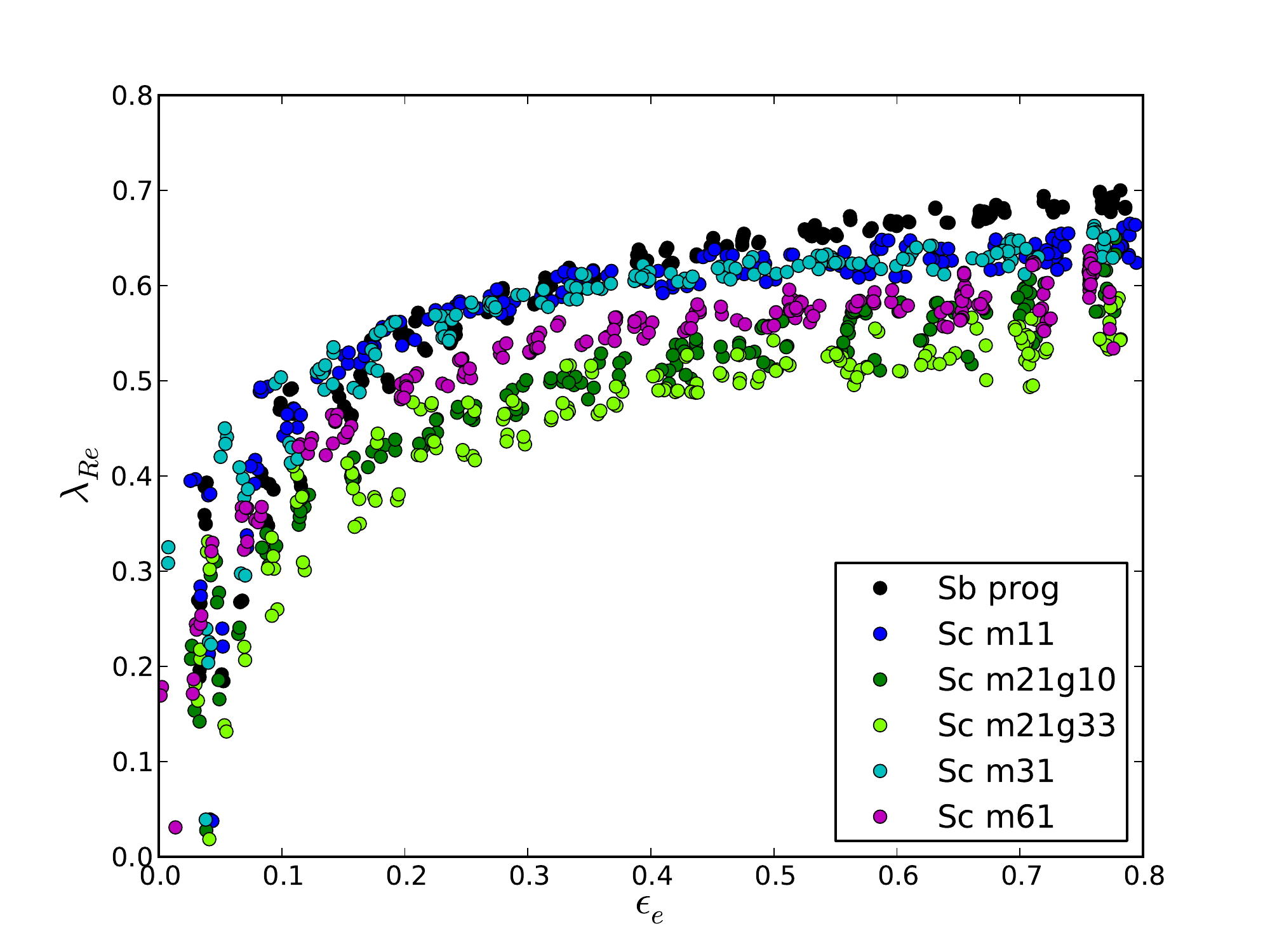}
 \caption{Left: Projected stellar velocity maps for the Sb progenitor and the different mass ratio Sc companions, the black contours represent the iso-magnitude contours, they are equally spaced in magnitude and are the same for all progenitors. Each subpanel is labelled with the name of the progenitor and with the maximum velocity V$_{max}$, the colorbar goes from -V$_{max}$ to +V$_{max}$. The field of view is $15\times15$~kpc$^2$ (15~kpc $\simeq$ 6~$R_e$). The white rectangle indicates a typical field covered by the instrument \sauron{} and corresponds to a field of $41" \times 33"$ for a galaxy at a distance of 20~Mpc, its orientation follows the photometric position angle taken at 3$R_e$. Right: \lreps{} diagram for the progenitors of binary mergers. The Sb progenitor (black points) is the same in all simulations, the other colours correspond to the Sc companion at different mass ratios (m11 in blue, m21g10 in green, m21g33 in light green, m31 in cyan, m61 in magenta). All projections (200 per remnant) are plotted.}
 \label{fig:simprog}
\end{center}
\end{figure*}

\section{Simulations and analysis} \label{sec:simus}

  \subsection{Method \label{sec:method}}

    \subsubsection{Code \label{sec:code}}
We use the particle-mesh code described in \citet{B08}, \citet{boisresol} and references therein. This code uses a Cartesian grid on which the particles are meshed with a "Cloud-In-Cell" interpolation. The gravitational potential is computed with an FFT-based Poisson solver and particle motions are integrated with a time-step of 0.5~Myr. All simulations in this paper were performed with a softening length (\textit{i.e.} spatial resolution) of 58~pc.

Interstellar gas dynamics is modelled with the sticky-particle scheme with elasticity parameters $\beta_t=\beta_r= 0.6$. This scheme neglects the thermal pressure of the gas, assuming it is dominated by its turbulent pressure, which is the case for the bulk of the gas mass in the star-forming interstellar medium \citep{ES04,burkert06}. The velocity dispersion of the gas particles models the turbulent speed, and their collisions model the turbulent pressure. The collisions are inelastic to ensure that the modelled turbulence dissipates in about a vertical crossing time in discs \citep{ml99}.

Star formation is modelled with a local Schmidt law \citep{schmidt,kenni}: the star formation rate in each resolution element is proportional to the gas density to the exponent 1.5. Energy feedback from supernovae is accounted for with the kinetic scheme initially proposed by \citet{MH94}. Each stellar particle formed has a number of supernovae computed from the fraction of stars above  8 M$_{\odot}$ in a Miller-Scalo IMF. A fraction $\epsilon$ of the typical $10^{51}$~erg energy of each supernova is released in the form of radial velocity kicks applied to gas particles within the closest cells. We use $\epsilon = 2 \times 10^{-4}$ , as \citet{MH94} suggested that realistic values lie around $10^{-4}$ and less than $10^{-3}$. Simulations with parsec-scale resolution have to use much higher values for the $\epsilon$ parameter \citep[\textit{e.g.}][]{bournaud-lmc}: the low value of the $\epsilon$ parameter used here reflects the fact that a large fraction of the released energy is radiated away (through dissipation and gas cooling processes) before spreading on scales of $50-100$~pc (i.e., the resolution of our present models). Larger values of $\epsilon$ would thicken the gas discs of isolated spirals to unacceptable proportions \citep{MH94,EBE08}.

    \subsubsection{Set-up for initial parameters \label{sec:setup}}
We have simulated binary mergers of discs galaxies (i.e., ``spiral-spiral'' mergers) with mass ratios of 1:1, 2:1, 3:1 and 6:1. The main parameters are summarized in Table~\ref{tab:tabsetup} for the progenitor galaxies and Table~\ref{tab:taborbit} for the orbital parameters. Table~\ref{tab:propremergers} summarizes the parameters used for re-mergers using the result of a first merger simulation as an input for a new merger simulation (i.e., ``spiral-ETG'' or ``ETG-ETG'' mergers).

    \paragraph*{Initial progenitor galaxies}
Our main sample consists of binary mergers of disc galaxies, i.e. ``spiral-spiral'' mergers. The first progenitor, which is defined as the most massive for unequal-mass mergers, has a baryonic mass of $1.3 \times 10^{11}$~M$_{\odot}$. The bulge fraction is $B/T=0.20$ and the gas fraction in the disc is usually 10 per cent (33 per cent in some m21 simulations). The initial disc has a Toomre profile\footnote{The profile rapidly evolves into a quasi-exponential profile with a slightly smaller exponential scale length.} with a scale length of 4~kpc, consistent with observations of nearby disc galaxies \citep{fathi09}, truncated at 10~kpc. The gas has a scale length of 8~kpc and a truncation radius of 20~kpc. The bulge has a \citet{hernquist-profile} profile with a scale length of 700~pc. The dark matter halo is modelled with a Burkert profile \citep{burkert-profile} with a 7~kpc scale-length and a truncation radius of 70~kpc, inside which the dark matter mass is $3\times 10^{11}$~M$_{\odot}$. This initial galaxy is representative for an Sb spiral galaxy, given its bulge fraction in particular, and is denoted as the {\em Sb} spiral progenitor throughout the paper.

The other progenitor, which has the lower mass in unequal mass mergers, has its total mass determined by the mass ratio. Apart from the bulge which was reduced to $B/T=0.12$, all components have their mass scaled by a factor equal to the mass ratio, and all sizes and scale lengths scaled by the square root of the mass ratio, thus keeping the central density of their discs constant. The gas fraction is 10 per cent, except for the lower-mass companions used in 6:1 mergers where a gas fraction of 15 per cent is used. The main difference with the first progenitor is thus the lower bulge fraction, and this progenitor galaxy is denoted as the {\em Sc} spiral progenitor.

The Sb spiral is modelled with $2 \times 10^6$ particles for each component (stars, gas, and dark matter). The Sc spiral is modelled with  a number of particles scaled by the mass ratio. Equal-mass mergers thus use a total of 12 million particles.

\medskip

The progenitors are initialized as perfectly axisymmetric disc galaxies. Such galaxies will unavoidably develop substructures such as spiral arms, bars, etc. We need to avoid this spontaneous, intrinsic evolution to take place during the merger, otherwise the effects of the merger itself can not be disentangled. We also need to avoid the artificial consumption of gas that could result from applying the Schmidt law during the transition from an axisymmetric gas disc to a spiral disc. To this aim, each progenitor galaxy was evolved in isolation and without star formation for about two rotations of the outer stellar disc, so that a reasonable steady state in its structure and gas density distribution is reached. At this point, we remark that the Sb progenitor is more concentrated and dynamically stable than the Sc galaxy. The merger simulation is then started, with star formation and feedback, using these pre-evolved progenitors --~see \citet{martig08} for further discussion of this technique.

The initial properties of our pre-merger spiral galaxies, after their initial relaxation in isolation, are shown in Fig.~\ref{fig:simprog}. We in particular compare the isophotal ellipticity $\epsilon$ to their angular momentum tracer $\lambda_R$ at the effective radius $R_e$ for 200 different projections (see Sections~\ref{sec:projmaps} and \ref{sec:physparam} for details on the $\epsilon$ and $\lambda_R$ parameters and on the projection effects). Small differences between the progenitors arose during the initial relaxation as galaxies with different sizes and bulge fractions behave somewhat differently in isolation; these properties are essentially stabilized (on time-scales of a few $10^8$~yr) when the merger simulations are started.

    \paragraph*{Merger orbits \label{sec:orbits}}
The different orbits used for the merger simulation are described in Table~\ref{tab:taborbit}. They are all parabolic or hyperbolic, with an initial total energy $E > 0$ or $E \simeq 0$, corresponding to initially unbound galaxy pairs. Such orbits are representative of the most common mergers in $\Lambda$CDM cosmology \citep{khochfar06}.

The fiducial orbit "0" has a velocity at infinite distance of V$_{\infty}$=120~km~s$^{-1}$, an inclination $i$=45$^{\circ}$ for each galaxy with respect to the orbital plane, and an impact parameter (\textit{i.e.} the perpendicular distance between the two velocity vectors of the progenitors at the beginning of the simulation) of $R$=60~kpc. The six other orbits correspond to one of the three parameters $V$, $R$, $i$ being changed compared to the fiducial values, one at a time. The nomenclature denotes the parameter varied and whether it is increased (p: plus) or decreased (m: minus), as detailed in Table~\ref{tab:taborbit}. We have then a total of 70 simulations of galaxy mergers.

In addition of the three listed parameters, we identify the spin of the progenitors which can be either direct (prograde) \textbf{d} or retrograde \textbf{r} with respect to the orbital angular momentum. The first letter refers to the spin of the first progenitor (Sb) and the second letter to the Sc companion.

Our nomenclature also indicates the type (mass ratio) of the modelled merger, as defined in Table~\ref{tab:tabsetup}. For instance, the simulation labelled {\em m31rdVm} refers to a 3:1 merger with a retrograde orientation for the main Sb progenitor and a direct orientation for the Sc companion on an orbit with lowered initial velocity.

The chosen orbits do not cover all the parameter space of initial conditions: the goal of this paper is to understand the formation of slow and fast rotators, and not to build a full library of binary merger remnants.

\begin{table}
\begin{center}
\caption{The different initial conditions for the mergers. (1) The relative incoming velocity V$_{\infty}$ (in km~s$^{-1}$) of the Sc companion computed at infinite distance. (2) Inclination $i$ (in $^{\circ}$) for each galaxy with respect to the orbital plane. (3) Impact parameter $R$ at infinite distance (in kpc). \label{tab:taborbit}}
\begin{tabular}{|c|c|c|c|}
\hline
Name         & V$_{\infty}$ $^{(1)}$     & $i$ $^{(2)}$  & $R$ $^{(3)}$        \\
\hline
0            & 120                       & 45            & 60                  \\
im           & 120                       & 25            & 60                  \\
ip           & 120                       & 75            & 60                  \\
Rm           & 120                       & 45            & 35                  \\
Vm           & 70                        & 45            & 60                  \\
Vp           & 200                       & 45            & 60                  \\
\hline
\end{tabular}
\end{center}
\end{table}

  \subsection{Analysis \label{sec:analysis}}
We analysed all merger remnants about 600-800~Myr after the central coalescence. The central body of the ETG-like systems formed in the mergers are thus relaxed within several effective radii when the analysis is performed, even if tidal debris is present at larger radii. We checked in several cases that the measured morphological and kinematical parameters were stabilized by comparing the ellipticity and the $\lambda_R$ parameter to earlier snapshots.

    \subsubsection{Projected velocity moment maps \label{sec:projmaps}}
Intrinsic and apparent properties of the merger remnant, such as the apparent ellipticity, are directly linked with its orbital structure \citep{JNB05}. To probe the intrinsic properties of the relaxed merger remnants in a way that is comparable to observations, we have therefore built velocity moment maps indicating the stellar surface density, stellar velocity field and stellar velocity dispersion for various projections. Two-dimensional maps are useful to reveal the wealth of photometric or kinematic structures associated with a galaxy merger remnant, such as young stellar clusters \citep[potentially globular clusters, ][]{boisresol}, kinematic misalignments \citep{BB00,JNPB07,kraj08}, and were also used to measure various parameters with definitions finely matching those used for the \atlas{} observations (see below).

The intensity and velocity moment maps are built to cover a $20\times20$~kpc$^2$ field of view around the stellar density peak of each system (figures show the inner $15\times15$~kpc$^2$ area). This covers at least about four effective radii $R_e$ to enclose most of the baryonic mass of early-type galaxies, the average effective radius of our merger remnants being 2.3~kpc. Each projection was computed on a $100\times100$ pixel grid, giving an effective resolution consistent with the \sauron{} data of the \atlas{} survey. The maps have been Voronoi binned \citep{CC03} to have a minimum number of 400 particles per bin.
 
To obtain statistically representative results rather than analyzing a particular projection, we have produced velocity moment maps and performed the subsequent analysis for 200 isotropically distributed viewing angles for each merger remnant. The choice of 200 projections ensures a sampling smaller than 10 degrees in any direction, so that intermediate viewing angles would not show significant differences. 

 \subsubsection{The presence of KDCs \label{sec:defkdc}}
The merger remnants analysed in the next Sections often exhibit a central stellar component with an apparent rotation axis distinct from the one of the outer parts. In the present study, using the definition given by \citet{kraj08}, we have conducted a visual inspection of the velocity maps to detect such central structures and flag them as KDCs. The use of this objective criterion leads to clear signatures of the KDCs which are easily recognizable.

In the two-dimensional velocity maps of the merger remnants presented in Appendix~\ref{sec:appprojmaps}, we indicate the typical field covered by the instrument \sauron{} (a field of view typically of $41" \times 33"$) for a galaxy at a distance of 20~Mpc. We note that a few KDCs in the merger remnants (see Section~\ref{sec:binary} for the properties of these galaxies and the formation processes of the KDCs) are larger than that typical \sauron\ field. This obviously depends on the distance of the system, as \sauron\ would fully cover these KDCs were the galaxy to be at distances larger than \textit{e.g.}, 30~Mpc. In the next Sections, we show that the merger remnants with a KDC --~even those which are larger than the \sauron{} field-of-view~-- are all classified as slow rotators and have properties which are distinct than those of the fast rotators.

With numerical simulations, we can easily investigate the stellar kinematics of galaxies at large radii to probe, \textit{e.g.}, large KDCs. It is observationally much more challenging with \textit{e.g.}, the \sauron{} field-of-view being limited to approximately one effective radius for local ETGs. This may lead to (a few) mis-identifications. New promising techniques are emerging which should allow to explore the faint outer stellar structures of larger samples of ETGs \citep[see \textit{e.g.},][or the SAGES project]{Weijmans09,Coccato09,proctor09,murphy11}.

 \subsubsection{Extracted parameters \label{sec:physparam}}
Our analysis is mainly based on two simple morphological and kinematic parameters, a choice motivated by the fact that these parameters are often used as standards in studies of nearby ETGs \citep[\textit{e.g.}][]{jesseitlr,boisresol}. The analysis in this paper assumes a constant stellar mass-to-light ratio, which should be a reasonable approximation for relatively old ETGs. Furthermore, our main results will show that the classification of merger remnants into slow and fast rotators, and the presence of KDCs, do not significantly depend on the youngest stellar populations formed during the merger, but rather on how the populations from each progenitor galaxy are distributed with respect to each others.

\medskip

To quantify the global morphology, we measured the ellipticity $\epsilon = 1 - b/a$, where $a$ and $b$ are the semi major- and minor-axes, respectively. To measure a profile of $a$, $b$ and $\epsilon$ as a function of the radius $r$ (and subsequently an $\epsilon (a)$ profile), we selected all pixels encircled by a given isophote and computed the inertia matrices as in \textit{e.g.} \citet{cap07}. The diagonalisation of these matrices provides $\epsilon$ and the Position Angle (PA) of the projected density map at different radii $r$. We also derive the effective radius $R_e$ from the stellar surface density as the radius encompassing half of the total stellar light (or stellar mass).

\medskip

To quantify the global kinematics of each system, using the velocity and velocity dispersion maps, we measure the $\lambda_R$ parameter, which is a robust proxy for the stellar projected angular momentum defined in \citet[][hereafter E+07]{ems07} :
\[
\lambda_R \equiv \frac{\langle R \, |V| \rangle}{\langle R \, \sqrt{V^2 + \sigma^2} \rangle}
\]
We hence compute $\lambda_R$ profiles for all projections of all merger remnants. This is achieved by selecting all pixels enclosed within the ellipse defined by the position angle and axis ratio of the photometric inertia matrice mirrorin the procedure described in E+07:
\[
\lambda_R = \frac{\sum_{i=1}^{N_p} F_i R_i \left| V_i \right|}{\sum_{i=1}^{N_p} F_i R_i \sqrt{V_i^2+\sigma_i^2}} \, ,
\]
where $F_i$ is the flux inside the $i^{th}$ pixel inside the ellipse, $R_i$ its distance to the centre, and $V_i$ and $\sigma_i$ the corresponding mean stellar velocity and velocity dispersion values of the pixel.

In E+07, $\lambda_R$ was used to reveal two families of early-type galaxies, the slow rotators with $\lambda_R \leq 0.1$ and the fast rotators with $\lambda_R > 0.1$ at one effective radius $R_e$. \citet{jesseitlr} have simulated binary disc mergers to further investigate the $\lambda_R$ parameter, and have shown that $\lambda_R$ is a good indicator of the true angular momentum content in ETGs, as confirmed via the knowledge of the full phase-space distribution in the corresponding remnants. As emphasised in E+07, \citet{cap07} and \citet{kraj08}, fast and slow rotators exhibit qualitatively and quantitatively different stellar kinematic properties. Our following analysis will lend further support to the fact that $\lambda_R$ can really separate two classes of galaxies and that the distinction of fast and slow rotators does not simply result from an arbitrary $\lambda_R$ threshold.

The \sauron{} survey was based on a representative sample of ETGs but biased towards the upper end of the luminosity function. The \atlas{} survey probes a complete sample of ETGs in the Local volume (Paper I). This led to a improved criterion for the separation between fast and slow rotators with $\lambda_R = 0.31 \cdot \sqrt{\epsilon}$ at one effective radius $R_e$ (see Emsellem et al. 2011, hereafter Paper III).

The main analysis performed in this paper is then based on the \lreps{} diagram plotted at one $R_e$. When the merger remnant presents a strong bar and/or strong spiral arms, the ellipticity is computed at 3$R_e$ to better account for the outer ellipticity and avoid being contaminated by the flattening of the bar (the difference being significant only for a small fraction of the projections), and $\lambda_R$ always being computed at $R_e$. This is the same procedure applied to the real \atlas{} galaxies in Paper III and makes our results directly comparable to the observations.

\medskip

The right panel of Fig.~\ref{fig:simprog} shows the values of $\epsilon$ and $\lambda_R$ for the 200 projections of the various disc progenitors. A few generic statements, valid for any of our spiral progenitors as well as for our fast rotator merger remnants (see Section~\ref{sec:binary}), can be made \citep[see also][]{jesseitlr}:
\begin{itemize}
  \item The edge-on view (high $\epsilon$) of the progenitors has, as expected,  the highest $\lambda_R$ value;
  \item as the galaxy is inclined (lower $\epsilon$) the value of $\lambda_R$ slightly decreases, and 
  \item as the viewing angle gets nearly face-on, the value of $\lambda_R$ drops abruptly, reaching $\lambda_R \simeq 0$ for $\epsilon \simeq 0$ as expected.
\end{itemize}

\section{Binary mergers of disc galaxies: slow and fast rotators} \label{sec:binary}
This section is devoted to the analysis of the remnants of binary mergers of disc galaxies, which in our sample are Sb+Sc mergers with various orbits and mass ratios. We study under which conditions slow and fast rotators are formed, according to the criterion defined in Section~\ref{sec:physparam}. We will provide a more direct comparison between the properties of the simulated slow and fast rotators with the galaxies of the \atlas{} sample of nearby ETGs in Section~\ref{sec:compa}. We also study which of the merger remnants harbour a KDC, and study the influence of the internal and orbital parameters of the merger in determining such properties.

\begin{figure}
  \includegraphics[width=0.98\columnwidth]{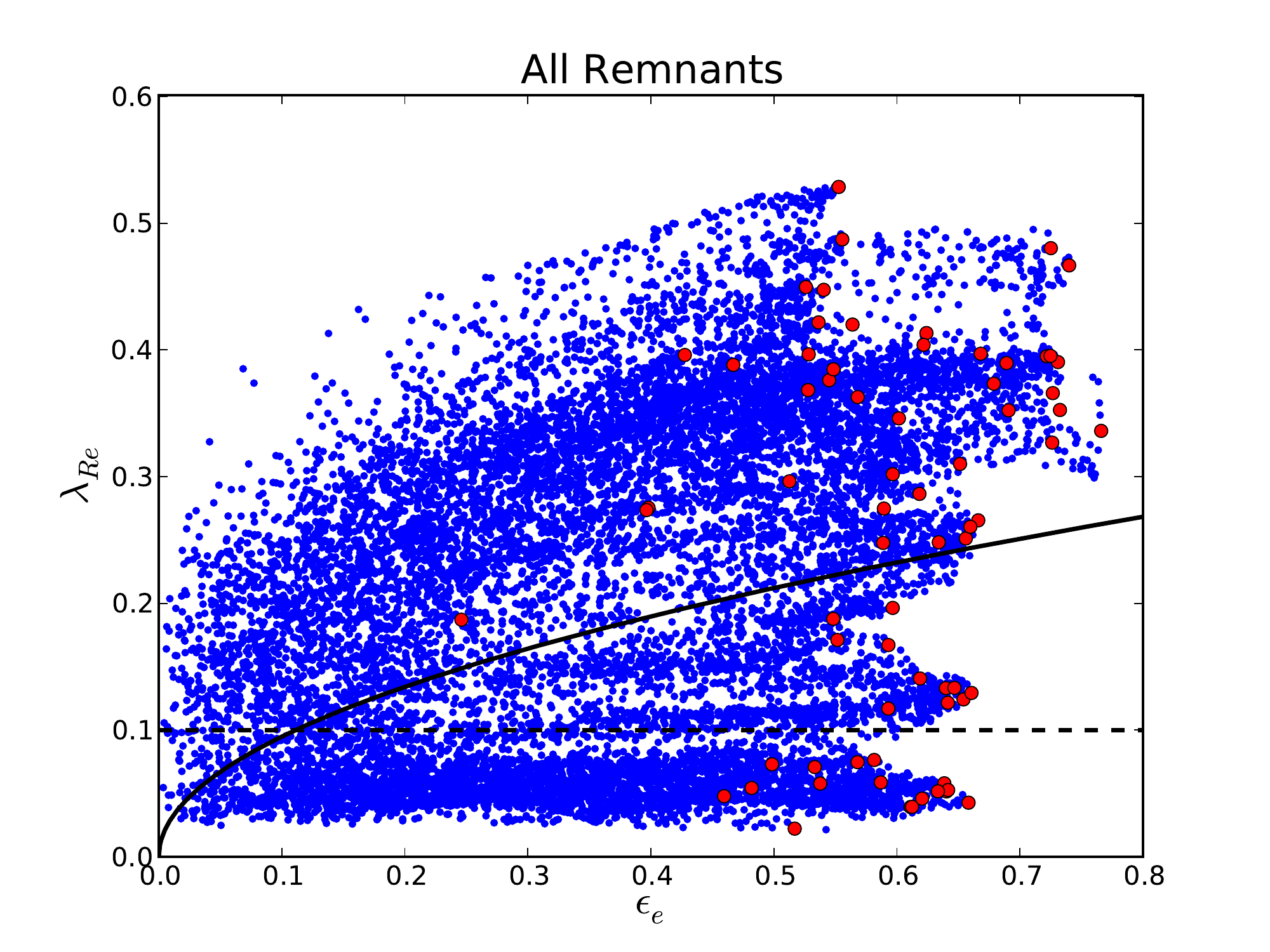} \\  
  \includegraphics[width=0.98\columnwidth]{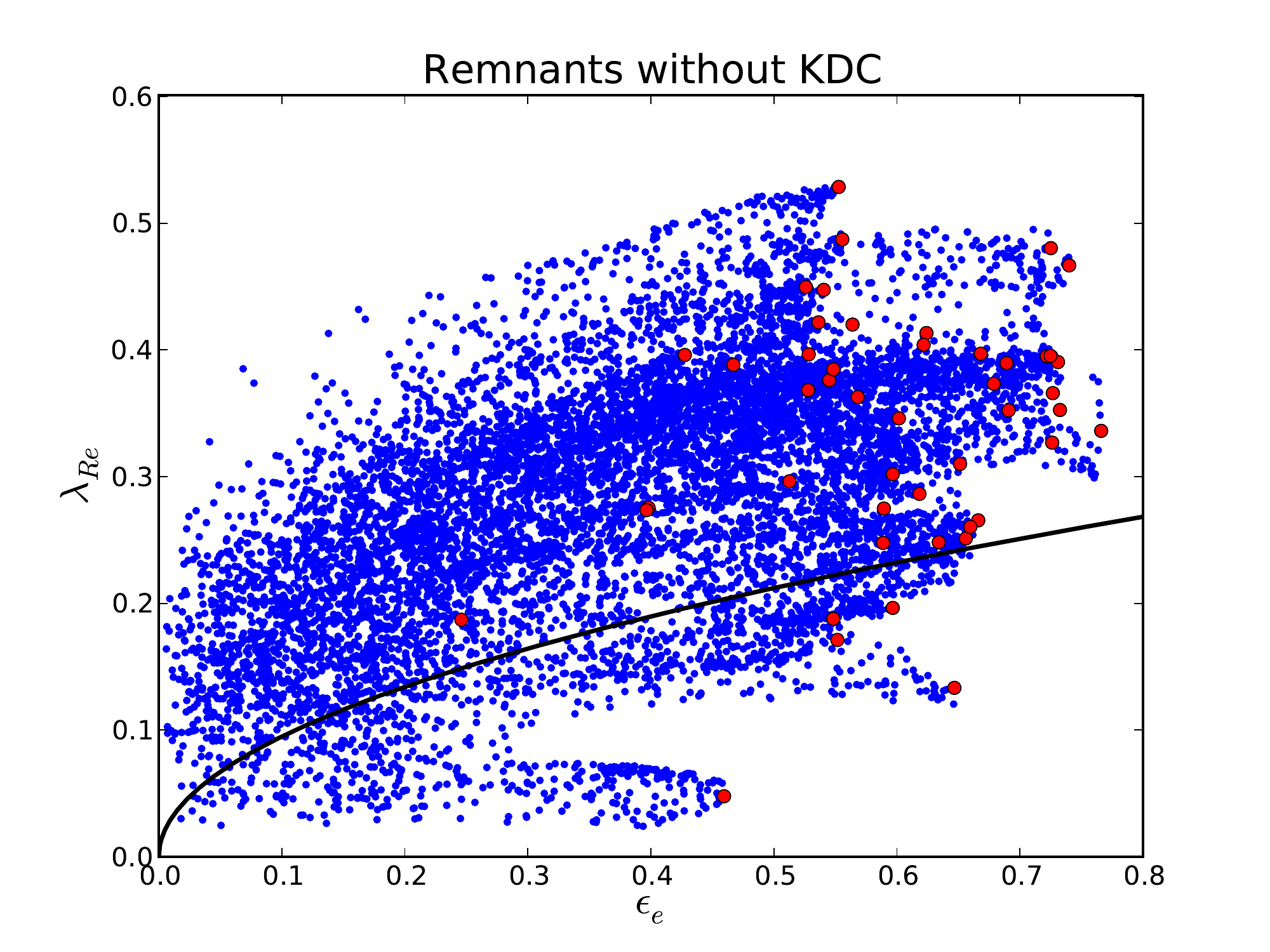} \\
  \includegraphics[width=0.98\columnwidth]{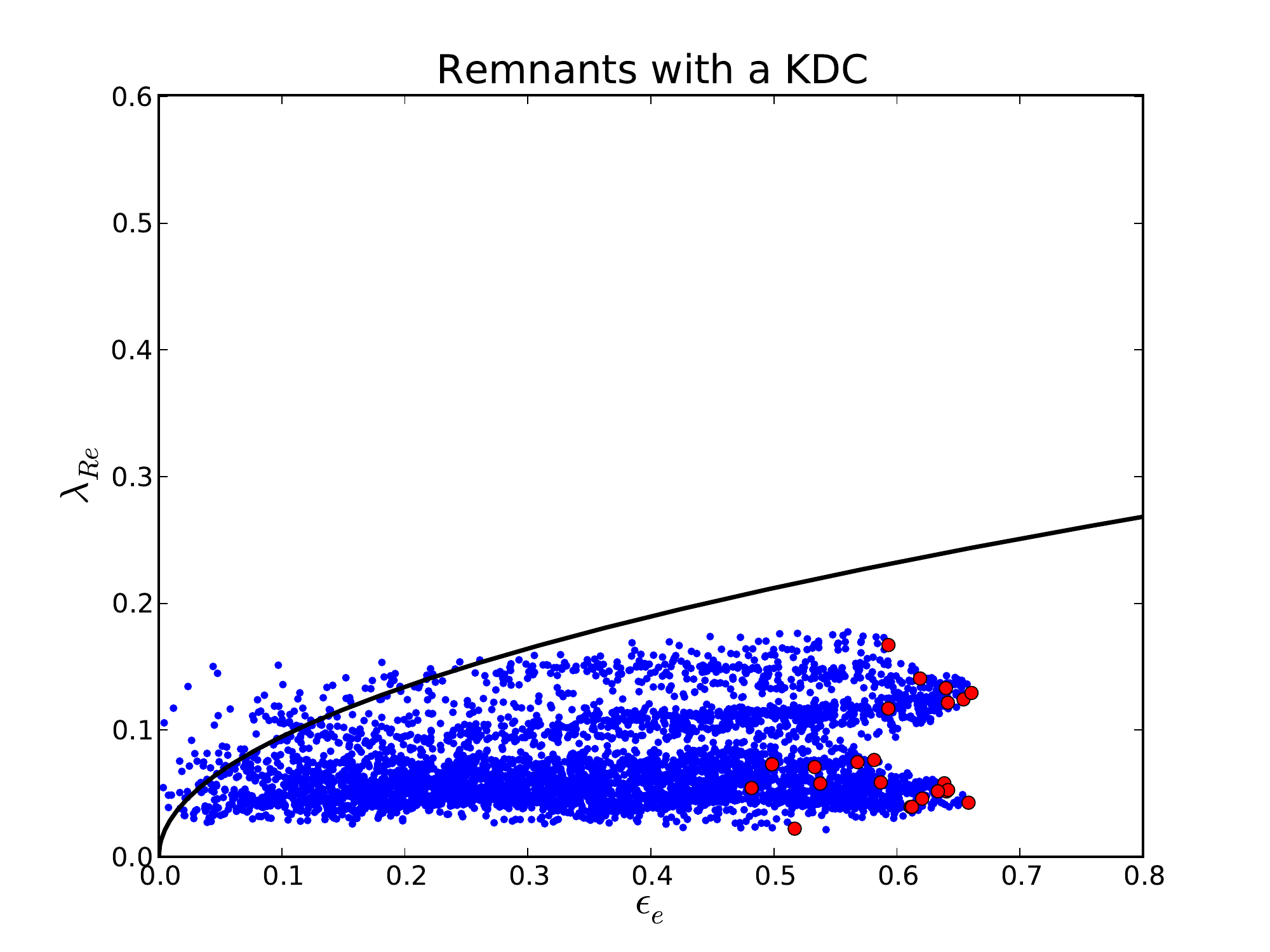} \\
  \caption{Top: \lreps{} diagram for all simulations of binary mergers of disc galaxies. All projections (200 per remnant) are plotted in blue, the red symbols corresponding to the maximum ellipticity (\textit{i.e} edge-on view) of a remnant. The limits defining the slow/fast categories from \atlas{} (solid) and from \sauron{} (dashed, only for the top panel) are plotted in black. Middle: Same diagram for the simulations which do not present a KDC in their velocity fields. Bottom: Same diagram for the simulations which present a KDC in their velocity fields.}
  \label{fig:lrepsmergerall}
\end{figure}

  \subsection{General results: populations of slow and fast rotators \label{sec:binresults}}
\begin{figure*}
\begin{center}
\begin{tabular}{ccc}
 \includegraphics[width=0.945\columnwidth]{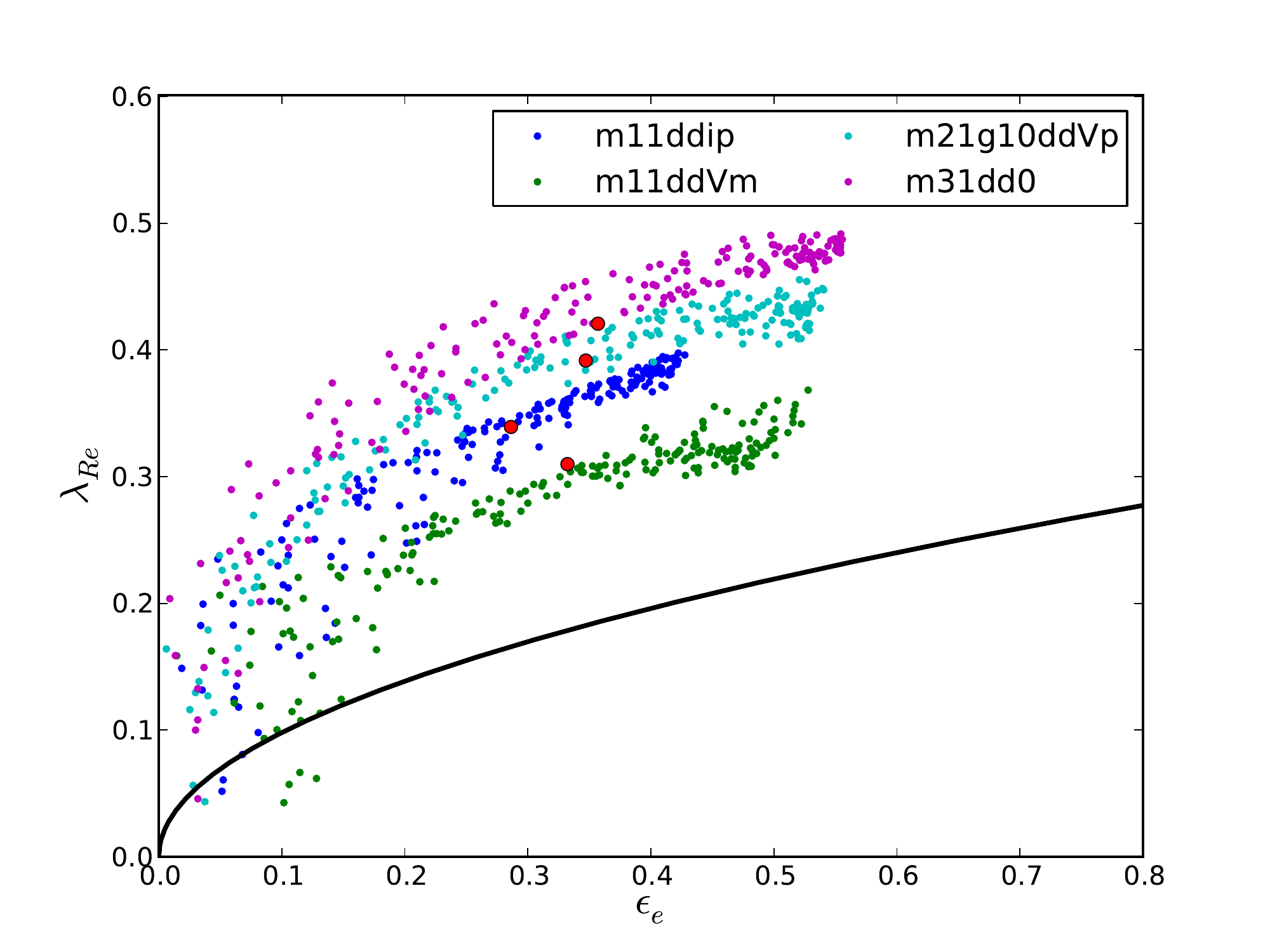} & \includegraphics[width=0.95\columnwidth]{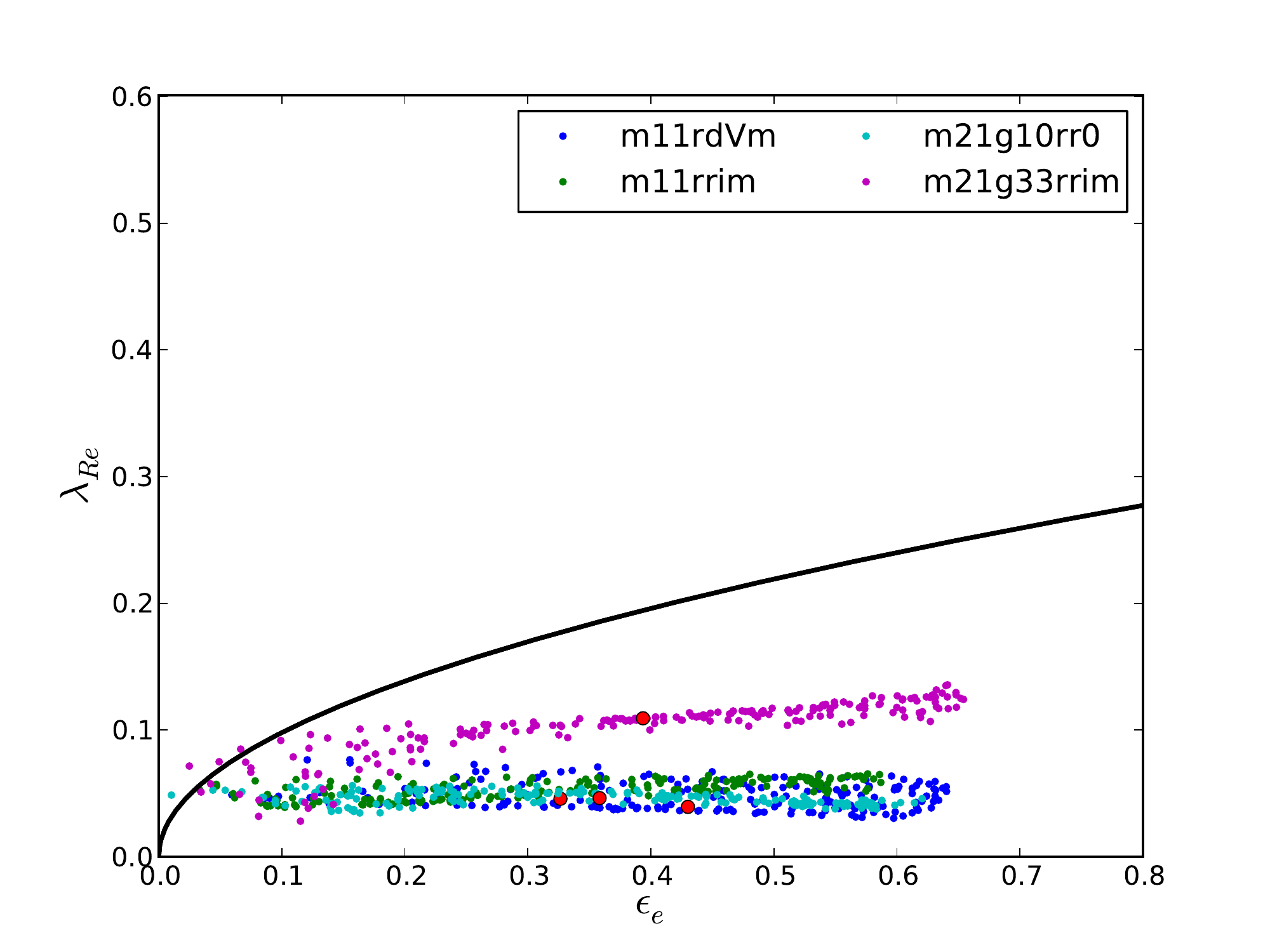} &\\
\includegraphics[width=0.77\columnwidth]{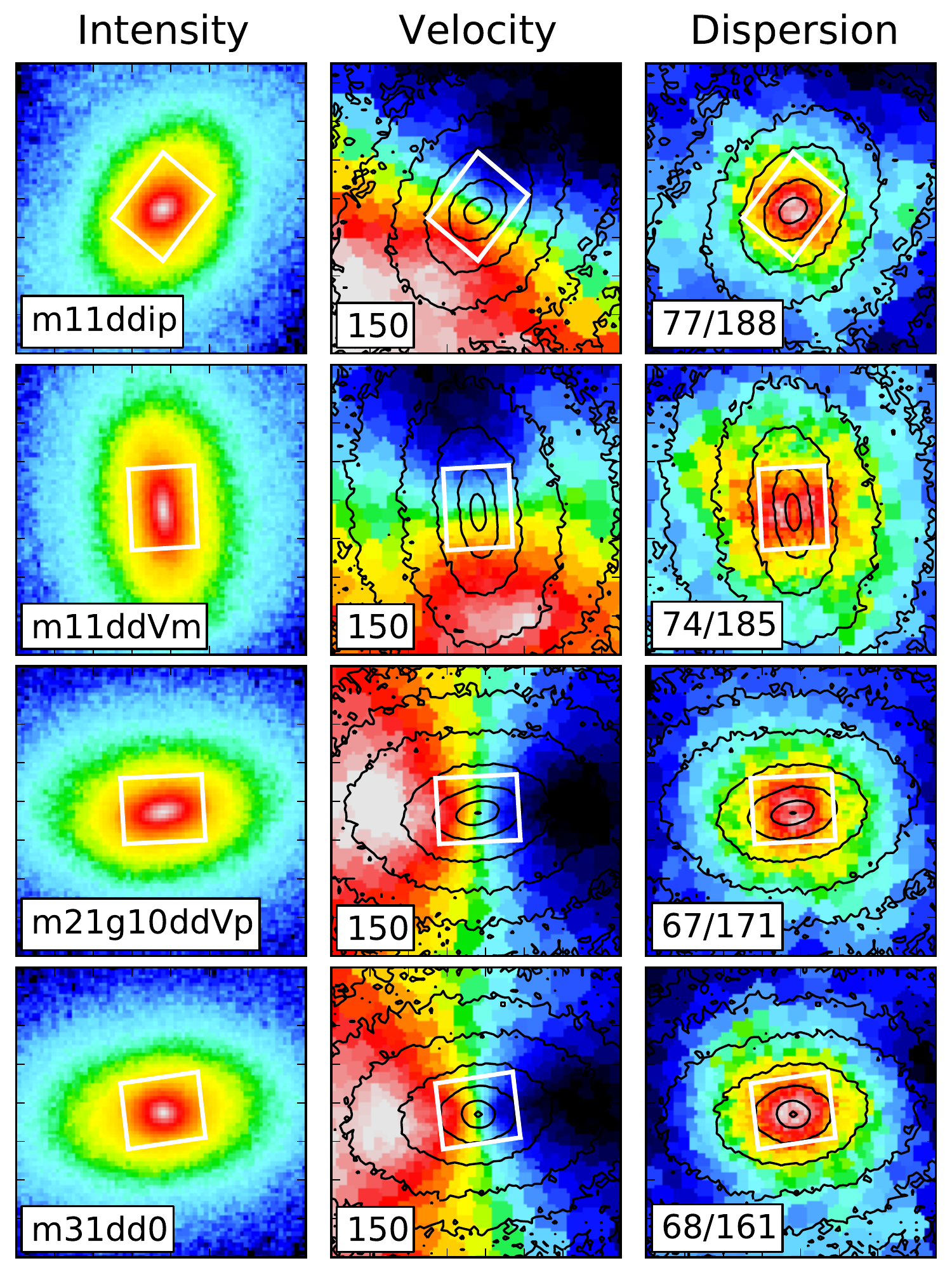} & \includegraphics[width=0.78\columnwidth]{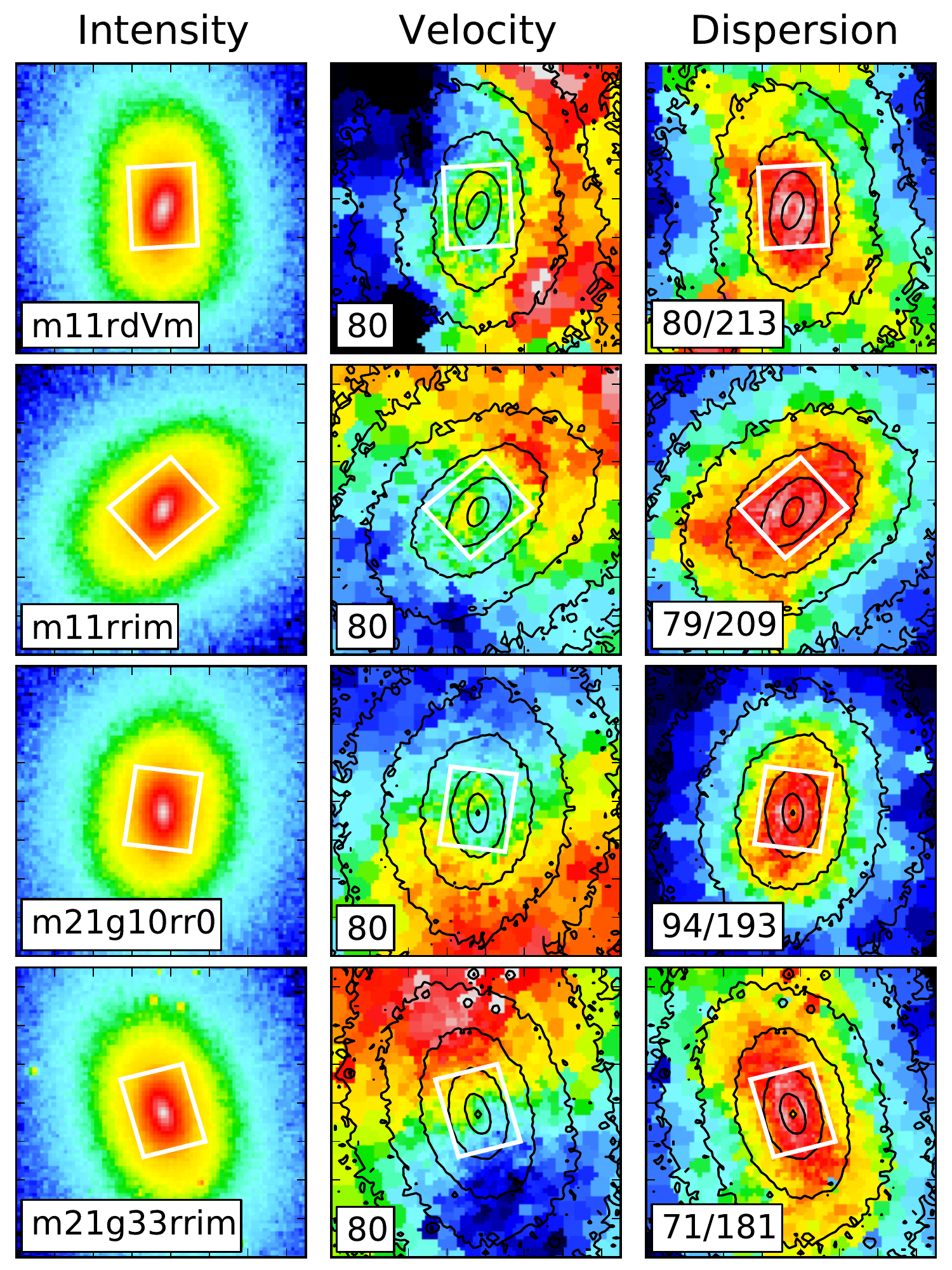} & \includegraphics[width=0.165\columnwidth]{pdffigures/proj_mergers/colorbar_only.pdf} \\
\end{tabular}
 \caption{\lreps{} for four fast rotators (left panels) and for four slow rotators with a KDC (right panels). The 200 projections of each merger are represented by a coloured point (see the legend in the figure for the name of the merger), the red point corresponds to the projection with an ellipticity closest to the mean ellipticity. The intensity, velocity and velocity dispersion fields of the previous red point projections are plotted below the \lreps{} diagram. The name of the merger is written in the intensity map, the maximum value of the velocity V$_{max}$ and the minimum/maximum values of the velocity dispersion are written in their respective maps. The colorbar goes from -V$_{max}$ to +V$_{max}$ and can also be used as an indicator for the intensity and the velocity dispersion ("-1" corresponding to the lower value, "1" to the maximum value). The field of view is $15\times15$~kpc$^2$ (15~kpc $\simeq$ 6$R_e$). The white rectangle indicates a typical field covered by the instrument \sauron{} and corresponds to a field of $41" \times 33"$ for a galaxy at a distance of 20~Mpc, its orientation follows the photometric position angle taken at 3$R_e$.}
 \label{fig:exslowfast}
\end{center}
\end{figure*}

\begin{figure*}
  \includegraphics[width=2\columnwidth]{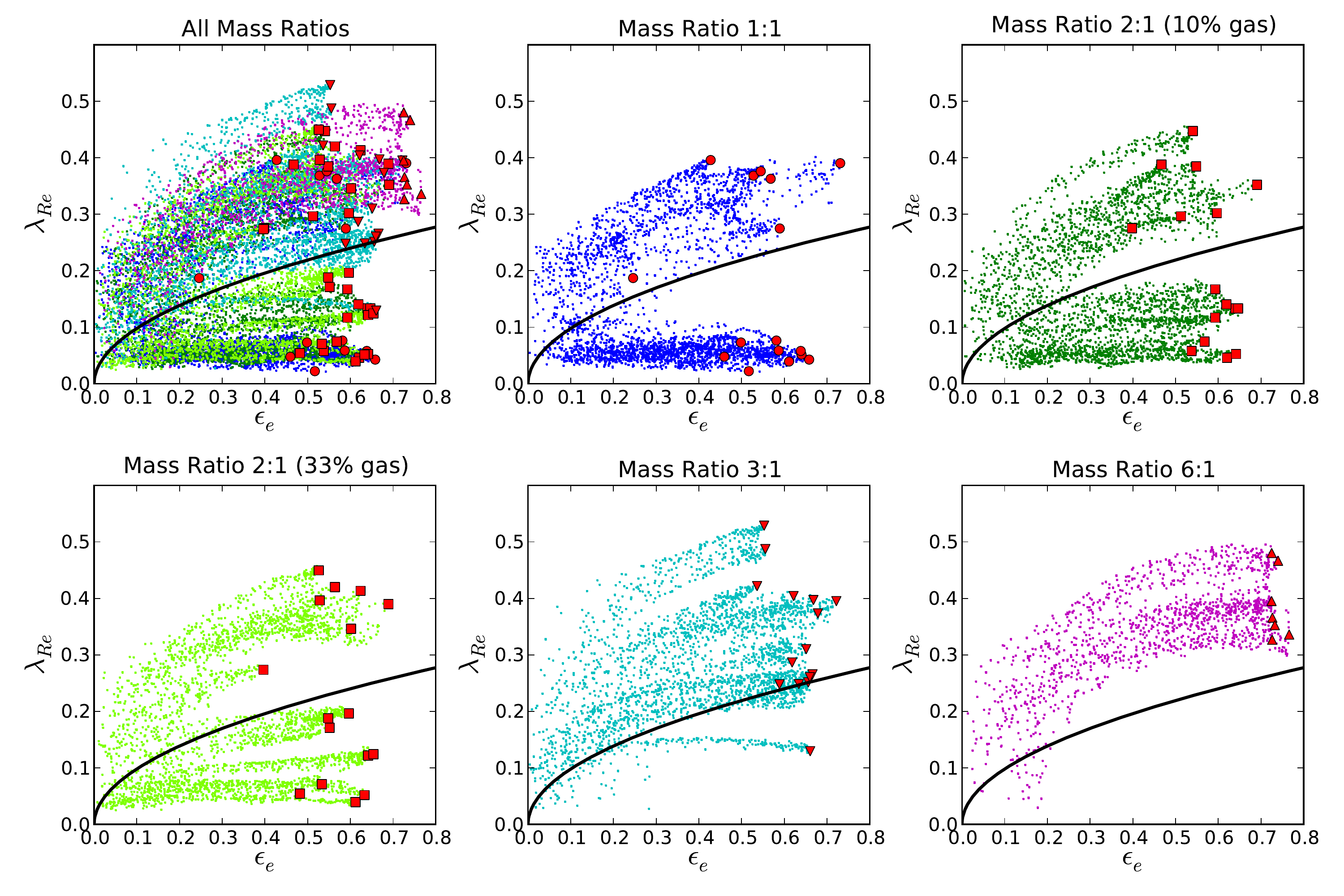}
  \caption{\lreps{} diagram for all simulations of binary mergers of disc galaxies. Top left: All projections of all mass ratios as in Fig.~\ref{fig:lrepsmergerall}, colours correspond to the different mass ratios. Top middle: All projections of mergers remnants for mass ratio 1:1. Top right: All projections of mergers remnants for mass ratio 2:1 with 10 per cent of gas. Bottom left: All projections of mergers remnants for mass ratio 2:1 with 33 per cent of gas. Bottom middle: All projections of mergers remnants for mass ratio 3:1. Bottom left: All projections of mergers remnants for mass ratio 6:1. The red symbols are for the projection which maximizes the ellipticity for a given remnant. The limit defining the slow/fast categories from \atlas{} is plotted as the solid black line.}
  \label{fig:lrepsmergermr}
\end{figure*}

The \lreps{} diagram is a useful tool to quantify the global properties of ETGs: it relates the angular momentum and flattening at one effective radius $R_e$, and disentangles fast and slow rotators. Fig.~\ref{fig:lrepsmergerall} shows this diagram for our complete sample of 70 binary disc merger simulations, including the 200 independent projections analysed for each relaxed merger remnant. The threshold defined by $\lambda_R = 0.31 \cdot \sqrt{\epsilon}$ separating slow and fast rotators as suggested by the \atlas{} observations is shown (Paper III). Each set of projections for a given simulation span a range of \lreps{} values: this is shown in Fig.~\ref{fig:lrepsmergerall} and individual examples are presented in Fig.~\ref{fig:exslowfast} as well as in Fig.~\ref{fig:lrepsmergermr} for various mass ratios. Fig.~\ref{fig:exslowfast} also shows examples of velocity moment maps for several fast and slow rotators formed in 1:1, 2:1 and 3:1 mergers.

The \lreps{} distribution for our entire sample of binary merger remnants is shown in Fig.~\ref{fig:lrepsmergerall} for galaxies with and without a KDC, respectively. The merger remnants can be seen in Appendix~\ref{sec:appprojmaps}, the colour code in the different panels presents in black the fast rotators, in red the slow rotators with KDC and in green the slow rotators without KDC. Fig.~\ref{fig:lrepsmergerall} shows evidence for a tight relation between the presence of a KDC and the slow versus fast rotator classification. All fast rotators presented in Fig.~\ref{fig:exslowfast} show a clear and regular rotation pattern at all radii, and most slow rotators exhibit KDCs. 

\medskip

We can then sketch the following general results regarding the formation of fast and slow rotators and the presence of a KDC from our sample of binary disc mergers:
\begin{itemize}
   \item We find both fast and slow rotators in our sample of relaxed remnants of binary disc mergers. An {\em apparent} bimodality is observed in the global \lreps{} distribution of the sample and in the presence of a KDC in the remnants. This could likely result from the specific choices of simulated mass ratios and the limited number of simulated incoming orbits.

  \item In our sample of merger remnants, fast rotators all have aligned kinematic and photometric axes, while slow rotators frequently exhibit kinematic misalignments (see Section~\ref{sec:alignobs} and examples in Fig.~\ref{fig:exslowfast}), and obviously weak rotational support. This is similar to what is observed in the \atlas{} sample (Paper~II). This suggests that slow rotators are not velocity-scaled down versions of members of the fast rotator population.

 \item Slow rotators remain slow for nearly all viewing angles, and fast rotators remain fast. Even nearly face-on projections of fast rotators are rarely classified as ``slow''. This result was already pointed out by \cite{jesseitlr} while using a constant threshold for $\lambda_R$ between slow and fast rotators as in E+07. The simulations allow here to further explore various projections of a given merger remnant: they confirm that the combination of the $\lambda_R$ and $\epsilon$ parameters, together with the refined definition for fast and slow rotators recently proposed in Paper III (see Section~\ref{sec:analysis}) allow a robust classification of ETGs, which is almost independent of the viewing angles.

 \item Another intrinsic difference between fast and slow rotators is that none of the fast rotators harbours a KDC, while the majority of slow rotators do harbour a clear KDC.  Only for a few slow rotators we could not detect a clear KDC (with sufficient rotation amplitude or misalignment angle). This strong trend brings further support to the relevance of the definition of fast and slow rotators as emphasising two families with different intrinsic properties. With the low noise level and very high velocity resolution of our simulations, we notice that KDCs in slow rotators are visible for the vast majority of projections, but sometimes with rather low velocity amplitudes which would be hard to detect in observations.

  \item Slow rotators have relatively high edge-on ellipticities (0.45 to 0.65), with the ones for fast rotators being only slightly larger (generally 0.5 to 0.75). The mean ellipticities over random projections show a somewhat more significant though still mild difference, being around 0.35 for slow rotators and 0.45 for fast rotators. The fast rotators with the highest edge-on ellipticities ($\epsilon > 0.7$) are remnants of some of the less violent 3:1--6:1 mergers; those which tend to appear rounder formed with high orbital inclinations. We also note that the majority of fast rotators have a bar, and 6:1 merger remnants often have weak spiral arms associated with the bars.

  \item  The typical $\lambda_R$ profile of a fast rotator (see Fig. \ref{fig:Vmerg11} to \ref{fig:Vmerg61} in the Appendix) increases rapidly within 1 to 2 effective radii, and then exhibits a shallower positive slope. As for slow rotators, we observe two general types of $\lambda_R$ profiles. (1) It can remain constant with a low value within 1.5 or 2~$R_e$ and then starts to slightly increase outwards or (2) $\lambda_R$ shows a local maximum around 0.5~$R_e$, decreasing from there to $\sim 1 R_e$, with a subsequent outward increase: such a $\lambda_R$ profile is the clear signature of large-scale KDC as mentioned in E+07 \citep[see also][]{mcdermid06}. In both cases, $\lambda_R$ never reaches the level of fast rotators at very large radii: the mean value of $\lambda_R$ over the edge-on projections at $3R_e$ is 0.6 for the fast rotators and 0.25 for the slow rotators. The radial $\lambda_R$ profiles of slow and fast rotators are therefore significantly different both at small and large radii. 
\end{itemize}

The definition of slow and fast rotators seems to robustly disentangle two families of ETGs with different intrinsic properties, with \textit{e.g.}, the formation of KDCs in slow rotators only. To further understand the influence of various initial parameters, we now explore the impact of the mass ratio, orbital parameters, gas fraction and nature of the spiral progenitors, on the properties of merger remnants.

   \subsection{Influence of the mass ratio \label{sec:binmr}}
Fig.~\ref{fig:lrepsmergermr} shows the \lreps{} diagram for different mass ratios, and different gas fractions in the cases of 2:1 mergers. 1:1 and 2:1 mergers form slow rotators in about 60 per cent of our simulations, the lowest $\lambda_R$ systems being formed in equal-mass mergers (the fraction of slow rotators directly reflecting the initial orbital parameters, see Section~\ref{sec:binic} and Fig.~\ref{fig:mergerIC}). The formation of slow rotators is unlikely in 3:1 mergers, and does not occur in 6:1 mergers. This is consistent with the conclusions of previous samples of merger simulations \citep[\textit{e.g.}][]{naab03,bour04, jesseitlr}. \citet{bour04} showed that spheroid dominated galaxies with little rotating disc components could be formed only during the so-called ``major'' mergers (1:1--3:1), while ``intermediate'' mergers with mass ratios between 4:1 and 10:1 did not perturb the galaxies much from the initial disc-like morphology and kinematics. \citet{naab03} found that mass ratios of 3:1 and above formed discy systems. Our present conclusions therefore confirm these results, but with a larger and higher-resolution sample of simulations and new analysis parameters.

   \subsection{Influence of the gas fraction \label{sec:bingas}}
2:1 mergers were simulated with gas fractions of 10 and 33 per cent. Both sub-samples show similar distributions of $\lambda_R$, $\epsilon$, and presence of a KDC. The gas fraction thus seems to have no major impact on the global properties of the merger remnants, although we did not explore the context of very high-redshift mergers ($z \sim 2$ and above) which could involve extreme gas fractions, e.g., above 50 per cent \citep{daddi10, tacconi10}. Mergers with very low gas fractions ($f_{gas} \sim 0$ per cent) should not be associated with spiral-like progenitors, which is the topic of the present section, but with ETG-like progenitors (see Section~\ref{sec:remergers}).

A possible cause for the observed similarities between the remnants with 10\% and 33\% is early gas consumption in the spiral progenitors. If the spirals with 33\% of their mass in gas transform most of this gas into stars before the coalescence, this could potentially lead to an actual merger with roughly 10\% of gas. We thus investigated how gas consumption proceeds in the progenitors before the merger for both samples (10\% and 33\% of gas). We first check whether or not the gas consumption is altered by differences in the duration of the merger in the two samples. This is not the case as only the gas fraction is varied, not the orbits: for both samples, the mean time t$_{coal}$ from the start of the simulation to the coalescence is $t_{coal} \simeq 7 \times t_{dyn}$ with the dynamical time $t_{dyn} \sim$ 100~Myr. We then explore the consumption of the gas before the passage at the first pericentre, \textit{i.e.} when the gravitational force did not have time yet to strongly perturb the progenitors. During that period, the spirals with 10\% and 33\% of their baryonic mass in gas respectively consume on average 11\% and 14\% (with a maximum of 17\%) of their gas (representing $\sim 1.1$\% and 4.6\% with a maximum of 5.6\% of the total initial baryonic mass). The available gas after the pericentre is then, on average, $\sim 9$\% and $\sim 28$\% for spirals with initial gas fractions of 10 and 33\%. The spiral progenitors are thus still very different when the merger occurs and the similarities between the two sub-samples cannot be attributed to a different consumption of gas before the merger.

One could a priori expect that a relatively high gas fraction eases the immediate re-building of massive rotating disc components during or soon after the merger. A few studies suggested that gas-rich mergers could form discy fast rotating systems, or even spiral galaxies \citep[\textit{e.g.},][]{SH05,rob06,H09}. The relatively stable morphology and kinematics of the 2:1 merger remnants with 10 and 33 per cent of gas could a priori seem conflicting with these earlier results: the value of $\lambda_R$ at 1~$R_e$ averaged over all 2:1 mergers varies only from 0.17 to 0.19 when the gas fraction is increased from 10 to 33 per cent and the average value of the ellipticity $\epsilon$ varies from 0.39 to 0.38. Actually, a separate study by \citet{bournaud-mergers-2010} shows that a high gas fraction does not easily reform massive and extended disc components in major mergers. This occurs provided that most of the ISM is modelled as a cold medium supported by supersonic turbulence, which is the case in our present sticky particle simulations\footnote{A difference being that turbulence dissipation in local shocks is ``sub-grid'' in sticky-particle models, while it is explicitly captured in high-resolution AMR models.} and in the AMR models of \citet{bournaud-mergers-2010}, rather than being supported by a high thermal pressure as it is often the case in existing SPH simulations. This probably explains the differences between our results and the published works mentioned above.

   \subsection{Influence of orbital parameters \label{sec:binic}}
We now examine, for mergers with mass ratios 1:1 and 2:1 which form the majority of slow rotators, how the orbital parameters influence or determine the slow/fast nature of the merger remnant and the associated presence/absence of a KDC.



\begin{figure}
  \includegraphics[width=\columnwidth]{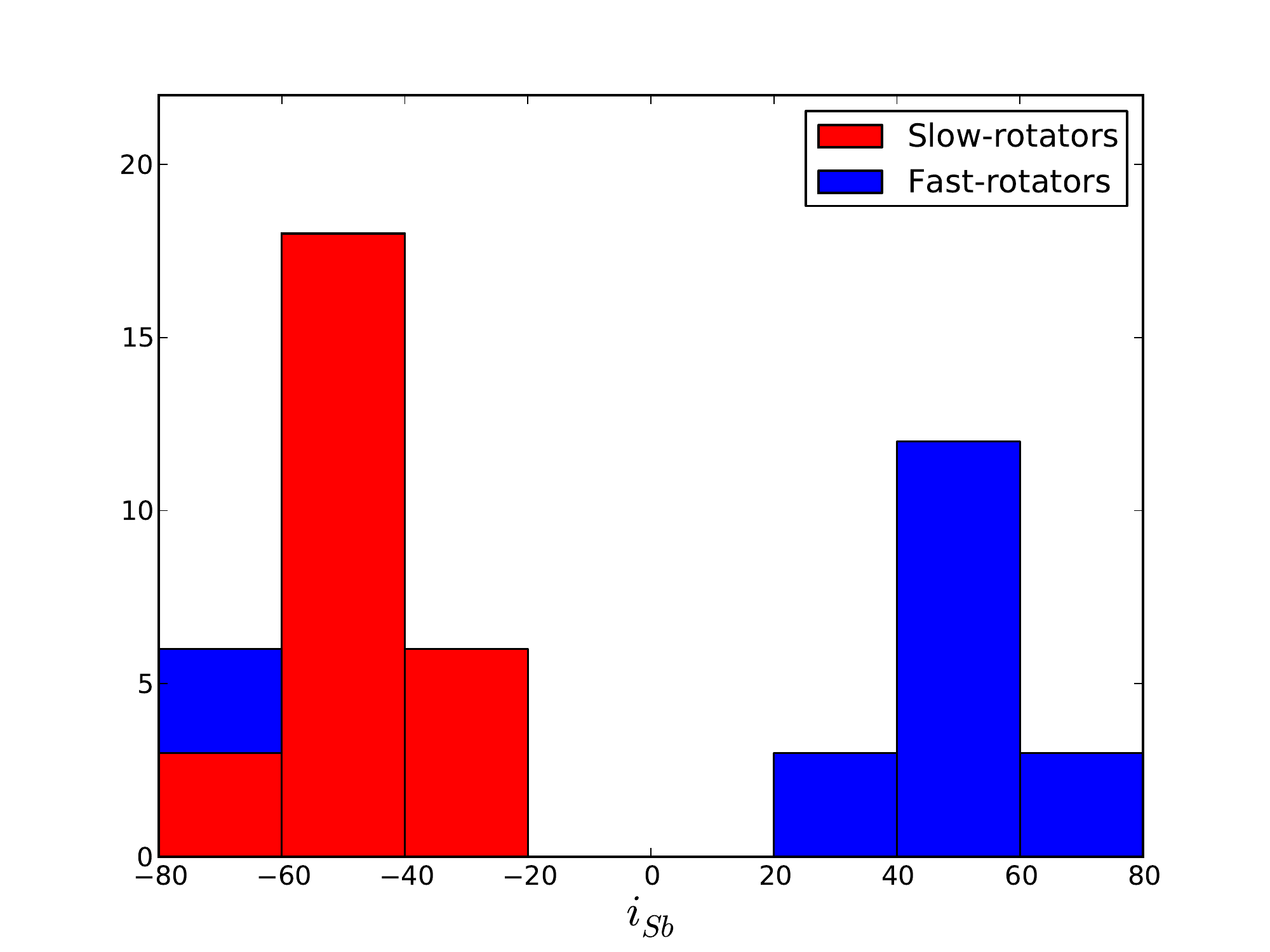} \\
  \caption{Fraction of slow/fast rotators as a function of the inclination of the main progenitor with respect to the merger orbital plane for mass ratio 1:1, 2:1g10, 2:1g33. The negative inclination occurs when the spin of the Sb progenitor and the spin of the orbital angular momentum are anti-parallel, so that -25$^{\circ}$ and 25$^{\circ}$ represent the same inclination for the galaxy but on retrograde orbit.}
  \label{fig:mergerIC}
\end{figure}

Fig.~\ref{fig:mergerIC} shows the distribution of fast and slow rotators as a function of the orbital inclination/orientation with respect to the Sb progenitor galaxy for the 1:1 and 2:1 merger remnants. We remind the reader that the Sb progenitor is the ``main'' (most massive) progenitor in the 2:1 mergers. Mergers with a direct (prograde) orbit with respect to the Sb progenitor all produce fast rotators. Mergers with a retrograde orbit with respect to the Sb progenitor almost exclusively produce slow rotators regardless of the spin of the Sc galaxy. The formation of a slow rotator may then be a likely event in a 1:1 or 2:1 merger of spiral galaxies: if the main requirement is to have a retrograde orbit for the main progenitor, the relative fraction of slow and fast rotators should directly reflect the spin distribution of the progenitors (with respect to the orbit, Fig.~\ref{fig:lrepsmergermr}).

\begin{figure}
  \includegraphics[width=\columnwidth]{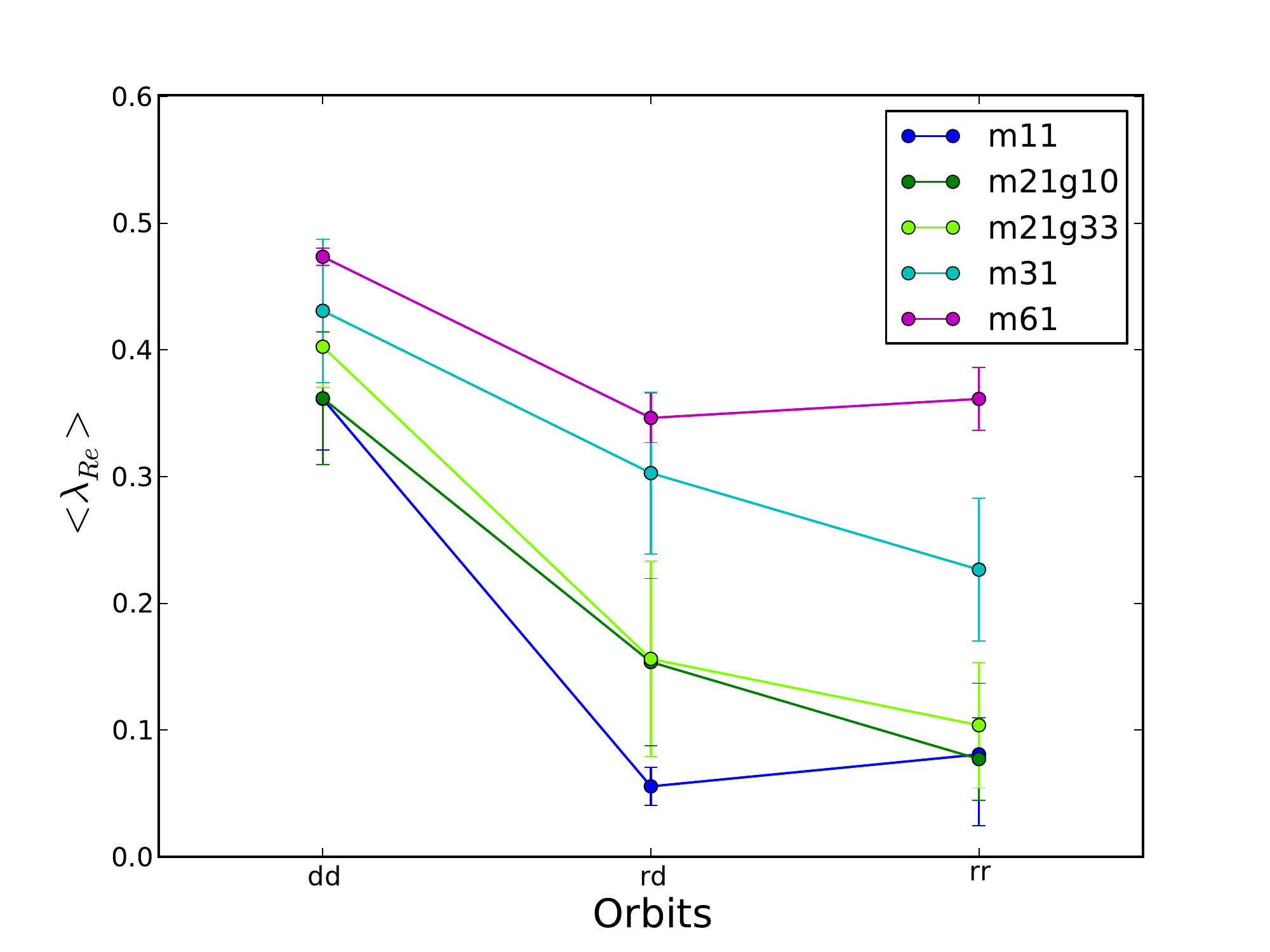} \\
  \caption{Average $\lambda_R$ as a function of the spin of the progenitors for the different mass ratios.}
  \label{fig:mergerIC2}
\end{figure}

These conclusions are further assessed by Fig.~\ref{fig:mergerIC2}: it presents the average $\lambda_R$ value as a function of the orbital orientation for both progenitor galaxies (\textit{i.e.}, \textbf{dd}, \textbf{rd} and \textbf{rr} orbits) for all mass ratios. It shows that, while \textbf{rr} orbits result in a somewhat lower angular momentum at 1~$R_e$ than \textbf{rd} orbits (or similar for 1:1 and 6:1 mass ratios), most of the angular momentum removal is already obtained for \textbf{rd} orbits, where the orbit is \textbf{r}etrograde for the main Sb progenitor but \textbf{d}irect (prograde) for the Sc companion.

Depending on the initial orbital parameters of the system, the redistribution of angular momentum (orbital momentum plus the internal momenta of each progenitor) via violent relaxation occurs differently. Ideally, we could derive a function describing how this redistribution proceeds in a binary galaxy merger, taking into account the various ingredients of that merger (orbital and internal spins, gas fractions, Hubble type of the progenitors, etc). This would have predictive power in terms of the dynamical status of the remnant (slow or fast rotators) and would be a major asset for implementation in \textit{e.g.}, semi-analytic models of galaxy formation. Considering that the number of initial parameters is vast, one should need more simulations to constrain and determine such function. This is clearly beyond the scope of the present paper.

   \subsection{Influence of the Hubble type of the progenitor spiral galaxies \label{sec:hubbletype}}
\begin{figure}
  \includegraphics[width=\columnwidth]{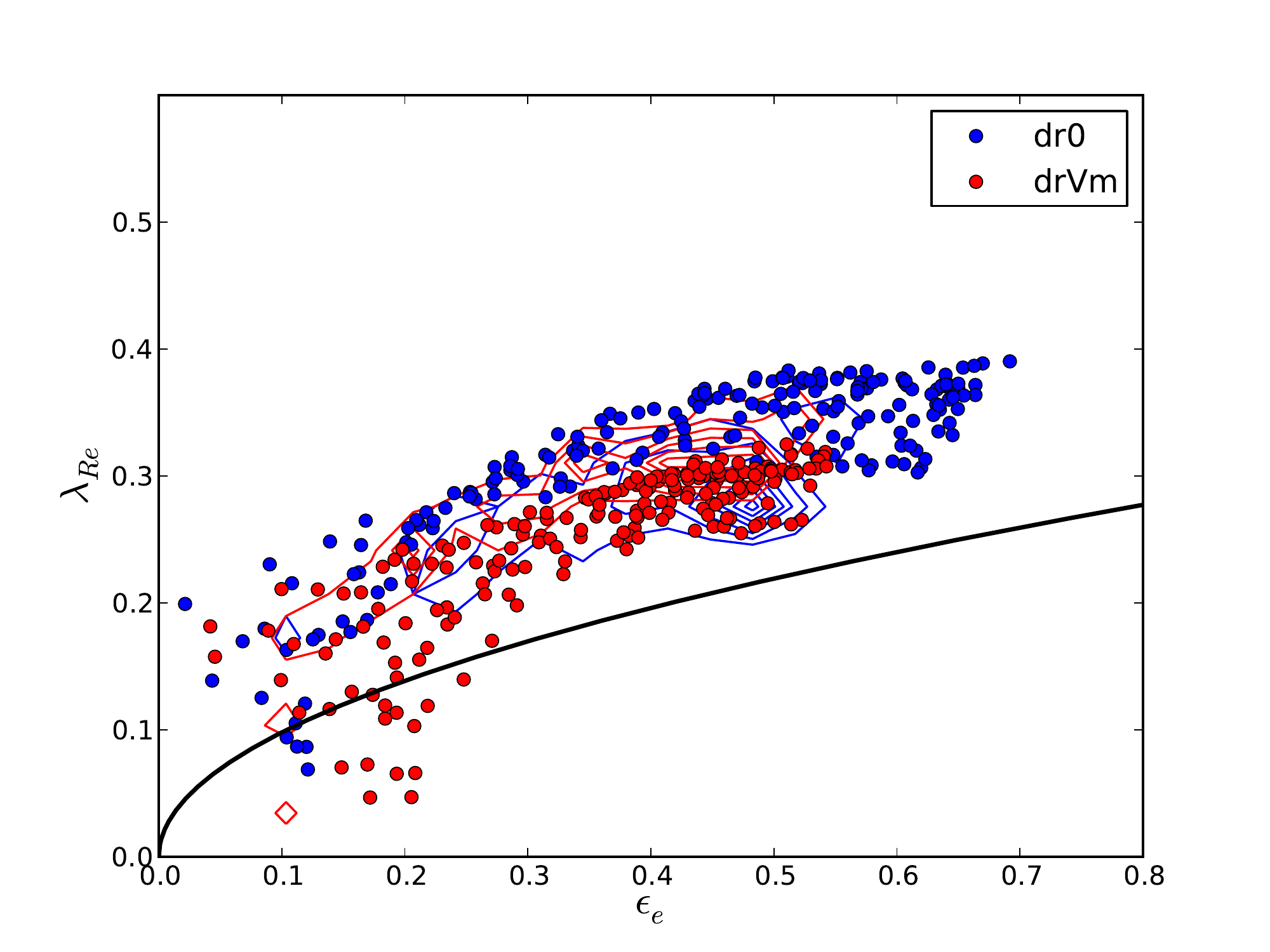} \\
  \caption{\lreps{} diagram for the \textit{m11dr0} (in blue) and for the \textit{m11drVm} (in red). The contours represent the distribution of the 200 projections of the \textit{m11dd0} (in blue) and \textit{m11ddVm} (in red) simulations.}
  \label{fig:drmergers}
\end{figure}

We noted above from Fig.~\ref{fig:mergerIC} that having the main progenitor prograde or retrograde largely determines the fast or slow rotator nature of the merger remnant, while the orbit orientation with respect to the companion has a weaker effect. This is expected for unequal mass mergers, and in particular for 6:1 mass ratios, where the companion mass and momentum are naturally much lower than those of the main progenitor. This is however more surprising for 1:1 mergers. Indeed, these equal-mass mergers end-up with a relatively similar ${\lambda}_R$ as long as the Sb progenitor has a retrograde orientation, regardless of the orientation of the Sc progenitor.

This could be the consequence of a ``saturation'' effect: the final angular momentum of the merger remnant could be dominated in the centre by one retrograde galaxy with the other (retrograde) galaxy contributing very little. However, the central properties of merger remnants which exhibit relatively high angular momentum obtained via a prograde orientation of the Sb progenitor show little variation whether the Sc progenitor is on a prograde or retrograde orientation. This suggests that the Sb and Sc progenitors do not have symmetric roles in determining the properties of the merger remnant. We first examined whether or not this is an artefact from the initial positioning of the two spiral galaxies in the simulation box and/or the initial orientation of their spin axis with respect to the numerical grid. To test this, we reversed the initial positioning of the Sb and Sc progenitors and changed the grid orientation in a re-simulation of the \textit{m11rd0} merger: we observe no significant difference in the merger remnants. This confirms that our results do not depend on the simulation box size and grid orientation.

We then simulated mergers in which the Sb progenitor is on a direct orbital orientation, and the Sc progenitor is on a retrograde orientation, \textit{i.e.} {\bf dr} mergers --~so far our sample comprised only {\bf rd} mergers with a retrograde Sb and a direct Sc progenitor. Given that the \textit{m11rd0} and \textit{m11rdVm} orbits resulted in quite typical slow rotators, we performed the corresponding \textit{m11dr0} and \textit{m11drVm} simulations. The \lreps{} profiles are shown on Fig.~\ref{fig:drmergers}. The \textit{m11dr0} and \textit{m11drVm} remnants are fast rotators, relatively typical compared to the fast rotators produced on {\bf dd} orbits. The outputs of \textbf{dr} and \textbf{rd} orbits are therefore widely different. This shows that while the angular momentum of a merger remnant largely depends on the total available (baryonic) angular momentum, \textit{i.e.} the sum of the orbital angular momentum and the internal spin of each progenitor galaxy, the Hubble type of the progenitor galaxies --~\textit{i.e} their central concentration and dynamical stability~-- has a significant impact. A retrograde orbit around an early-type spiral, such as our so-called Sb galaxy, is efficient in producing slow rotators (for 1:1 and 2:1 mergers). A retrograde orbit around a late-type spiral, such as our so-called Sc galaxy, does not significantly impact the angular momentum remaining in the central body of the merger remnant (see Fig.~\ref{fig:drmergers} and Section \ref{sec:hubexplan} for the interpretation).

\section{Galaxy remergers}
\label{sec:remergers}

\begin{table}
\begin{center}
\caption{Details of the characteristics of the remergers: (1) Spin of the progenitors with respect to the orbital angular momentum, (2) Galaxy remnants of binary merger of spirals as progenitors for the remerger, (3) Slow/Fast classification from Section~\ref{sec:binresults}} \label{tab:propremergers}
 \begin{tabular}{|c|c|c|c|}
\hline 
Name  & Orbit$^{(1)}$ & Progenitors$^{(2)}$ & Prog Type$^{(3)}$ \\ 
\hline
rem2x11    & dd & rd0 - rr0   & Slow - Slow \\
rem2x11    & dr & rrim - ddip & Slow - Fast \\
rem2x11    & rd & rrVp - ddVm & Slow - Fast \\
rem2x11    & rr & rd0 - dd0   & Slow - Fast \\
\hline
rem2x21g10 & dd & rd0 - rr0   & Slow - Slow \\
rem2x21g10 & dr & rrim - ddip & Slow - Fast \\
rem2x21g10 & rd & rrVp - ddVm & Slow - Fast \\
rem2x21g10 & rr & rd0 - dd0   & Slow - Fast \\
\hline
rem2x21g33 & dd & rd0 - rr0   & Slow - Slow \\
rem2x21g33 & dr & rrim - ddip & Slow - Fast \\
rem2x21g33 & rd & rrVp - ddVm & Slow - Fast \\
rem2x21g33 & rr & rd0 - dd0   & Slow - Fast \\
\hline
rem21g10+S & dd & rr0 - S     & Slow - Fast \\
rem21g10+S & dr & rdim - S    & Slow - Fast \\
rem21g10+S & rd & rdVp - S    & Slow - Fast \\
rem21g10+S & rr & dd0 - S     & Fast - Fast \\
\hline
 \end{tabular}
\end{center}
\end{table}

\begin{figure}
 \includegraphics[width=\columnwidth]{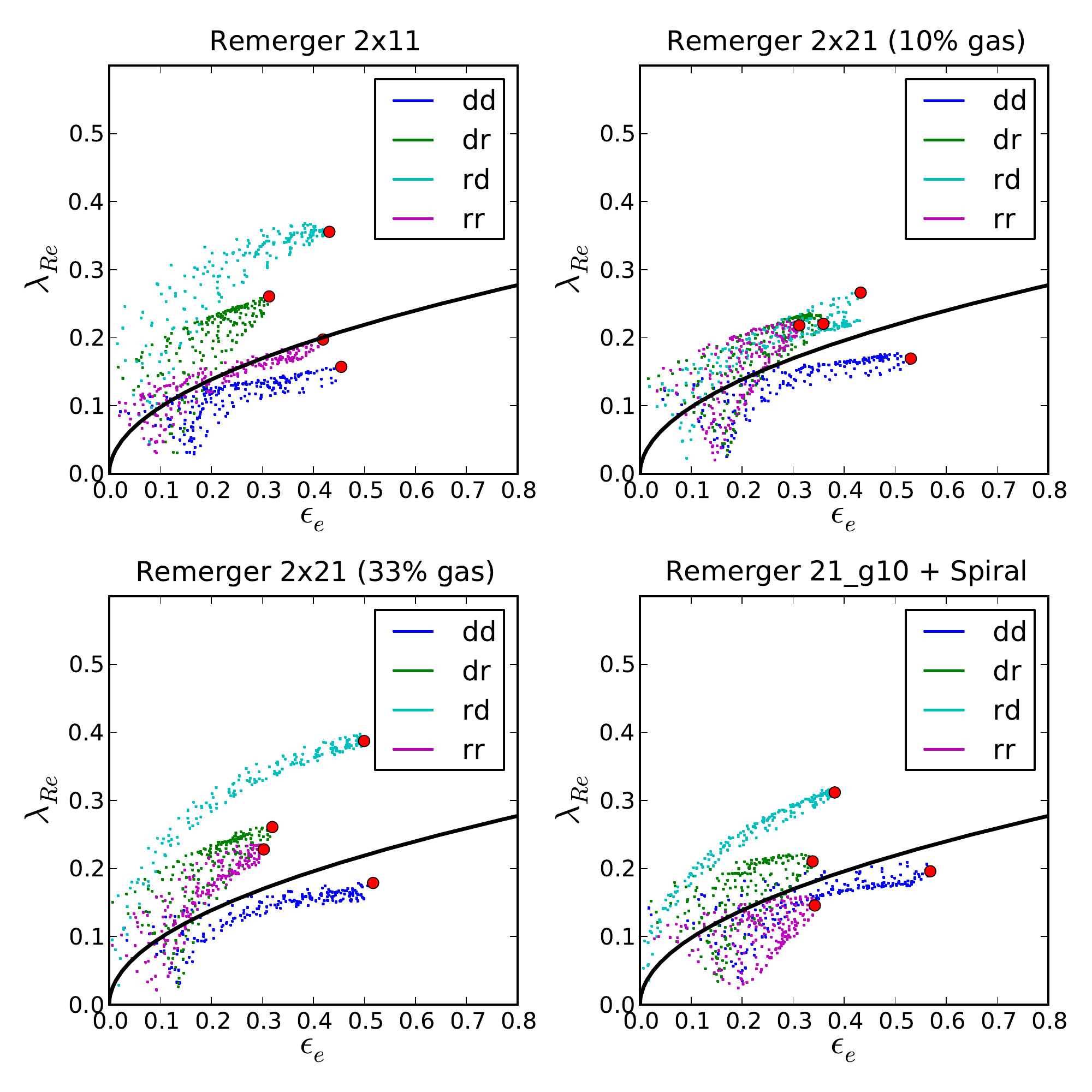} \\
 \caption{\lreps{} diagram for all projections for all simulations of remergers. The 4 panels represents the 4 different remergers, the four colours in each subplot correspond to the orbit of remerging (\textbf{dd}, \textbf{dr}, \textbf{rd}, \textbf{rr})}
 \label{fig:lrepsremerger}
\end{figure}

\citet{NKB06} suggested that mergers of ETGs play an important role in the assembly of massive galaxies. \citet{khochfar03} have also shown, via simulations in a cosmological context, that the last major merger of bright present-day ETGs (M$_B \lesssim 21$) was preferentially between bulge-dominated galaxies, while those with M$_B \simeq 20$ have mainly experienced last major mergers between a bulge-dominated and a disk-dominated galaxy. This Section thus introduces binary galaxy mergers of ETGs, or "remergers", with the goal to examine the morphology and kinematics of such remnants. The progenitors we used for this analysis are themselves the remnants of binary disc mergers described in the previous Sections. 

Table~\ref{tab:propremergers} details the remergers we have performed. We run simulations for four pairs of progenitors with different orbital configurations (\textbf{dd}, \textbf{dr}, \textbf{rd}, \textbf{rr}). The first three pairs have been drawn from simulations with two initial mass ratios and two initial gas fractions, namely m11, m21g10 and m21g33. In these three cases, the orbital configurations that generated the progenitors were kept the same for a given remerger configuration, \textit{e.g} for a \textbf{dd} remerger orbit we always use progenitors coming from rd0 (primary) and rr0 (secondary) orbits. The progenitors for the fourth pair (fourth remerger set) are m21g10 and a spiral. Velocity maps of all remerger remnants are shown in Fig.~\ref{fig:Vremergers} to \ref{fig:Vrem+s} in the Appendix.

Fig.~\ref{fig:lrepsremerger} presents the \lreps{} diagram for the remergers, the four subplots correspond to the four different types of remergers (rem2x11, rem2x21g10, rem2x21g33, rem21g10+S) and the four colours in each subplots corresponding to the orbit of remerging (\textbf{dd}, \textbf{dr}, \textbf{rd}, \textbf{rr}). From the results of binary galaxy disc mergers (see previous Sections), we could presume that the \textbf{rr} orbit would produce the slowest rotators of this sample of remergers. In fact, the \textbf{dd} cases are the ones to produce the slowest systems for the four different types of remergers. In the \textbf{dd} cases of galaxy remergers, both progenitors are slow rotators and hold a KDC. The progenitors acquire angular momentum from the orbit, resulting in merged galaxies with large-scale rotation, and the KDCs have been destroyed. The remnant is classified as a slow rotator but is very close to the dividing line separating slow and fast rotators. The remnant does not present any sign of a KDC and has its photometric and kinematic axis aligned. 

In the other cases (\textbf{dr}, \textbf{rd}, \textbf{rr}), the main progenitor is a slow rotator with a KDC and the companion is a fast rotator. During the merger, the main progenitor --~with no global rotation~-- acquires angular momentum from the orbit while the companion --~similar to the behaviour of the spiral Sb progenitor~-- keeps the initial orientation of its spin, the contribution of which determines and dominates the final spin of the remnant. None of these remergers exhibit a stellar KDC: the KDCs which were present in the progenitors have been destroyed and none are created during the merger. The final remnants of these remergers are rounder than binary disc merger remnants and all have regular kinematics. As mentioned, a few of these (mostly the \textbf{dd}) lie just below the limit between the slow and fast rotators, emphasising the potential small overlap between the two families (Paper III).

As seen in our sample of galaxy remergers, and in agreement with the results of \citet{dimatteo3}, slow or non rotating galaxies can gain central angular momentum through mergers via a transfer of the orbital angular momentum and start to rotate. The final state of a merger remnant is the combination of internal+orbital angular momentum: starting with one or two slow rotators can thus leads to a rotating merger remnant. To form a round slow rotator via ETG mergers, one should need two fast rotators on a favourable (\textit{e.g} \textbf{rr}) orbit. Repeated minor mergers may preserve the initial KDC of the initial ETG and heat the external parts of the galaxy \citep{bour07,qu10}: this could end up with a round slow rotator, but this scenario has not been tested yet.

\section{Properties and formation mechanisms for KDC via major mergers} 
\label{sec:kdc}

In this section, we focus on the processes involved in the formation of a KDC in our simulations, and we thus examine the relative contribution of the different components, namely the old stars, the young stars and the gas associated with the KDC observed in the merger remnant.

   \subsection{The contribution from old stars \label{sec:kdc2types}}
\begin{figure*}
  \includegraphics[width=2\columnwidth]{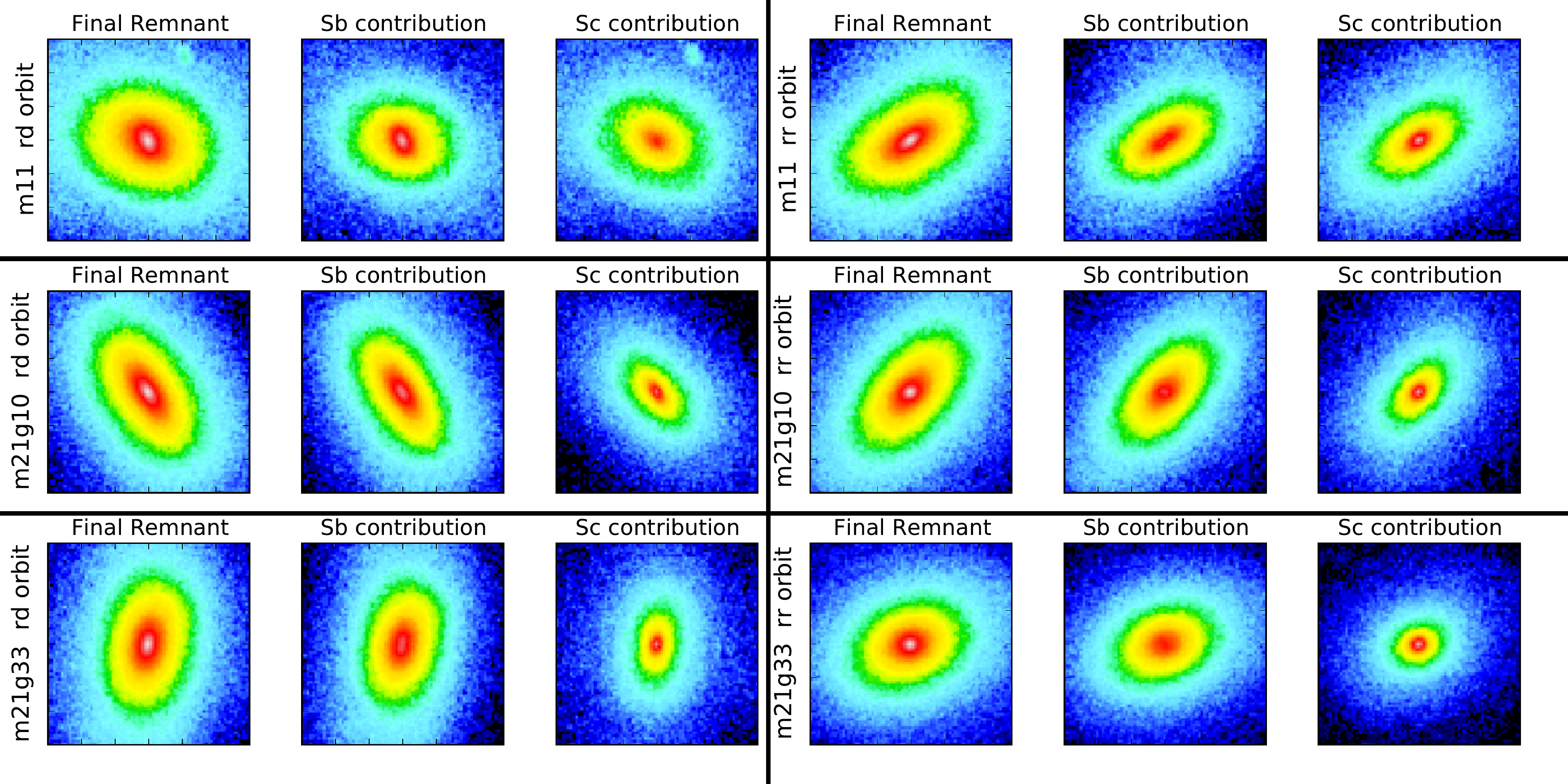} \\
  \hspace*{\fill} \\
  \hspace*{\fill} \\
  \includegraphics[width=2\columnwidth]{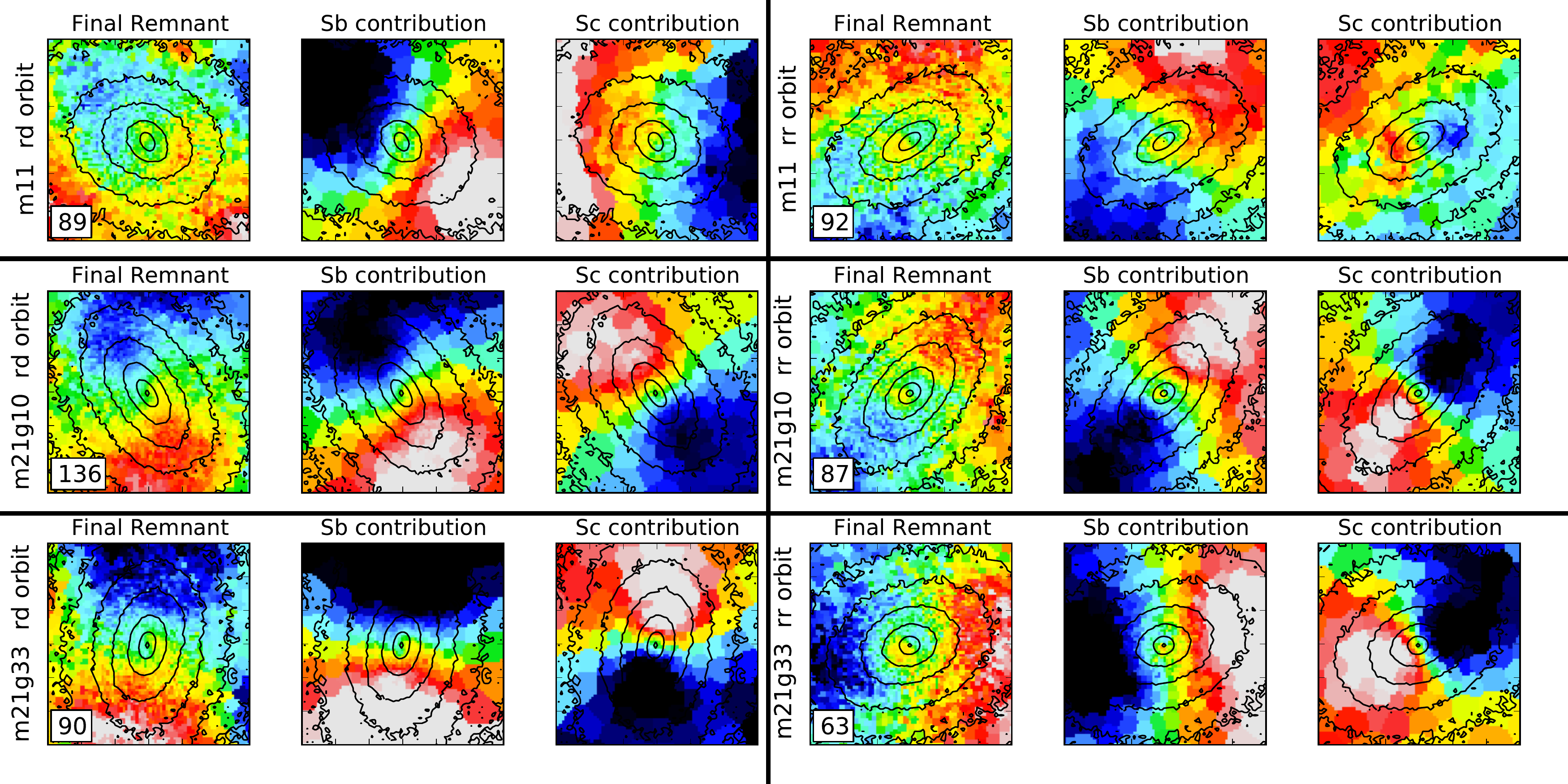} \\
  \caption{Intensity (three top panels) and velocity (three bottom panels) fields of the old stars for the final merger remnant, for the stars belonging to the Sb progenitor in the remnant, for the stars belonging to the Sc companion in the remnant. The cuts in magnitude and velocity are computed from the Final remnant and applied to the Sb and Sc progenitor contributions. The field of view for each panel is $15\times15$~kpc$^2$.}
  \label{fig:kdc2types}
\end{figure*}

To probe the contribution of the old stars in the slow rotator remnants with a KDC, we have separated the old stars which belong to the Sb progenitor from the ones of the Sc companion. The intensity and velocity fields of six analysed cases (m11, m21g10, m21g33 with \textbf{rd} orbit and m11, m21g10, m21g33 with \textbf{rr} orbit) are shown in Fig.~\ref{fig:kdc2types}. The projections used in that figure have been selected to emphasise the respective contributions of the progenitors in the final remnant and do not  specifically correspond to edge-on or face-on views of the galaxy.

There seem to be two qualitatively different types of KDCs which can in fact be associated with different initial orbits. In the \textbf{rd} cases, a KDC is visible within 1~$R_e$ and results from the luminosity weighted average of two counter-rotating stellar systems. In the \textbf{rr} orbits, the KDC is more prominent and extended, and is visible in the {\em individual} contributions of both progenitors, at least for the 1:1 mergers. This can be understood by following the Sb and Sc progenitor in turn.

        \paragraph*{The Sb progenitor} remains, within the central region, mostly unaffected by the merger, even for large mass ratios (1:1). This is true except at $R~<~R_e$ for the \textbf{rr} orbit: when the Sc companion has a negative spin (with respect to the orbit) the Sb progenitor is more severely affected. This is consistent with the picture developed by \citet{renaud} which suggests that retrograde galaxy orbits allow more material to be available for later interactions when galaxies get closer to each others.

        \paragraph*{The Sc progenitor} is, in stark contrast with the Sb progenitor, almost entirely disrupted during the merging event and will basically adopt part of the orbital angular momentum (the rest being transferred to the dark matter component) and its sign. The final contribution of the Sc progenitor in the remnant therefore mostly reflects that choice of orbit, which was assumed as prograde (positive) as a reference. When the Sc progenitor itself is on a prograde orbit, this will add up to produce a rapidly rotating stellar component to the remnant. When the Sc progenitor is on a retrograde orbit, and considering that the close interaction time between stars of the Sc and the global potential of the Sb is then shortened \citep{renaud}, some stellar material with the initial angular momentum sign will remain in the remnant potentially producing a rather large stellar KDC (from old stars).

\medskip

We can therefore naturally expect a KDC to form as soon as the Sb progenitor is on a retrograde orbit, due to the superposition of a positively rotating contribution from the companion, and the counter-rotating stellar contribution of the main Sb progenitor. When the Sc is on a retrograde orbit, it more violently affects the Sb progenitor which then exhibits a KDC. The Sc is then also transformed into a stellar system with rather disturbed and complex dynamics resulting from the superposition of the orbital wrapping (following the galaxy orbit) and the initial angular momentum of the stars (in the opposite sense).

This process is valid for both the 1:1 and 2:1 mergers. However, the Sc companion in an unequal mass mergers has obviously a smaller influence on the Sb progenitor and is itself more easily destroyed which explains why it appears as having fully adopted the sign of the orbital angular momentum vector. This is nicely illustrated in Fig.~\ref{fig:kdc2types}.  In the case of a 2:1 \textbf{rr} merger, the two contributions have roughly the same mass profiles within the central kpc. The velocity map of the remnant is then composed of the superposition of the counter-rotating component of the Sb progenitor and the contribution from the Sc companion. At larger radii (r $>$ 1~kpc), the Sb galaxy is dominant in mass and leads the velocity field of the merger remnant. An apparent KDC is then visible in the merger remnant. The small ($\sim$~1~kpc) decoupled core in the centre of the Sb progenitor is also smaller than in 1:1 mergers because the companion is lighter and does not affect the central stars of the Sb progenitor much. However, as also pointed out in the 1:1 merger cases, the decoupled core is larger than in the \textbf{rd} case as the mass of the companion falling in the centre is bigger.

We can now easily extend this analysis to the \textbf{dd} and \textbf{dr} cases which produce fast rotators. In the case of a \textbf{dd} orbit, the two spins of the progenitors are aligned with the spin of the orbital angular momentum. This maximises the (positive) angular momentum, which naturally produces a flattened fast rotator. The output of the case of the \textbf{dr} orbit can be extrapolated from the \textbf{rr} case: the Sc companion has a spin anti-parallel with respect to the orbital angular momentum vector. During the approach, the companion is violently disturbed and its stars (most of them in 1:1 and all of them in 2:1 mergers) merge with the Sb progenitor with the spin of the orbit (and of the Sb progenitor). The final merger remnant is then composed of two components with the same sense of rotation and is then a fast rotator, just slightly rounder than for a \textbf{dd} orbit.

   \subsection{The contribution from young stars and gas \label{sec:kdcstargas}}
\begin{figure}
  \includegraphics[width=\columnwidth]{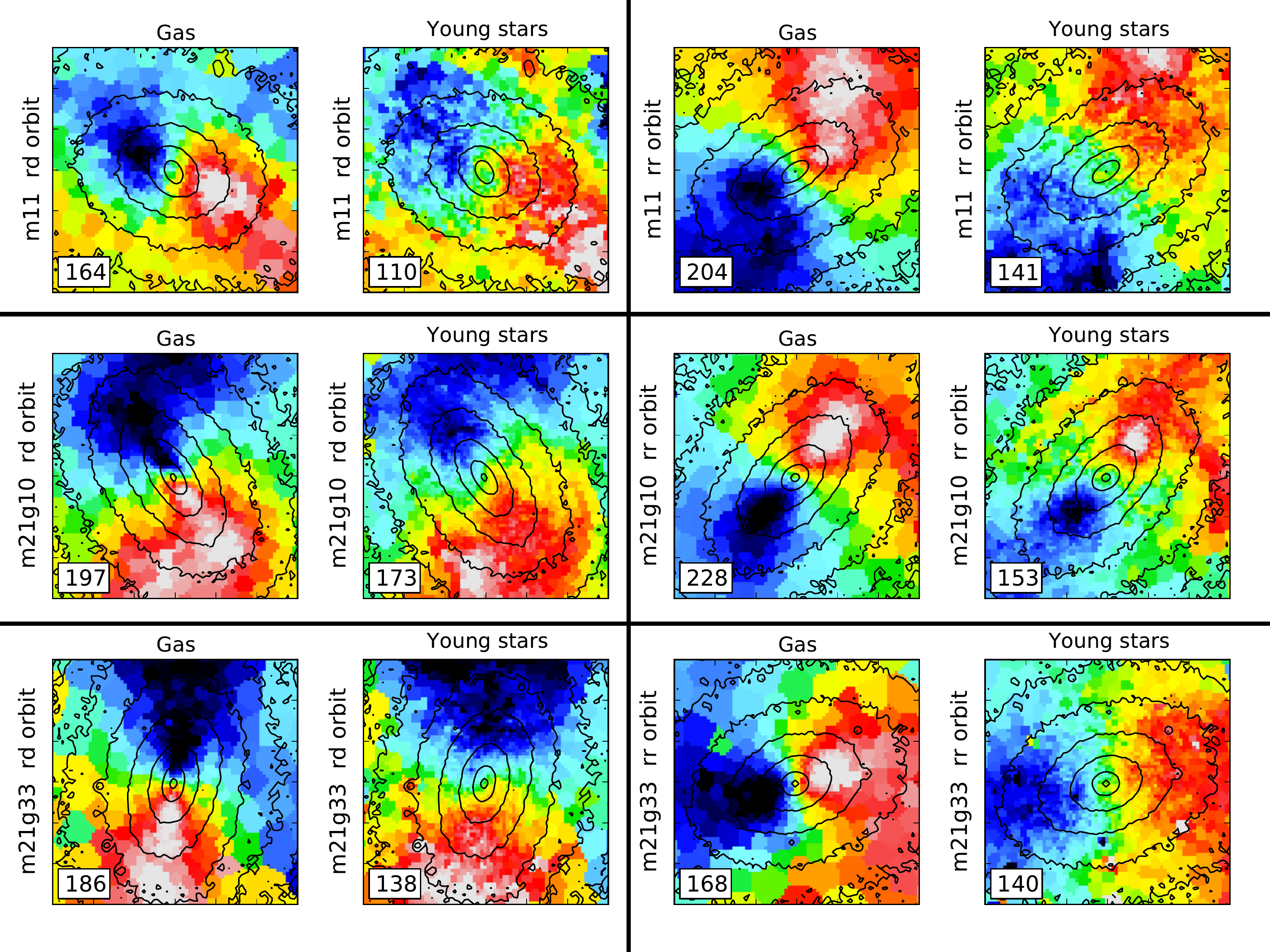} \\
  \caption{Velocity fields of the gas and the young stars in the final merger 
remnant, iso-magnitude contours of the old stars of the final remnant are 
superimposed. The field of view for each panel is $15\times15$~kpc$^2$.}
  \label{fig:kdcgasyoung}
\end{figure}

In Fig.~\ref{fig:kdcgasyoung} we present the velocity fields of the gas and the young stars of the projections used for Fig.~\ref{fig:kdc2types}, the iso-magnitude contours of the old stars being superimposed. We can see that the gas and the young stars share the sense of rotation with the outer part of the final remnants. The gas does not show much sign of a disturbed morphology or kinematics in the centre. The young stars exhibit some misalignments in the very centre of the galaxy, these misalignments being roughly at 90$^{\circ}$ with respect to the photometric major-axis of the remnant. The gas, and the associated star formation, thus do not play a major role in the KDC itself, although the concentration of gas and young stars in the central part of the merger remnant bring an additional central density which could influence the evolution of the KDC.

A small ($\sim$~1~kpc) KDC is observed in the map of the young stars for the m11rr case (Fig.~\ref{fig:kdcgasyoung}). These young stars, formed in a burst of star formation near the pericentre, are thus subjected to violent relaxation, as are the old stars. Part of the young and old stellar populations of the two progenitors end up in two counter-rotating discs (see previous Section): forming, under certain conditions, a central KDC. Galaxies with a young KDC counter-rotating with respect to the gas supply are actually observed in the \atlas sample (McDermid et al., \textit{in prep}).

   \subsection{Summary \label{sec:hubexplan}}
The KDCs in the merger remnants are mostly seen in the old stars, with only a weak signature in the young stars (none in the gas). Moreover, most of the KDCs formed in the major mergers (1:1, 2:1) do not correspond to physically distinct stellar components (see also the study of \citealt{remcokdc}), except in the case of 1:1 binary disc mergers on a \textbf{rr} orbit. These "apparent" KDCs result from the superposition of two counter-rotating stellar components where one disc is dominating the mass profile in the centre while the second disc is dominating outwards. This scenario seems to hold even when the two progenitor discs have the same sense of rotation but with a spin which is retrograde with respect to the orbital angular momentum. The formation of an apparent counter-rotating core (CRC) with two initial retrograde progenitors has also been probed by \citet{balcells90} in a 5:1 merger of oblate ETGs: the companion adopts the sign of the orbital angular momentum during the merger to form a CRC when the two galaxies merge. The physical process invoked to form CRCs in \citet{balcells90} appears consistent with the one witnessed in the present study, although we only form here CRCs in remnants of major (1:1 or 2:1) mergers.

As pointed out in Section~\ref{sec:hubbletype}, the two progenitors have a different influence on the output of the merger. Our so-called Sb progenitor is more concentrated and dynamically stable, and most of its stars keep their original orientation during the merger. The so-called Sc progenitor is colder, less concentrated and responds more efficiently to potential disturbances.

Our study on the formation of the KDCs in binary mergers of disc galaxies should therefore be completed by probing mergers with various morphological types e.g., Sb--Sb or Sc--Sc. The detailed role of the Hubble type for disc progenitors in galaxy mergers is an important issue which will be examined in a future study.

\section{Recent studies on simulated binary mergers and the formation of KDCs} \label{sec:discu}

We have focused our analysis on the central part of the merger remnants to study in details the formation of the slow and fast rotators (and the formation of the KDCs) and here, we specifically compare our results with two other sets of simulations, namely the work of \citet{jesseitlr} and \citet{hoff}. They have both performed a suite of simulations using the TreeSPH code Gadget-2 \citep{springel05} with a gravitational softening length of 140~pc. In these simulations, \citet{hoff} included radiative heating and cooling, star formation, feedback from supernovae and active galactic nuclei and used a total of $4 \times 10^{5}$ particles while \citet{jesseitlr} ``only'' included star formation and stellar feedback but with a total of $\sim 10^{6}$ particles. We discuss in the following some relevant differences between these studies and ours.

\medskip

\citet{jesseitlr} emphasised the formation of slow and fast rotators with their sample of simulations. They have simulated binary mergers of disc galaxies with mass ratios from 1:1 to 4:1 as well as galaxy remergers. We find some significant differences with their results:
\begin{itemize}
 \item 75 per cent of their 1:1 mergers, but only 10 per cent of their 2:1 are slow rotators. This can be compared with 60 per cent for both 2:1 and 1:1 mergers in our sample. 

 \item Overall, their merger remnants are rounder than ours. Their progenitors have $\epsilon < 0.5$ and their slow rotators have $\epsilon < 0.4$, their dry remnants are flatter but do not have $\epsilon > 0.6$. The distribution of $\lambda_R$ and $\epsilon$ for their fast rotator remnants are consistent with the fast rotators described in the present paper.

 \item The ellipticity and $\lambda_R$ distributions of their 1:1 disc-disc merger remnants are ``hardly distinguishable'' from their ETG-ETG merger remnants. From our sample, the difference is very clear, our ETG-ETG merger remnants are rounder and are classified as fast rotators, and we do not produce slow rotators with very low $\lambda_R$ values from re-merging of disc binary mergers.
\end{itemize}

These differences could be mainly explained by variations in the initial parameters of the disc galaxies. The progenitors in \citet{jesseitlr} have a more massive bulge ($B/T=1/4$, as compared to $B/T=1/5$ and $1/8$  for our study), and their bulges are non-rotating: this may produce rounder remnants.  These remnants are also classified as slow rotators, which may be again linked with the absence of rotation of the bulges in their progenitors. 

The distance between the two progenitors at the pericentre is about two disc scale-lengths ($\sim$ 7~kpc) in \citet{jesseitlr} while this distance is about 10 to 25~kpc (depending on the mass ratio) in our study. The lower angular momentum of orbits induced by the lower pericentric distance may also affect the merger remnants. In the present study, we have also simulated binary mergers with an impact parameter $R$ decreased from 60 to 35~kpc. In this range of $R$ values, there is no significant difference in the shapes and kinematics of the merger remnants we obtain, for all mass ratios and initial spins of the progenitors, when the value of $R$ is decreased. Other simulations with even lower pericentre distances \citep[\textit{e.g} with impact parameters $\lesssim$ 10~kpc as in][]{hoff} should be considered for further comparison, but this is outside of the scope of this paper. Orbits with large impact parameters and pericentric distances, as in our study, seem statistically more representative of hierarchical merging in $\Lambda$CDM context \citep{khochfar06}. However, the presence of an external gravitational field (in dense environment) may significantly impact the orbit of the two progenitors, decreasing the pericentric distance (Martig, private comm.): this will be examined in more detail in a forthcoming paper.

\bigskip

\citet{hoff} have simulated binary mergers of disc galaxies of mass ratio 1:1 at different gas ratio (from 0 to 40 per cent of gas). Again, there are several differences between their results and ours:
\begin{itemize}
 \item From 0 to 10 per cent of gas, their remnants are all slowly rotating. They do not have KDCs, the slow rotator galaxies are dominated by box orbits or by minor axis rotation. With 15 and 20 per cent of gas, most of their remnants have a KDC. When reaching 30 and 40 per cent of gas, their remnants are all fast rotators. They do not mention any trend between the initial condition of merging (\textit{e.g} the orientation of the discs) and the formation of slowly rotating early-type galaxies.

 \item Their KDCs are small discs of young stars coming from the gas in rotation in the centre. When the fraction of gas is increased, the size of the final disc of gas --~and thus the disc of young stars~-- is also increased and it creates a fast rotator. Their fast rotators are purely dominated by the young stars created in this large disc of gas in rotation. In our sample, the KDC is an apparent KDC (except for the 1:1 merger with \textbf{rr} orbits) and is seen only in the old stars. The gas does not show any sign of counter-rotation. Our KDCs do not depend on the gas ratio, as we do not find any significant difference between the remnants with 10 or 33 per cent of gas.
\end{itemize}

\citet{hoff} have used pure stellar discs (\textit{i.e.} without a stellar bulge or spheroid) as progenitor galaxies, and a low impact parameter of 7.1~kpc (and thus a lower pericentric distance). These initial conditions are far from ours and lead to a very different merging process. We speculate that, as the orbit leads to a more rapid and direct collision, the two discs are destroyed and strongly influenced by violent relaxion and are not expected to keep a trace of the original disc dynamics (except at large radii as pointed out in their paper). This may explain why none of their low gas fraction mergers shows signs of rotation around the short-axis of the remnants.

The treatment of the gas and the resolution used for the simulations are also different between this study and the one presented here. \citet{boisresol} showed that at high resolution, thinner gas structures are resolved during the merger, which can result in structured and clustered star formation. These local density peaks are accompanied by rapid variations of the gravitational potential, which help scatter stellar orbits and evacuate the angular momentum. A gas-free and a gas-rich merger should differ not only with the reformation of a disc of gas in the centre of the remnant as seen in \citet{hoff}, but also at all radii with different stellar orbits (see also \citealt{bournaud-mergers-2010}). These results emphasized the need for high spatial and mass resolution \citep[see also][]{boisresol, powell10,teyssier10} and for realistic physical inputs: \textit{e.g.} the treatment of the gas with models capable of resolving the main dense gas clouds/SF regions \citep[see ][in a cosmological context]{governato09}.

\section{Comparison with observations} \label{sec:compa}
We here compare directly our sample of merger remnants (binary mergers and ETG remergers) with the sample of 260 galaxies observed in the context of the \atlas{} project (Paper I). For that purpose, we first use the \lreps{} diagram, and also compare the distribution of alignments between the photometric and kinematic axes.

  \subsection{The \lreps{} diagram \label{sec:lrepsobs}}
\begin{figure}
 \includegraphics[width=\columnwidth]{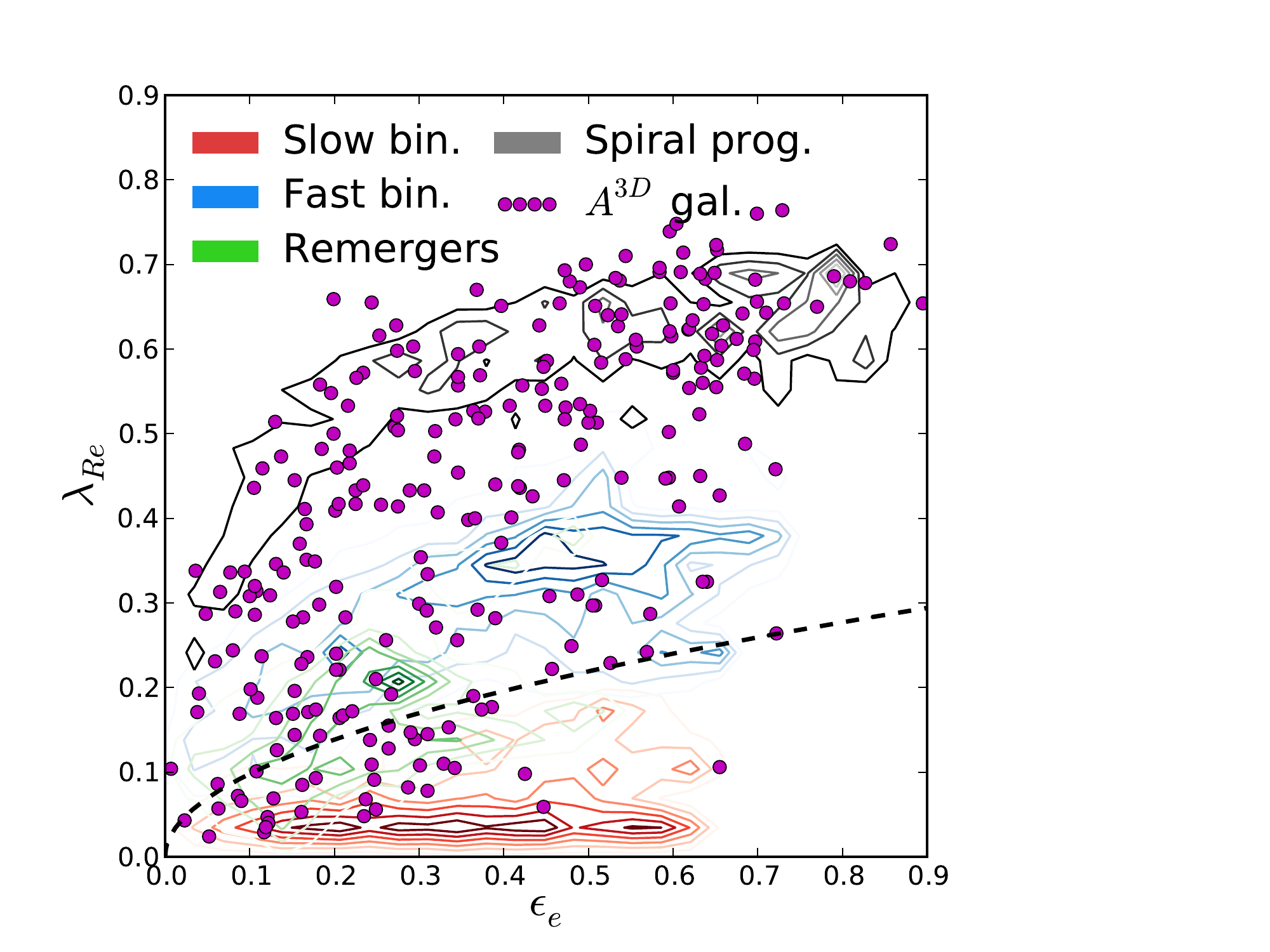} \\
 \includegraphics[width=\columnwidth]{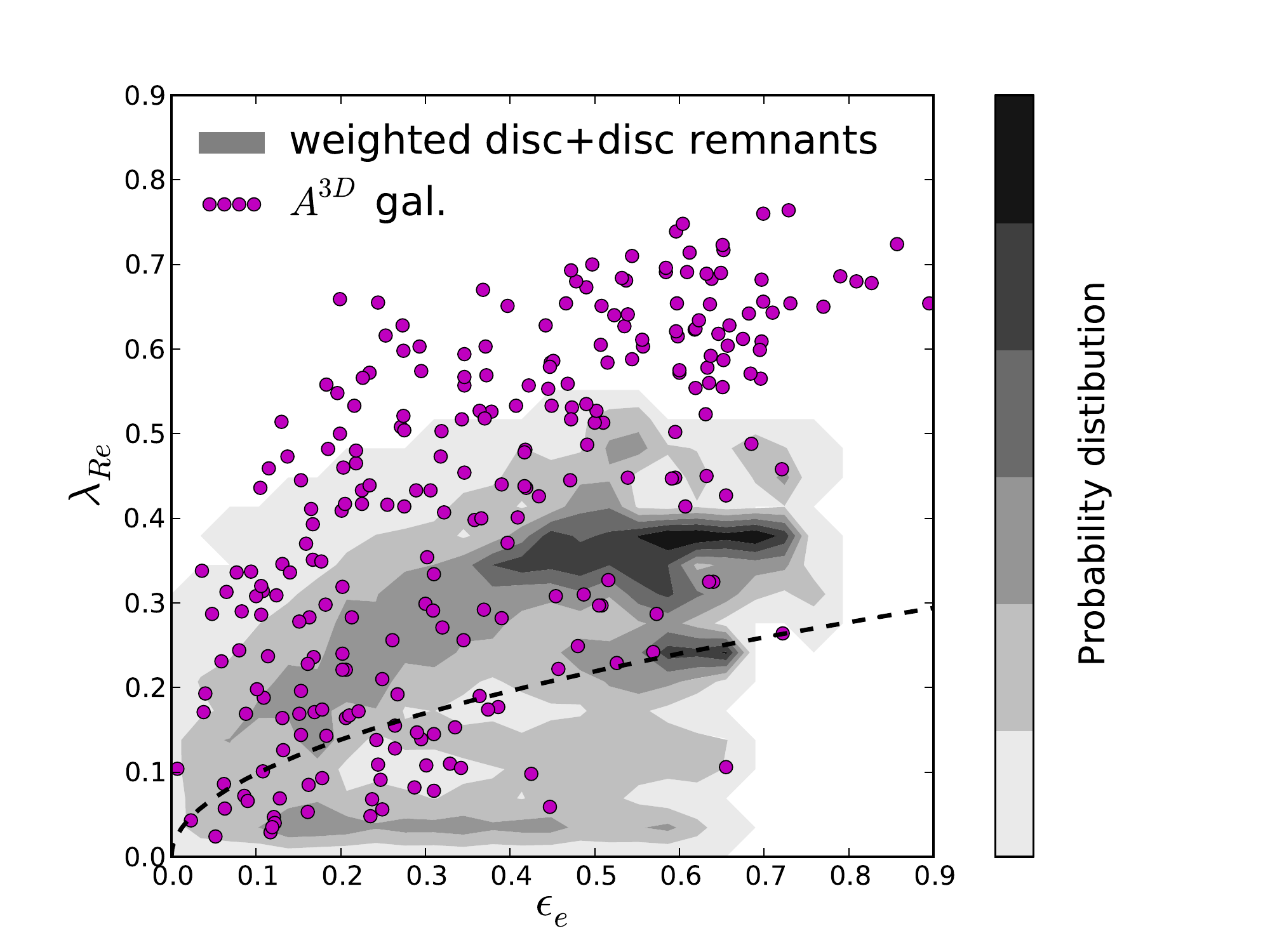} \\
 \caption{\lreps{} diagram for the 260 galaxies of the \atlas{} sample (shown as dots) compared to our simulation sample. The dashed line corresponds to the limit defining the slow and fast rotators. Top: The contours represent the distribution of projections of the simulated fast rotator disc remnants in blue, slow rotators disc remnants in red, remerger remnants in green and spiral progenitors in black. Bottom: The contours represent the distribution of projected remnants (slow+fast) of disc mergers statistically weighted depending on the likelihood of its mass ratio (see text for details).}
 \label{fig:compaobs}
\end{figure}

\begin{figure*}
 \begin{center}
 \begin{tabular}{cccc}
 \includegraphics[width=0.62\columnwidth]{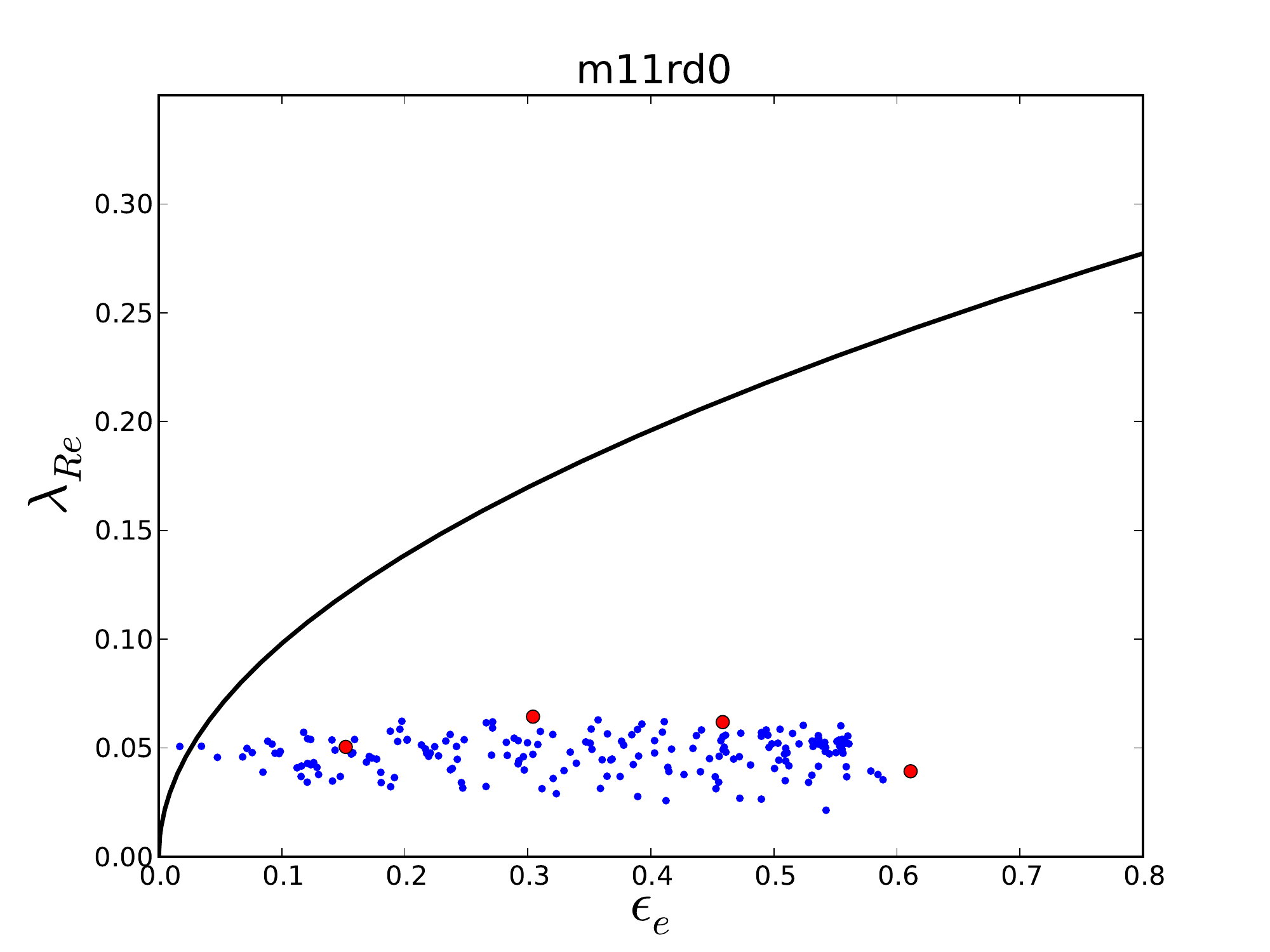} & \includegraphics[width=0.62\columnwidth]{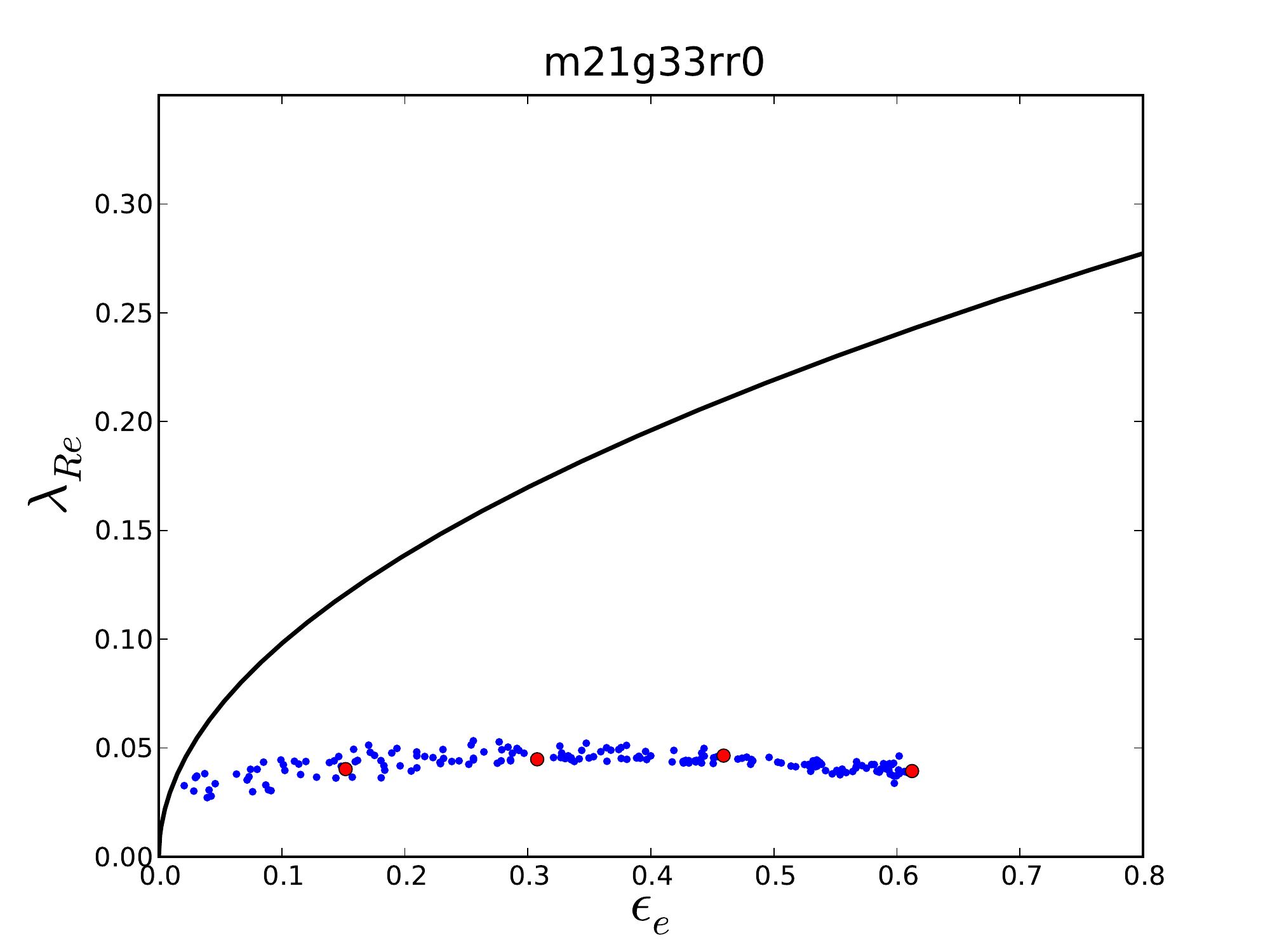} & \includegraphics[width=0.62\columnwidth]{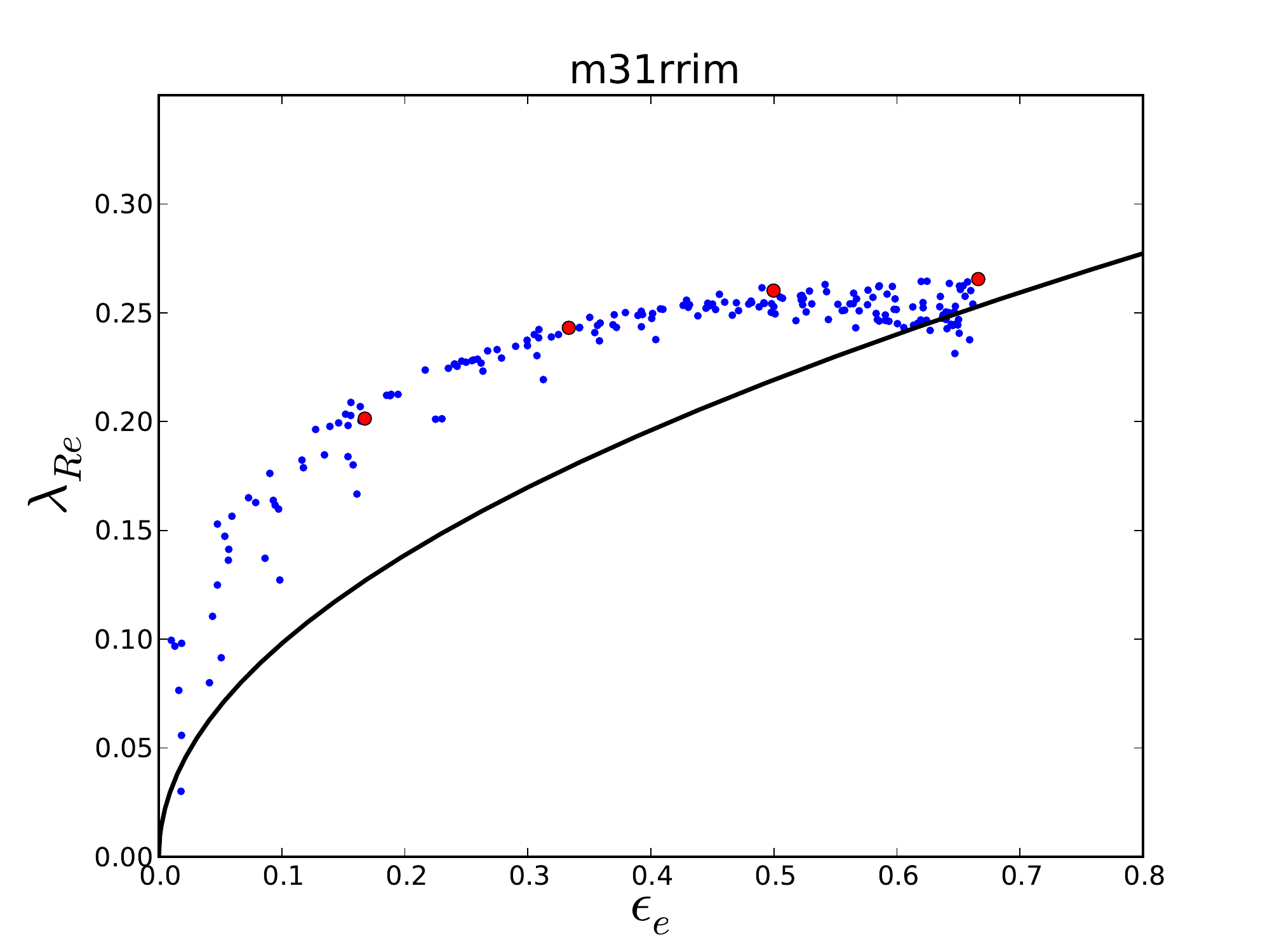} \\
 \includegraphics[width=0.6\columnwidth]{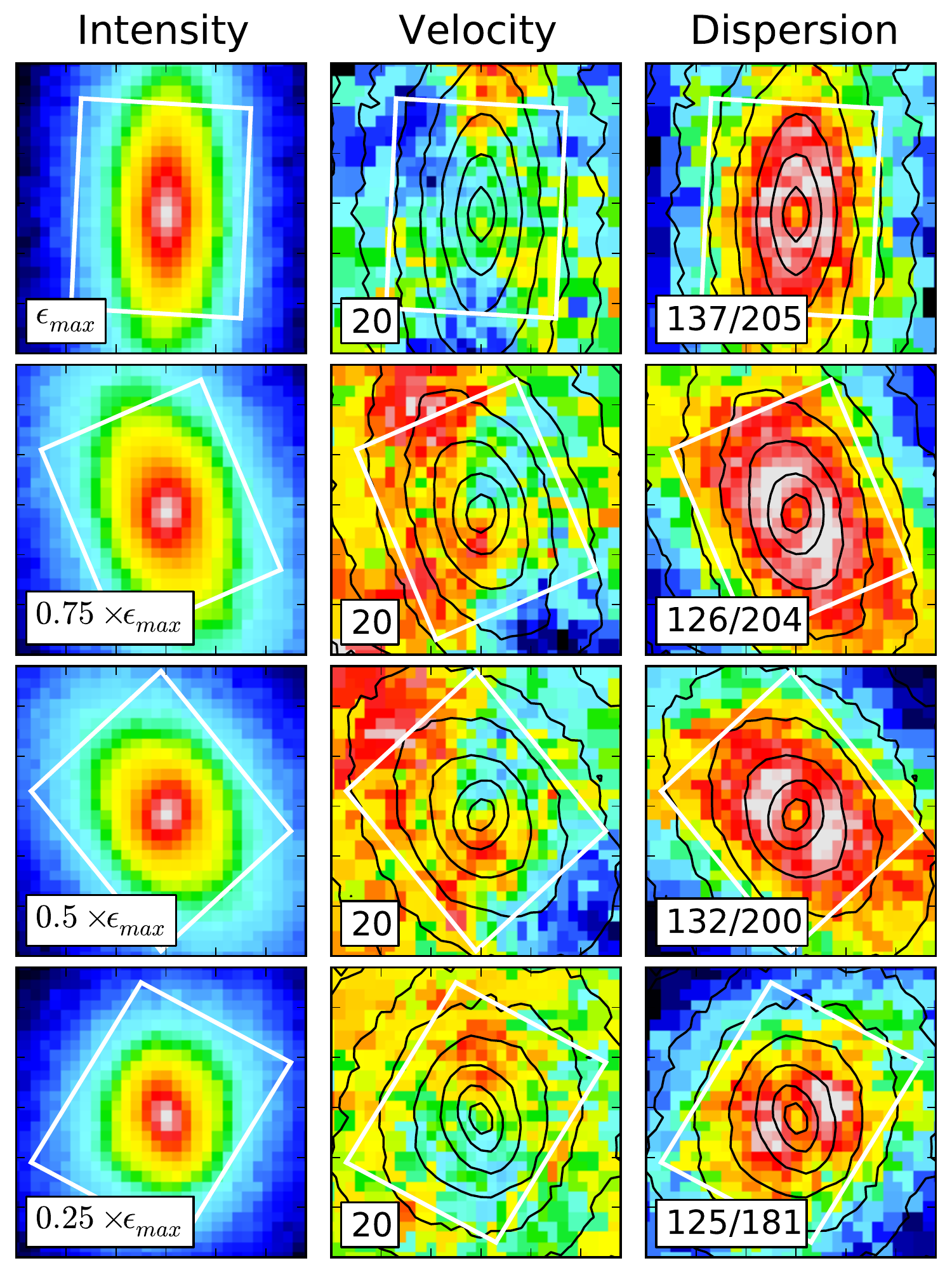} & \includegraphics[width=0.6\columnwidth]{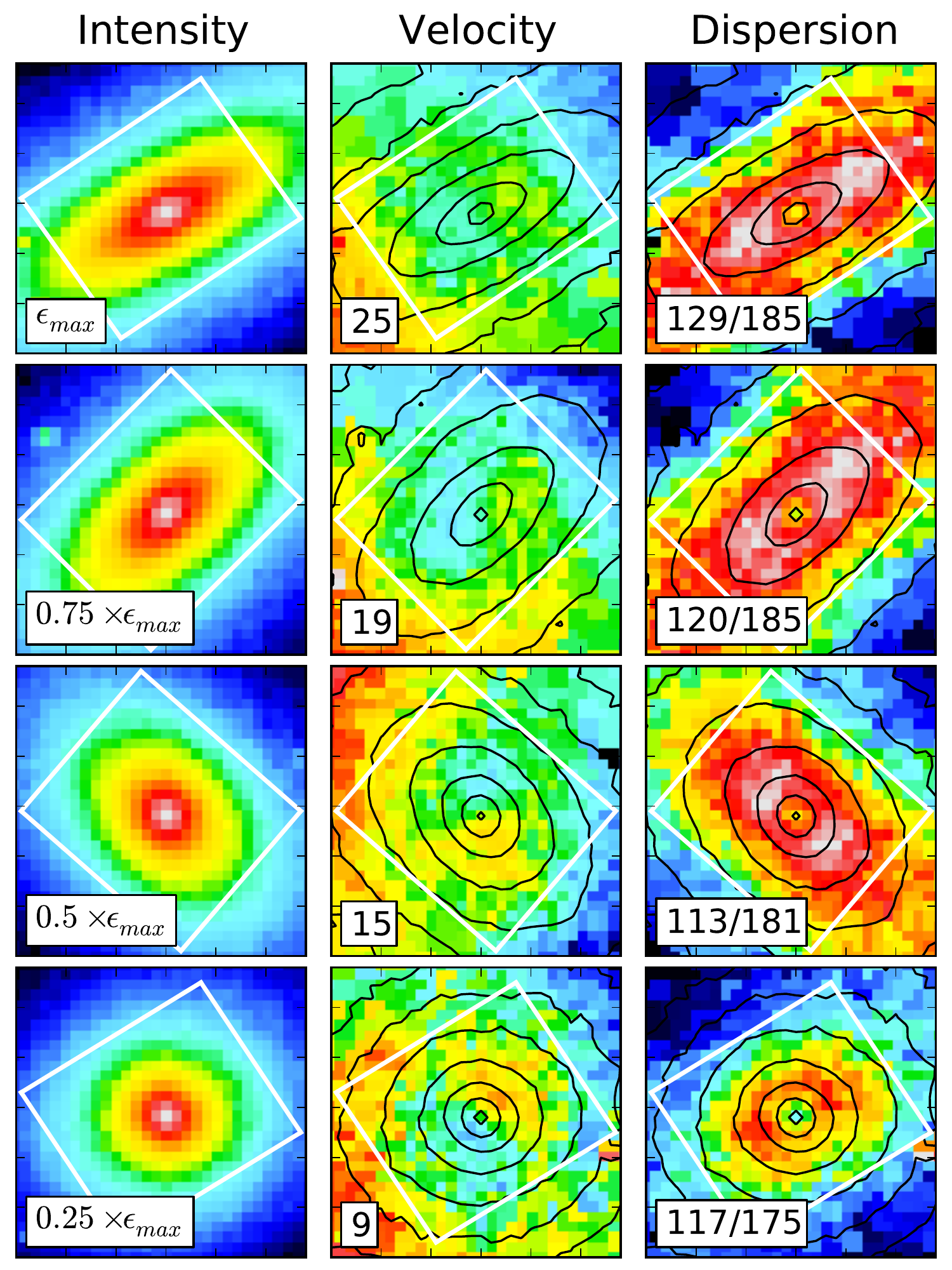} & \includegraphics[width=0.6\columnwidth]{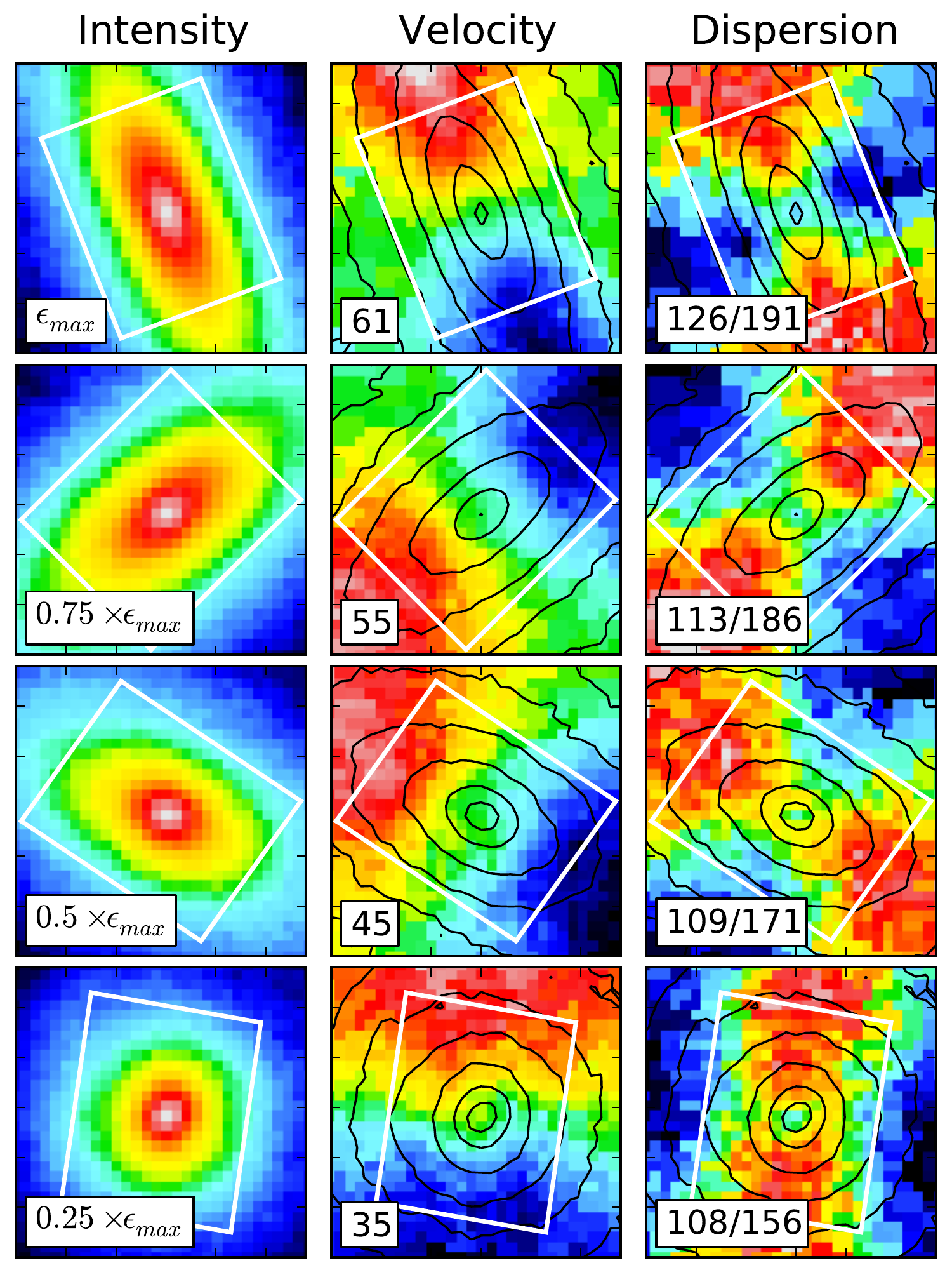} & \includegraphics[width=0.125\columnwidth]{pdffigures/proj_mergers/colorbar_only.pdf} \\
 \end{tabular}
 \caption{Three simulated 2-$\sigma$-like galaxies: a 1:1 slow rotator (left), a 2:1 (with 33 per cent of gas) slow rotator (middle) and a 3:1 fast rotator (right). The top panels represent the \lreps{} diagram for the 200 projections plotted as blue dots, the red points correspond to the projection with the maximum ellipticity $\epsilon_{max}$ and the projections corresponding to $0.75 \times \epsilon_{max}$, $0.5 \times \epsilon_{max}$ and $0.25 \times \epsilon_{max}$. The lower panels show the intensity, velocity and velocity dispersion maps of the red point projections. The projection is written in the intensity map, the maximum value of the velocity and the minimum/maximum values of the velocity dispersion are written in their respective maps. The colorbar goes from -V$_{max}$ to +V$_{max}$ and can also be used as an indicator for the intensity and the velocity dispersion. The field of view is $6\times6$~kpc$^2$ (6~kpc $\simeq$ 2.5$R_e$). The white rectangle indicates a typical field covered by the instrument \sauron{} and corresponds to a field of $41" \times 33"$ for a galaxy at a distance of 20~Mpc, its orientation follows the photometric position angle taken at 3$R_e$.}
 \label{fig:2sigma}
 \end{center}
\end{figure*}

Fig.~\ref{fig:compaobs} shows the \lreps{} diagram for the 260 galaxies of the \atlas{} sample (extracted from Paper III) and presents the distribution of projections for the simulated spiral progenitors, remerger remnants, and merger remnants of binary disc galaxies. It also presents the distribution of projections for our merger remnants of binary disc galaxies weighted by the probability of the galaxy-galaxy merger rate as a function of the mass ratio in $\Lambda$CDM. From \citet{hopkins10} for a galaxy of mass 10$^{11}$ M$_{\odot}$ (\textit{i.e.} with the mass of our main progenitor), if $P$ is the probability of having a 1:1 merger, the probability to have a 2:1 merger (\textit{resp.} 3:1 and 6:1) is $4P$ (\textit{resp.} $6P$ and $9P$). The lower mass ratios are favoured, so does the formation of fast rotators via a binary galaxy merger. 

By looking at where the observed \atlas{} galaxies and our sample of galaxy mergers lie in such a \lreps{} diagram, we can make a first rough assessment of the relevance of binary disc mergers and remergers for nearby ETGs:
\begin{itemize}
 \item For high values of $\lambda_R$ ($\sim$ 0.5 -- 0.8), the distribution of \atlas{} galaxies coincides in fact with the distribution of our progenitors, this has been also pointed out in \citet{jesseitlr}. These local galaxies often contain a bar and are consistent with disc galaxies. These could be disc galaxies which have evolved via various internal processes --~at high redshift \citep[\textit{e.g.}][]{EBE08,dekel09} and/or at low redshift \citep{athana99}~-- and through accretion of gas or very small companions \citep[][]{toth92,MB10,moster10}.

 \item For intermediate values of $\lambda_R$ ($\sim$ 0.25 -- 0.5), the \atlas{} galaxies are closer to the fast rotator merger remnants. These observed ETGs could thus be the remnants of a merger between a spiral galaxy and a companion with a mass ratio from 10:1 to 1:1 \citep[see \textit{e.g.}][]{BB00,cretton01,naab03,bour05,naabtru06}.

 \item In the \atlas{} sample, we find $\sim 30$ galaxies with $\lambda_R < 0.25$ classified as fast rotators. These galaxies have rather small ellipticities: some of these therefore may be consistent with the remnants of a binary merger of disc galaxies viewed face-on, but this cannot hold for the full set of low $\lambda_R$ fast rotators. As seen in Fig.~\ref{fig:compaobs}, the result of a major remerger can form a galaxy with a relatively low intrinsic ellipticity and low $\lambda_R$ values and can better account for these galaxies. A remnant of a major binary merger of spirals followed by mergers of smaller companions, or the fast evolution of galaxies in groups \citep[see \textit{e.g.}][]{konstan10} may also lead to a similar output.

 \item The comparison with the observed slow rotators is less evidential. In the context of the sub-classes of slow rotators defined in Paper~II, we can emphasise the following items. Observed non rotators are among the most massive objects in the \atlas{} sample (Papers~I and III) and very probably have a complex merger history \citep{nieto91,tremblay96}, clearly beyond the simple picture provided by the binary mergers described here. Galaxies observed to exhibit KDCs all have apparent $\epsilon <$ 0.4: as already emphasised, the slow rotator remnants obtained in our sample of simulations are much too flat comparatively. This implies that only a few of the observed slow rotators could have been simply formed via binary mergers of disc galaxies, and these would in addition need to be viewed at relatively high inclination. Slow rotators could therefore be the result of a single major merger but then with significantly more violent initial conditions (\textit{e.g.} with a very small impact parameter, Duc et al. 2011, \textit{in prep}) or must have experienced further interactions: a sequence of small satellite mergers may produce rounder remnants and at the same time preserve the KDCs. Such accretion of low-mass objects should be much more frequent than major mergers and are therefore expected \citep[see \textit{e.g.}][ $\Lambda$CDM models]{khochsilk06,naabresol,bour07,genel08,genel10,naab09,hopkins10,oser10,qu10,abadi10}.

 \item We can probably account for 2-$\sigma$ galaxies with our sample of simulations. All the observed 2-$\sigma$ galaxies have $\epsilon >$ 0.3 and their properties (see Papers II and III) indicate that they could have been formed via a single binary merger \citep[see also][]{balcells90,HB91,balcells98,barnes02,JNPB07,crocker09}. Fig.~\ref{fig:2sigma} shows three simulated merger remnants (formed with a 1:1, 2:1 and 3:1 mergers) which resemble the observed 2-$\sigma$ galaxies. The remnants formed via a 1:1 and 2:1 mergers are classified as slow rotators and present a KDC while the remnant formed with a 3:1 merger is a fast rotator (without KDC) but staying close to the limit defining these two populations. This picture is consistent with the \atlas{} observations where 2-$\sigma$ galaxies are found in both families of ETGs.
\end{itemize}

  \subsection{Photometric and kinematic alignments  \label{sec:alignobs}}
\begin{figure}
 \includegraphics[width=\columnwidth]{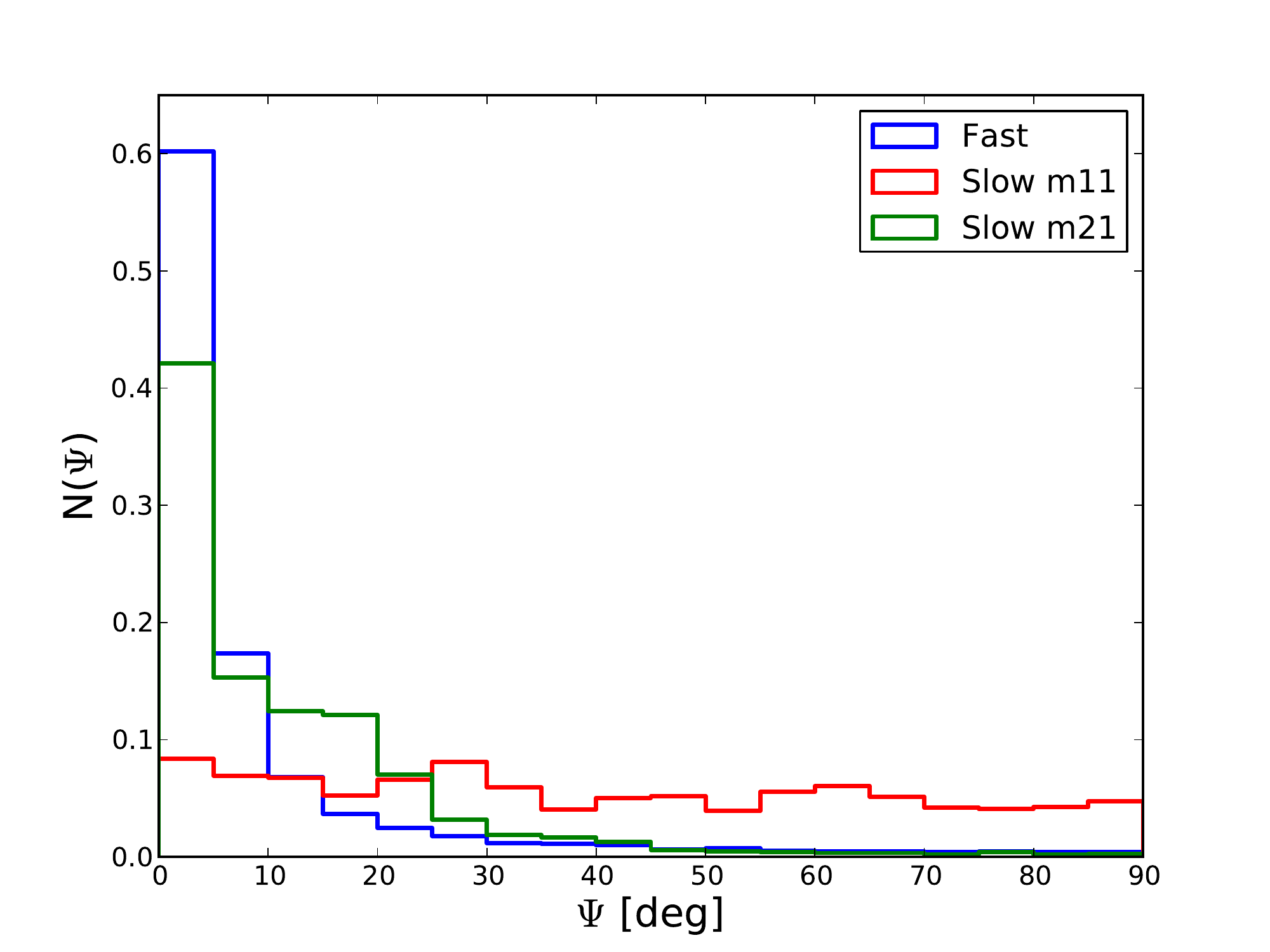} 
 \includegraphics[width=\columnwidth]{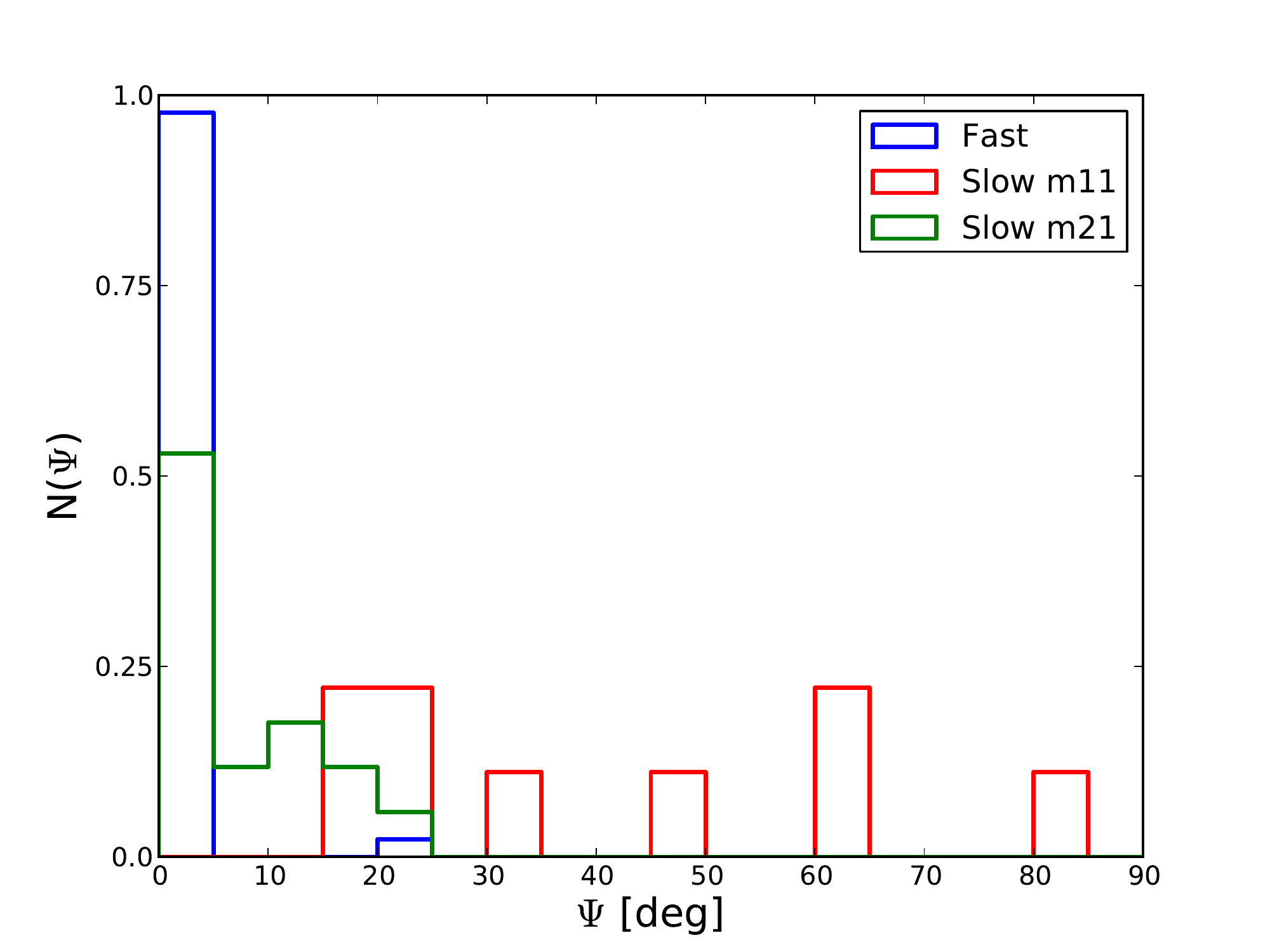}
 \caption{\textbf{Top:} Histogram of the kinematic misalignment angle (in degree) for all projections of the merger remnants. The y-axis is normalised to the total number of projections per categories: the fast rotators (for mass ratios 1:1 to 6:1) are in blue, the slow rotators of mass ratio 1:1 in red, the slow rotators of mass ratio 2:1 (both 10 and 33 per cent of gas) in green. The errors on $\Psi$ are within a bin size. \textbf{Bottom:} Same histogram but only for the projections which maximize the ellipticity (edge-on view) of the merger remnants.}
 \label{fig:compamis}
\end{figure}

Based on the \citet{franx} definition, we calculate the kinematic misalignment angle $\Psi$ as the difference between the measured photometric and kinematic position angles taken at three effective radii (see also Paper II) as:

\medskip
\noindent sin $\Psi$ = $|$sin (PA$_{phot}$ - PA$_{kin}$)$|$.
\medskip

The measurement of this quantity at large radii minimises the impact of central decoupled structures such as \textit{e.g.} a KDC and bar. $\Psi$ is defined between two observationally related quantities and it approximates the true kinematic misalignment angle which should be measured between the intrinsic minor axis and the intrinsic angular momentum vector. $\Psi$ can range from 0 to 90$^{\circ}$ (the use of $\sin$ removes additional differences of 180$^{\circ}$ between PA$_{phot}$ and PA$_{kin}$).

This analysis has been made for the sample of the 260 \atlas{} galaxies in Paper II. The regular velocity pattern galaxies (\textit{i.e} fast rotators) are mostly found at small $\Psi$ values, and the galaxies with complex kinematic structures (non rotators, NRV, 2-$\sigma$ and KDC galaxies) often exhibit strong misalignments between the photometry and kinematics.

Our simulations are in good agreements with these observational results. The top panel of Fig.~\ref{fig:compamis} shows the histogram of the kinematic misalignment angles for all projections of all binary merger remnants. The fast rotators have 60 per cent of their projections in the first bin with $\Psi <$ 5$^{\circ}$, with another 17 per cent with 5 $< \Psi <$ 10$^{\circ}$, and 88 per cent of the projections have $\Psi <$ 20$^{\circ}$. The slow rotators for the 2:1 mergers have respectively 42, 15, 82 per cent of their projections in these domains, respectively, while the slow rotators associated with 1:1 mergers are distributed homogeneously from 0 to 90$^{\circ}$. Our fast rotator remnants should thus be considered as perfectly aligned: the spin axis of the stellar component is the photometric short-axis. The slow rotators associated with 2:1 mergers are also relatively well aligned: the most massive progenitor (the Sb spiral in this study) dominates the velocity rotation, and the galaxy remnant remains flat and disc-like. The slow rotators formed in 1:1 mergers show significant misalignment: the two progenitors end up contributing to counter-rotating stellar components which almost cancel any central rotation around the photometric minor-axis. The rotation is then dominated by the kpc-size KDC and by rotation along the photometric minor-axis at larger radii \citep{hoff}. These results are confirmed when selecting the projections which maximize the ellipticity (\textit{i.e} the edge-on projection) of the merger remnants (see also Fig.~\ref{fig:Vmerg11} to \ref{fig:Vmerg61} in Appendix~\ref{sec:appprojmaps} for the corresponding velocity maps): the edge-on projections of the fast rotators are always aligned with $\Psi <$ 5$^{\circ}$ (except for the merger remnant \textit{m11rrip}), while some 2:1 slow rotators show significant misalignments and all 1:1 slow rotators do.

The same results are found by \citet{jesseitlr}, they have analysed misalignment angles for the oblate, prolate and triaxial simulated merger remnants described in \citet{thomas07}. Their disc-like oblate system is a fast rotator and has most of its projections with $\Psi <$ 5$^{\circ}$ while the more disturbed prolate and triaxial remnants show misalignments.

  \subsection{Summary}
The morphology and kinematics of the modelled fast rotators (including the spiral progenitors) are in good agreement with the observed local fast rotators, mainly composed of flat and rotationally supported discs with nearly aligned photometric and kinematic axes. In the context of our sample of simulations, we can exclude the formation of the non rotators via binary mergers of disc or early-type galaxies: these are intrinsically round and thus supported by velocity dispersion alone, properties which are not observed in our sample of merger remnants. The observed galaxies with a KDC shared some properties (misalignments and angular momentum content) with our simulated merger remnants but they are also intrinsically rounder: their formation history is then certainly much more complex than the simple picture of binary mergers of disc galaxies. We can possibly account for the formation of some observed slow rotator galaxies of the \atlas{} sample, namely the \textit{so-called} 2-$\sigma$ galaxies, with 1:1 or 2:1 mergers: these systems present signatures of the presence of two counter-rotating components \citep[see \textit{e.g.}][]{remcokdc}. A more detailed comparison, e.g., including $h_{3,4}-v/\sigma$ diagrams, or gas distribution and kinematics, would be required to confirm this result.

\section{Conclusions} \label{sec:conclu}

We have simulated 70 binary mergers of disc galaxies at an unprecedented resolution for such a sample of simulations. We have studied the effect of different initial parameters on the global properties of merger remnants, varying: the mass ratio (from 1:1 to 6:1), the initial conditions of the mergers (incoming velocity, impact parameter, inclination in the orbital plane), the spins of the progenitors. We have also simulated 16 binary remergers of galaxies between two binary spiral merger remnants, varying: the mass, the class (slow/fast) of the progenitors, and the spin of the initial ETG-like progenitors. We have then compared the properties of our sample of (re)merger remnants with the observed 260 ETGs of the \atlas{} survey. We have also compared our work with previous established samples of numerical simulations. Our main conclusions can be summarized as follows:

\begin{enumerate}
 \item We obtain both fast and slow rotators. In binary mergers, the fast rotators can be observed at all mass ratios while the slow rotators are formed only for mass ratios between 1:1 and 2:1, in agreement with \citet{jesseitlr}. We confirm that the limit separating the two families of ETGs defined by the \atlas{} survey is meaningful, as our simulated slow and fast rotators present distinct characteristics: the fast (\textit{resp.} slow) rotators have a high (\textit{resp.} low) angular momentum content, their photometric and kinemetric position angle are aligned (\textit{resp.} misaligned), they present (\textit{resp.} do not present) regular velocity patterns. An important difference is the presence (or absence) of a KDC: none of the fast rotators hold a KDC while most of the slow rotators do.

 \item The two parameters which constrain the formation of the KDCs are (1) the mass ratio between the initial spirals, as we form a KDC only with 1:1 and 2:1 mergers, and (2) the orientation of the initial spin axis of the earliest-type (Sb) disc progenitor with respect to the orbital angular momentum (its spin has to be anti-parallel). The spin of the later-type (Sc) progenitor has less importance as this galaxy is mostly disrupted during the merging process. For a 1:1 merger, if the two progenitors are retrograde, the KDC is intrinsically decoupled from the external part of the merger remnant. For all other initial conditions leading to the formation of a slow rotator, the KDC is only an apparent KDC formed via the superposition of two counter-rotating discs.

\item The Hubble type of the initial spiral progenitors seems to play a prominent part in the formation of slow rotators and additional simulations are needed to further constrain its role.

 \item To test the importance of the presence of gas, we have simulated binary mergers with mass ratio 2:1 with either 10 or 33 per cent of gas. We do not find major differences in the morphology and the kinematics of the merger remnants. There are some visible differences (\textit{e.g.} the number of newly formed globular clusters) but it does not impact much the kinematics of the remnants and the comparison with the \atlas{} sample. Higher gas fractions representative of mergers at high redshifts could have a larger impact, though.

\item  All the major remergers of ETG-like galaxy remnants are either classified as fast rotators or close to the boundary between the slow and fast families. Such remnants present clear rotation patterns and do not hold a KDC: all the KDCs in the progenitors have been destroyed during the remerger and none has been created during the merger event. 

 \item The properties of the fast rotators formed in our simulations are consistent with some observed fast rotators of the \atlas{} sample. Some of the simulated slow rotators may also be associated with a few observed 2-$\sigma$ galaxies which present clear evidence of an apparent counter-rotating stellar component. 

 \item Our simulations cannot, however, account for the other classes of slow rotators which are intrinsically rounder than any of our major merger remnants: these galaxies are generally massive and have certainly a more complex history. To simulate these galaxies, the full cosmological context of their formation history has to be considered, including major merger(s), repeated minor mergers, stellar mass loss and also smooth accretion of gas from the cosmic filaments.
\end{enumerate}

\section{Acknowledgments} \label{sec:acknow}
MBois and TN acknowledge support from the DFG Cluster of Excellence `Origin and Structure of the Universe'.
MC acknowledges support from a Royal Society University Research Fellowship.
This work was supported by the rolling grants `Astrophysics at Oxford' PP/E001114/1 and ST/H002456/1 and visitors grants PPA/V/S/2002/00553, PP/E001564/1 and ST/H504862/1 from the UK Research Councils. RLD acknowledges travel and computer grants from Christ Church, Oxford and support from the Royal Society in the form of a Wolfson Merit Award 502011.K502/jd. RLD also acknowledges the support of the ESO Visitor Programme which funded a 3 month stay in 2010.
SK acknowledges support from the the Royal Society Joint Projects Grant JP0869822.
RMcD is supported by the Gemini Observatory, which is operated by the Association of Universities for Research in Astronomy, Inc., on behalf of the international Gemini partnership of Argentina, Australia, Brazil, Canada, Chile, the United Kingdom, and the United States of America.
MS acknowledges support from a STFC Advanced Fellowship ST/F009186/1.
NS and TD acknowledge support from an STFC studentship.
The authors acknowledge financial support from ESO.

\appendix
\section{Projected velocity maps of binary mergers and remergers} \label{sec:appprojmaps}

All simulations used for this study are listed here. For each simulation, an edge-on view of the velocity field (with the iso-magnitude contours in black) and its associated $\lambda_R$ profile are plotted.

\bigskip

\begin{figure*}
 \begin{tabular}{cc}
  \includegraphics[width=1.05\columnwidth]{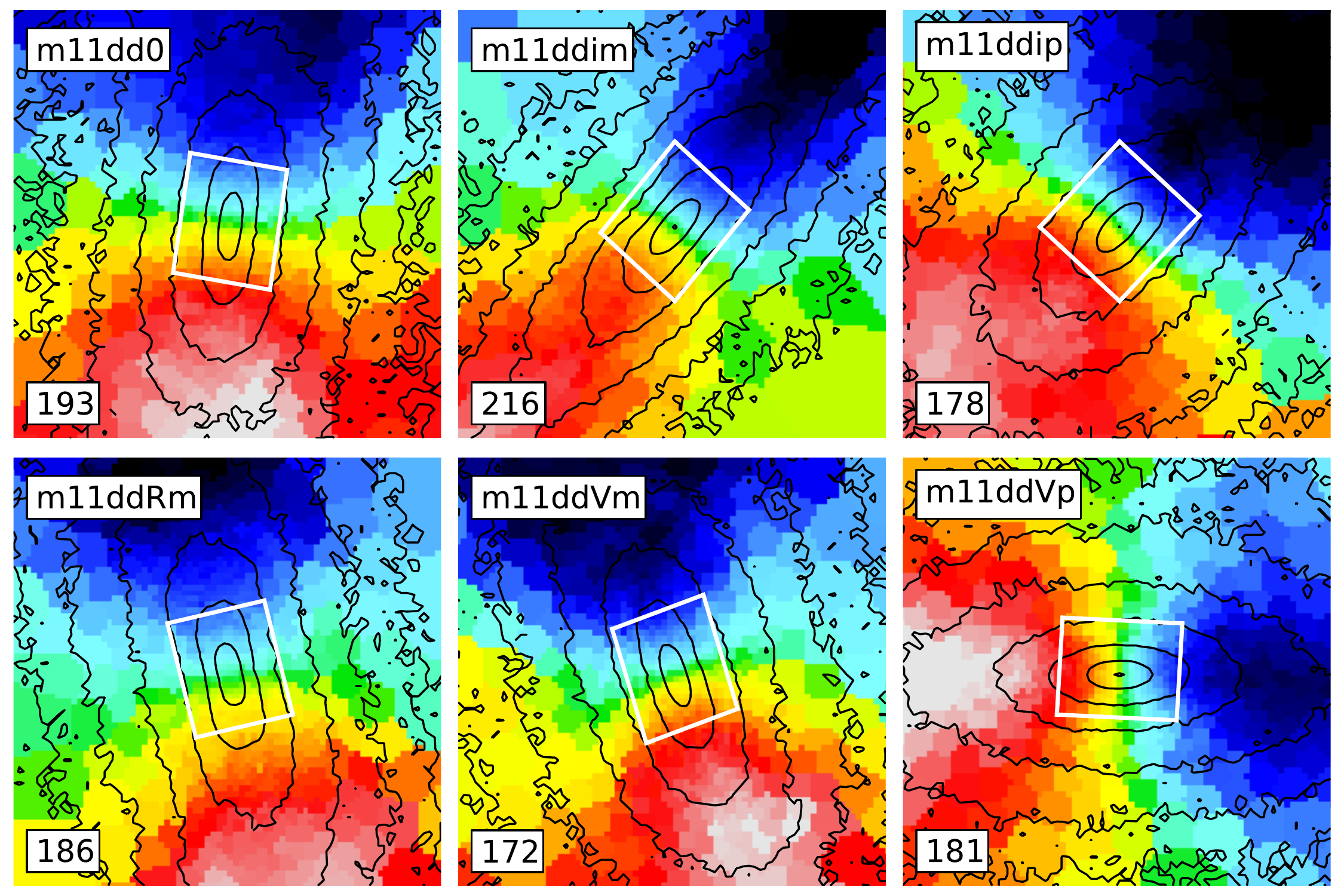} & \includegraphics[width=0.95\columnwidth]{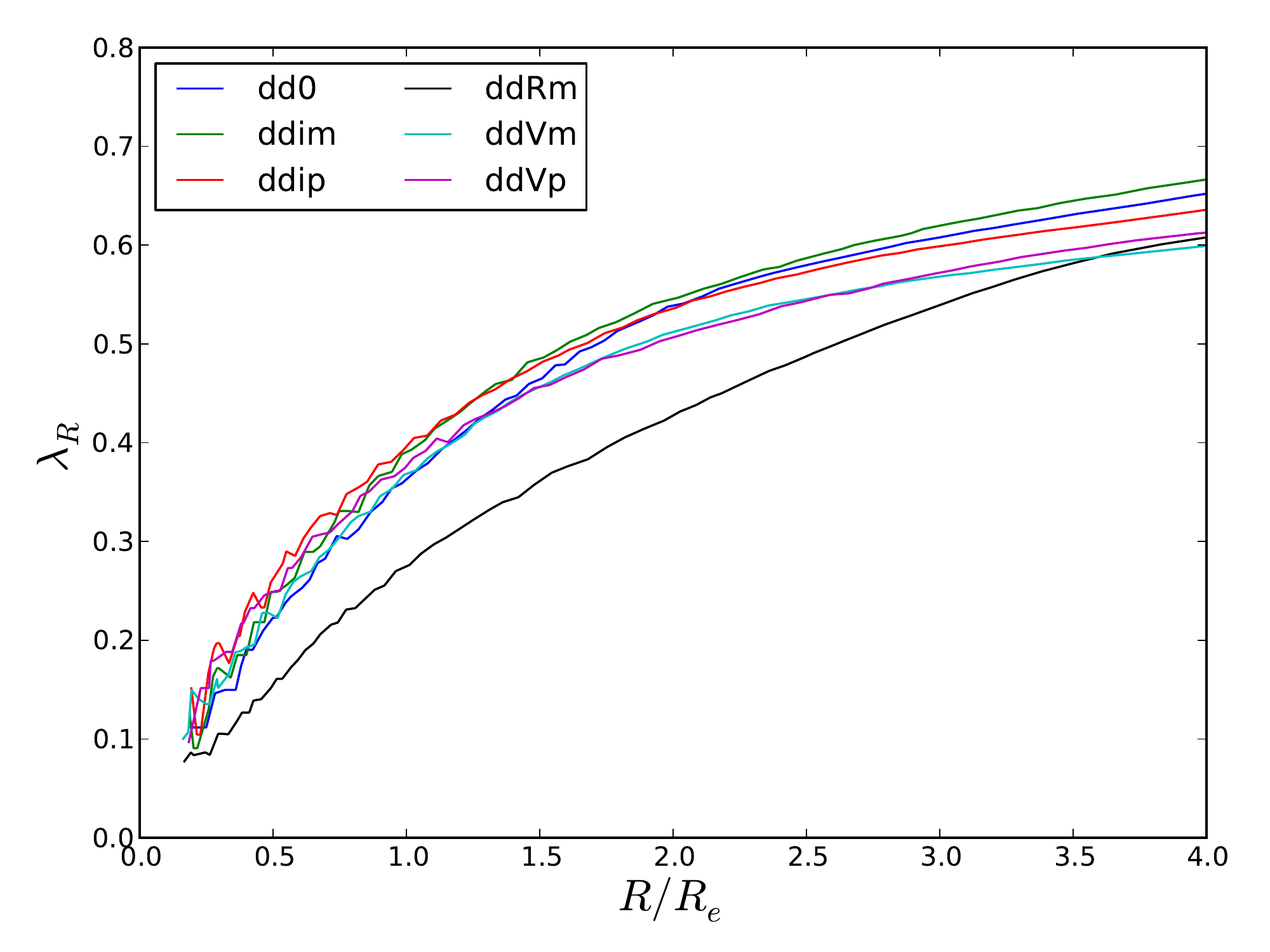}  \\
 & \\
  \hline
 &   \\
 \includegraphics[width=1.05\columnwidth]{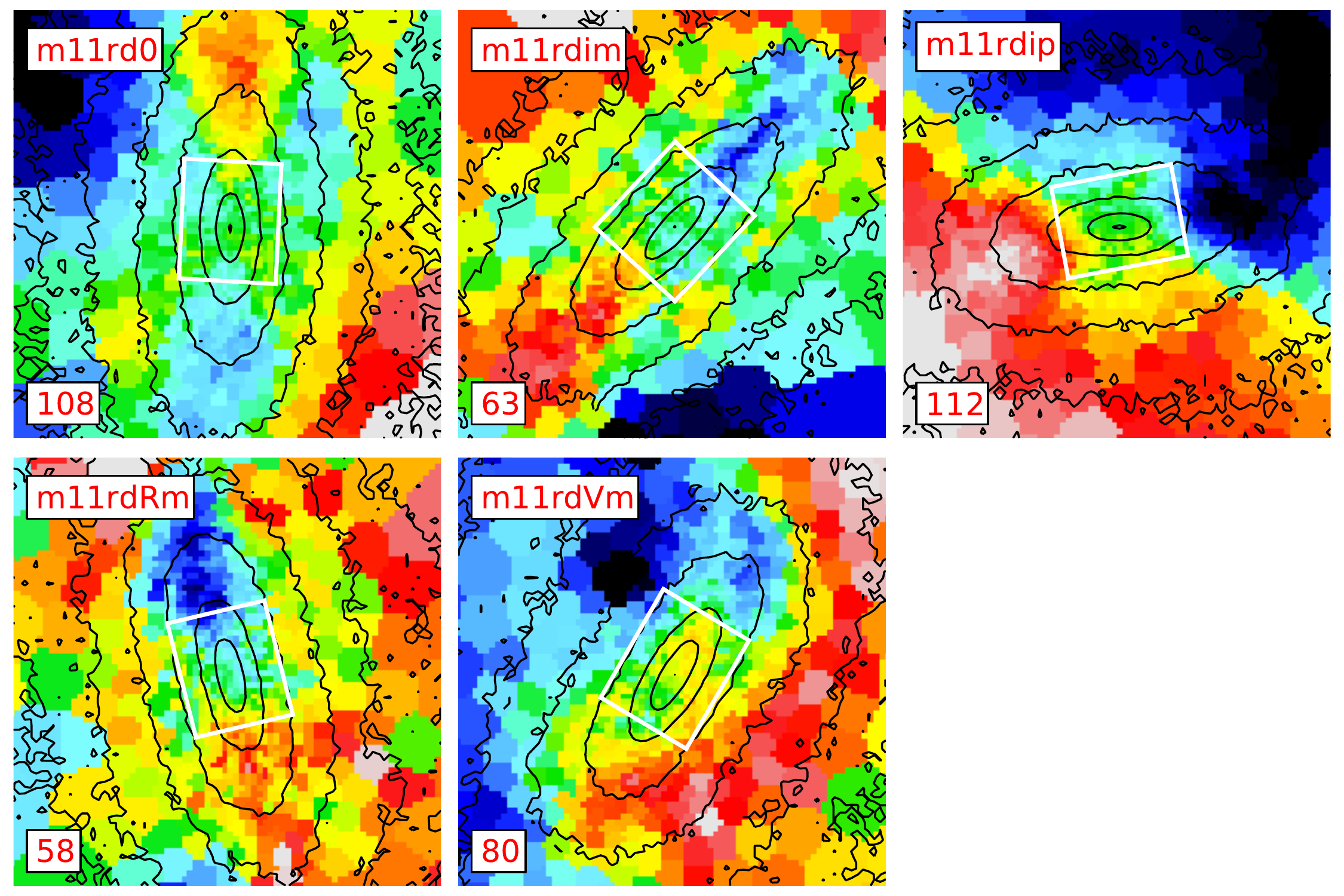} & \includegraphics[width=0.95\columnwidth]{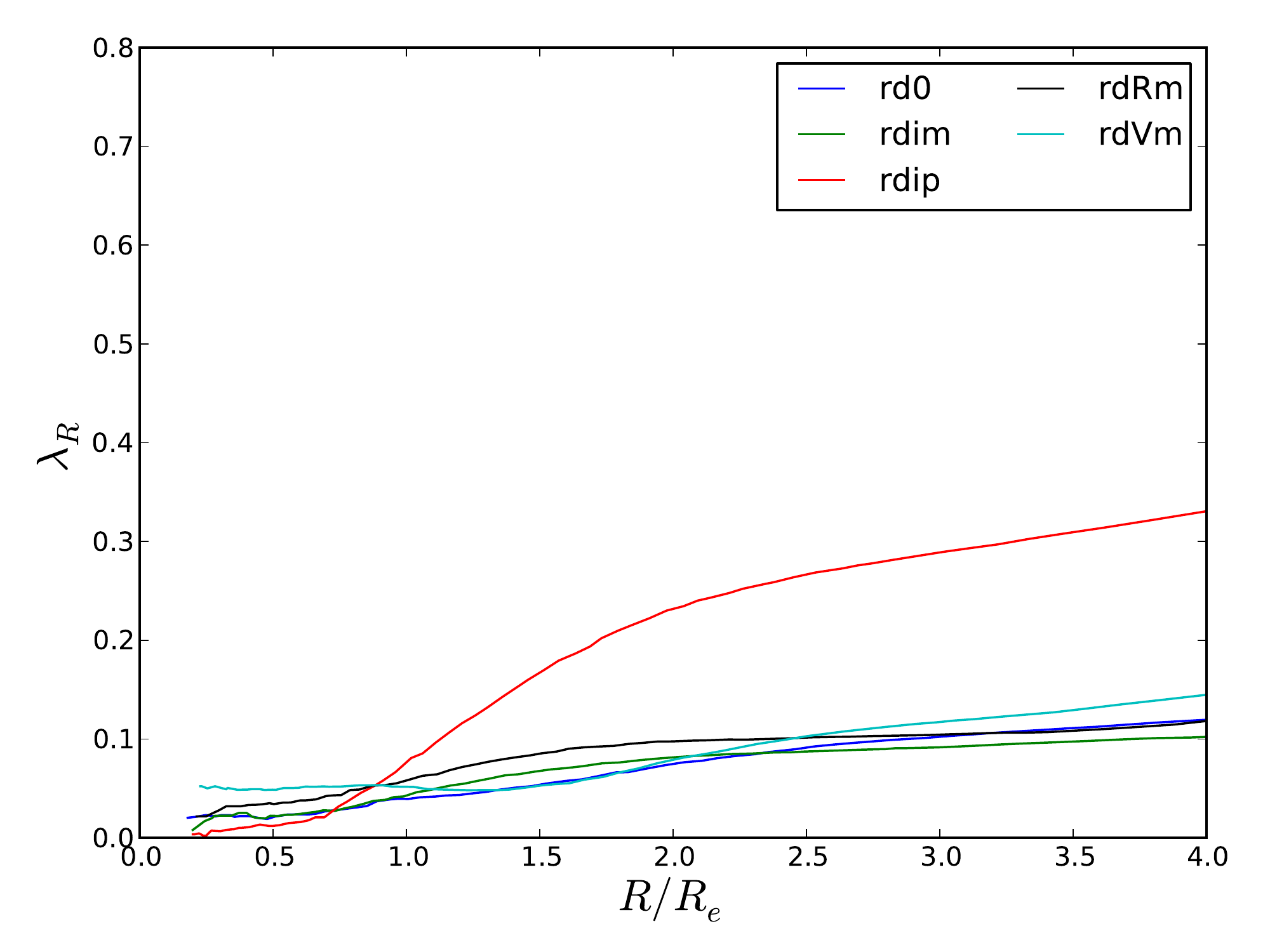}  \\
 &  \\
 \hline
 &  \\
  \includegraphics[width=1.05\columnwidth]{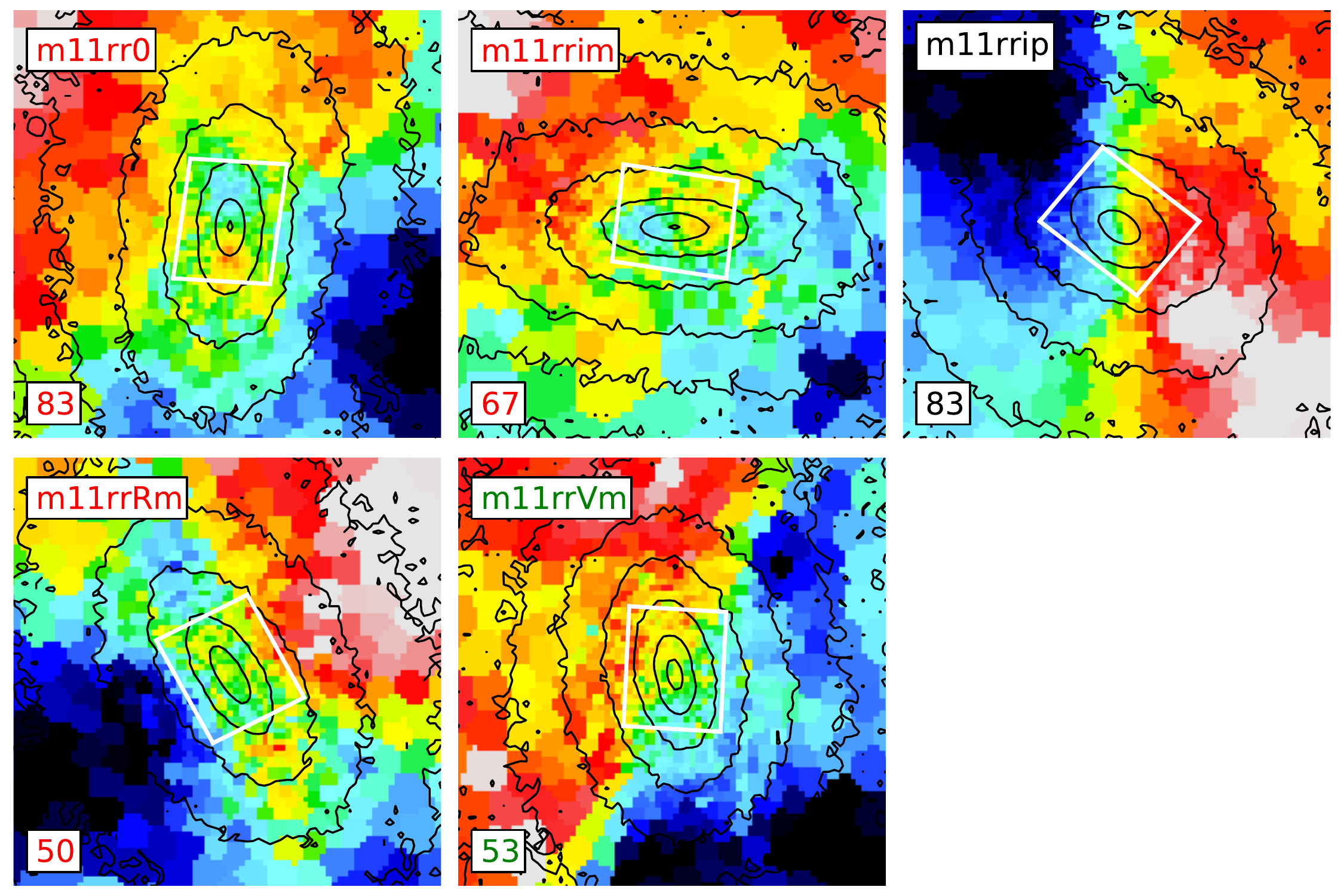} & \includegraphics[width=0.95\columnwidth]{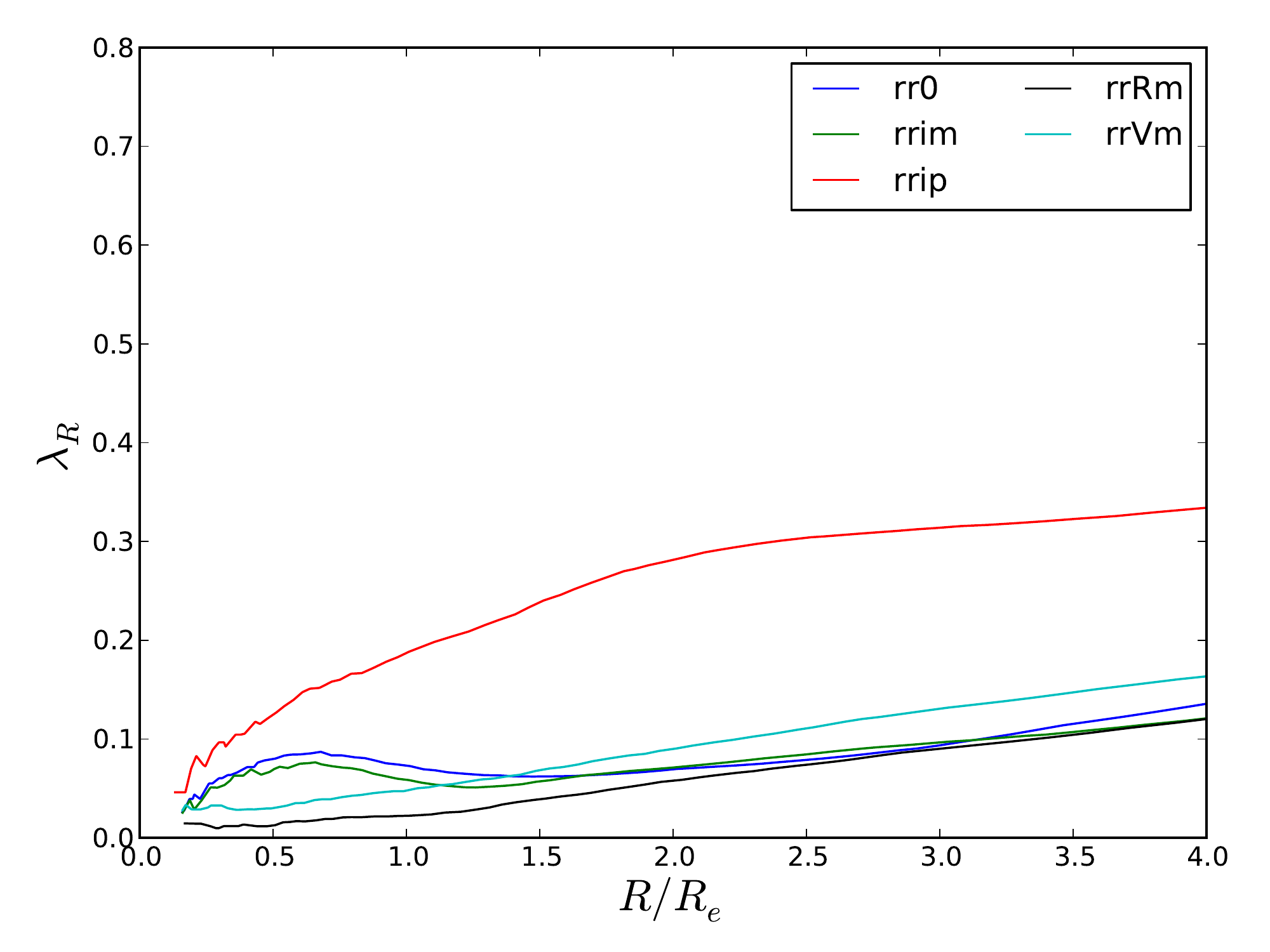}  \\
 &  \\
  \hline
 \end{tabular}
\caption{\textbf{Left:} Edge-on projection of the velocity field for the binary mergers of mass ratio 1:1 for the 3 type of orbits (top: \textbf{dd} orbit, middle: \textbf{rd} orbit, bottom: \textbf{rr} orbit). The black lines correspond to the iso-magnitude contours. The different initial conditions and velocity cuts are noted in the sub-panels: the color of the text indicates if the galaxy is classified as a fast rotator (black), a slow rotator with a KDC (red) or a slow rotator without KDC (green); the KDC may not be visible under the edge-on projection but is visible for most of the others. The field of view is $15\times15$~kpc$^2$. The white rectangle indicates a typical field covered by the instrument \sauron{} and corresponds to a field of $41" \times 33"$ for a galaxy at a distance of 20~Mpc, its orientation follows the photometric position angle taken at 3$R_e$. \textbf{Right:} The corresponding $\lambda_R$ profiles as a function of the radius $R$ divided by the effective radius $R_e$ of the edge-on projection.}
\label{fig:Vmerg11}
\end{figure*}

The binary merger simulations are classified in this way:
\begin{itemize}
 \item A figure corresponds to mergers with the same mass ratio, \textit{e.g} Fig.~\ref{fig:Vmerg11} shows the binary mergers of mass ratio 1:1
 \item A group of six projections associated to their $\lambda_R$ profiles correspond to a specific orbit of merging (\textbf{dd}, \textbf{rd} or \textbf{rr} orbits)
 \item The six projections correspond to the different initial conditions for a specific orbit of merging (\textit{0}, \textit{im}, \textit{ip}, \textit{Rm}, \textit{Vm}, \textit{Vp}, see Table~\ref{tab:taborbit})
 \item The label of the simulation (\textit{e.g} m31ddip) and the velocity cut are noted in the projected velocity maps.
\end{itemize}

\bigskip

\begin{figure*}
 \begin{tabular}{cc}
  \includegraphics[width=0.70\columnwidth]{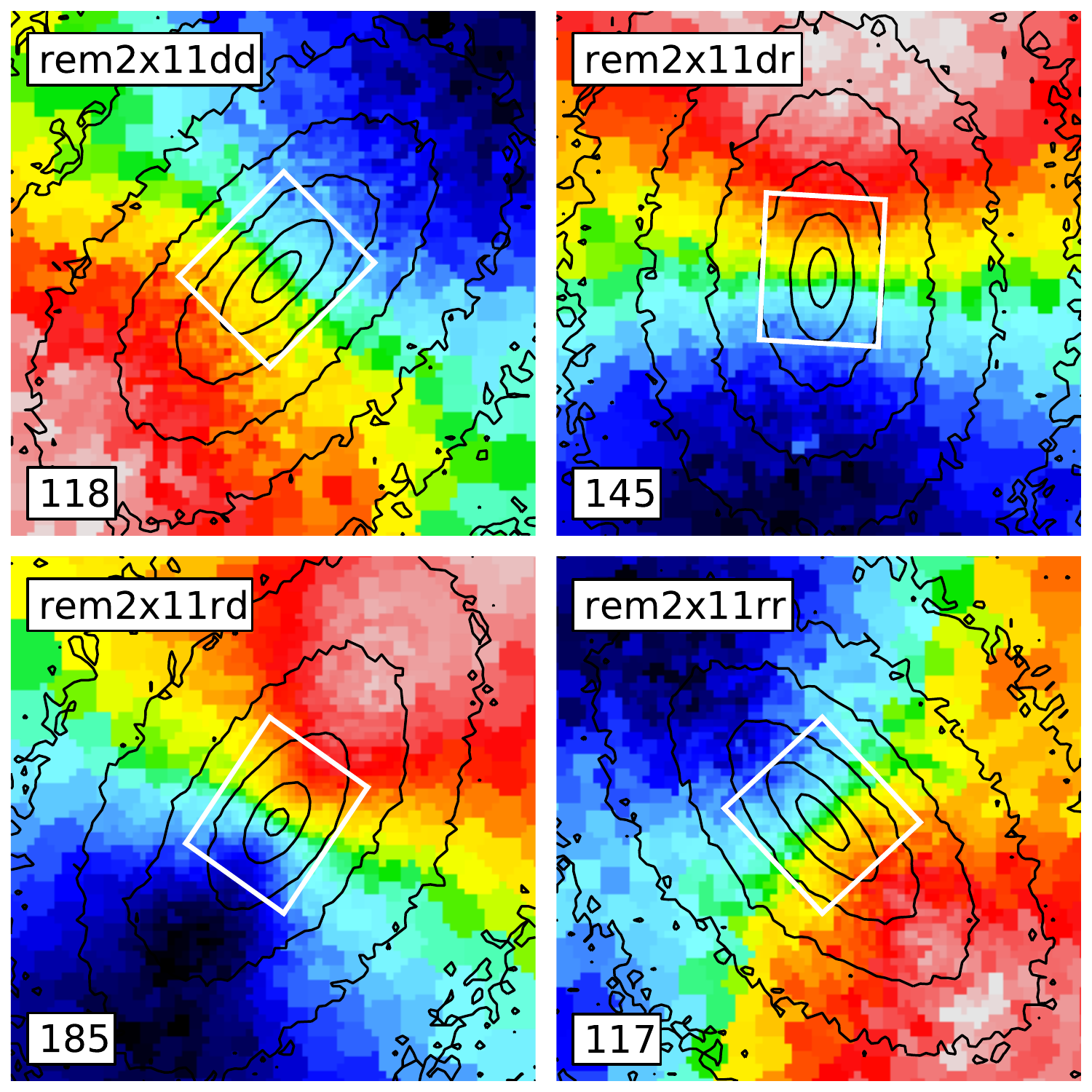} & \includegraphics[width=0.95\columnwidth]{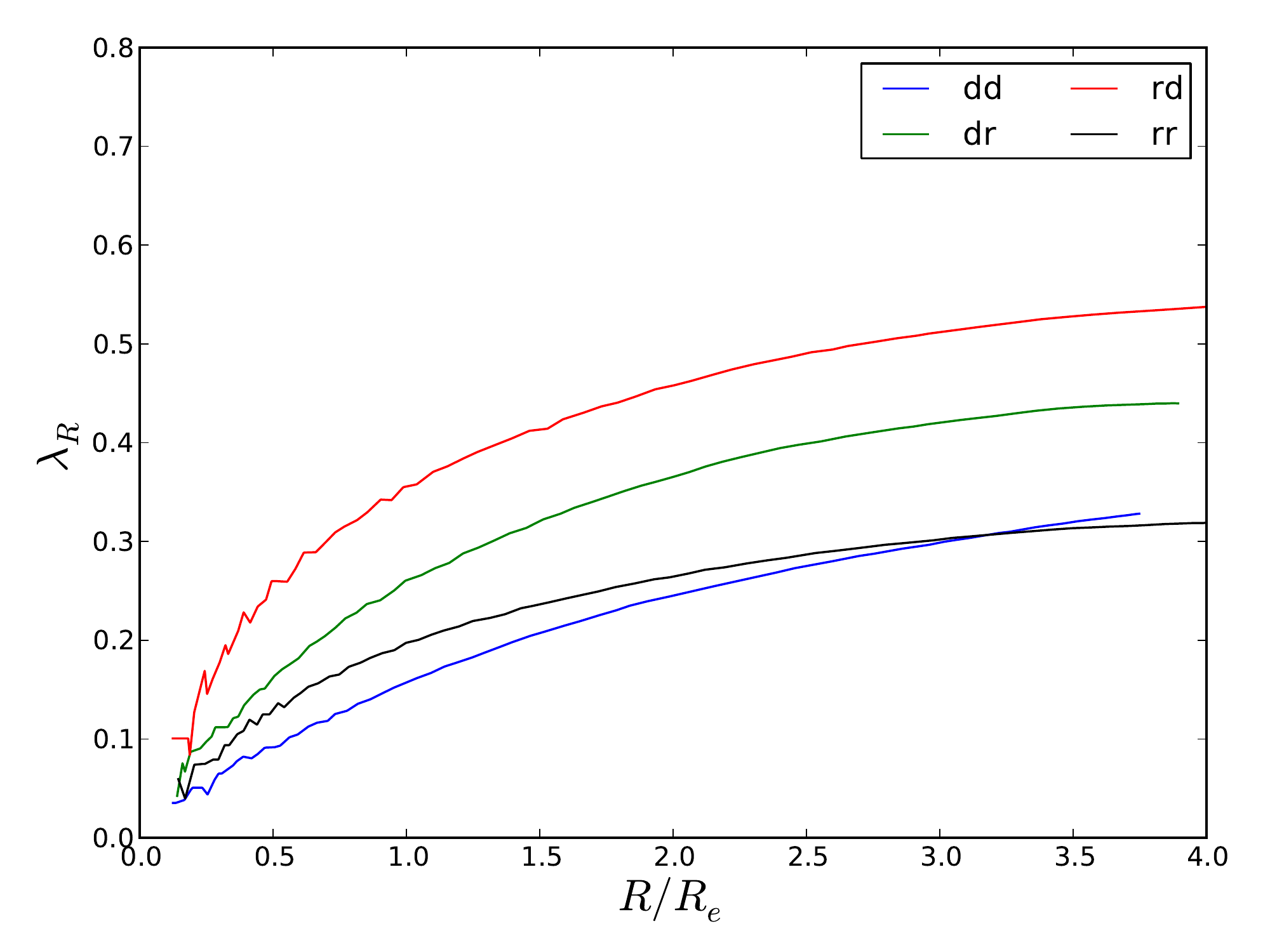}  \\
 \end{tabular}
\caption{\textbf{Left:} Edge-on projection of the velocity field for remergers of two remnants of binary mergers of mass ratio 1:1. The black lines correspond to the iso-magnitude contours. The different initial angular momentum spins and velocity cuts are noted in the sub-panels. The field of view is $15\times15$~kpc$^2$.  The white rectangle indicates a typical field covered by the instrument \sauron{} and corresponds to a field of $41" \times 33"$ for a galaxy at a distance of 20~Mpc, its orientation follows the photometric position angle taken at 3$R_e$. \textbf{Right:} The corresponding $\lambda_R$ profiles as a function of the radius $R$ divided by the effective radius $R_e$ of the edge-on projection.}
\label{fig:Vremergers}
\end{figure*}

The remergers are classified in this way:
\begin{itemize}
 \item A figure corresponds to remergers of two remnants of binary mergers with the same initial mass ratio, \textit{e.g} Fig.~\ref{fig:Vremergers} shows the remergers of two remnants of binary mergers of mass ratio 1:1.
 \item The four subpanels of a figure correspond to the four different orbit of merging (\textbf{dd}, \textbf{dr}, \textbf{rd} or \textbf{rr})
 \item The label of the simulation (\textit{e.g} rem21g10+Sdd) and the velocity cut are noted in the projected velocity maps.
\end{itemize}

\begin{figure*}
 \begin{tabular}{cc}
  \includegraphics[width=1.05\columnwidth]{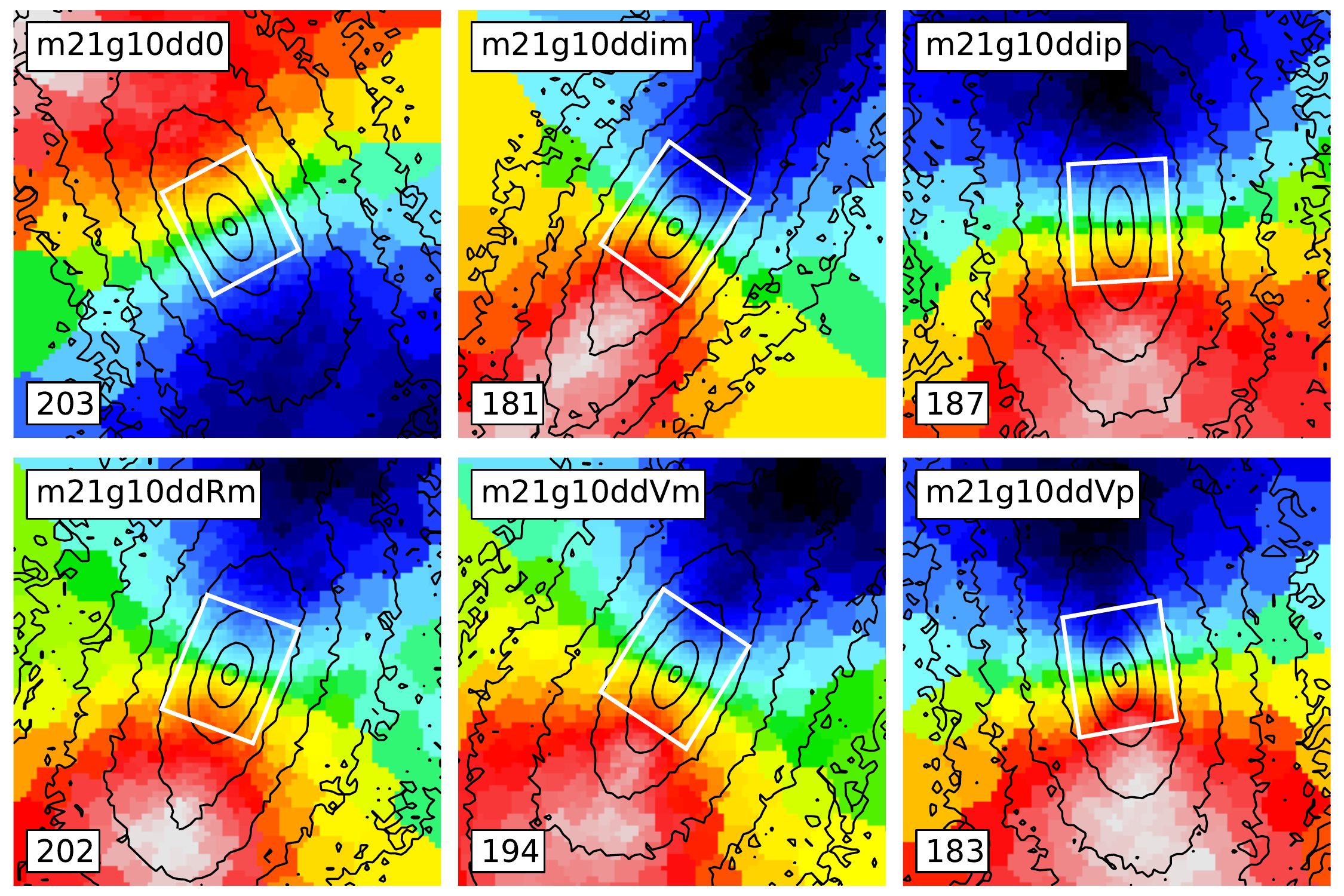} & \includegraphics[width=0.95\columnwidth]{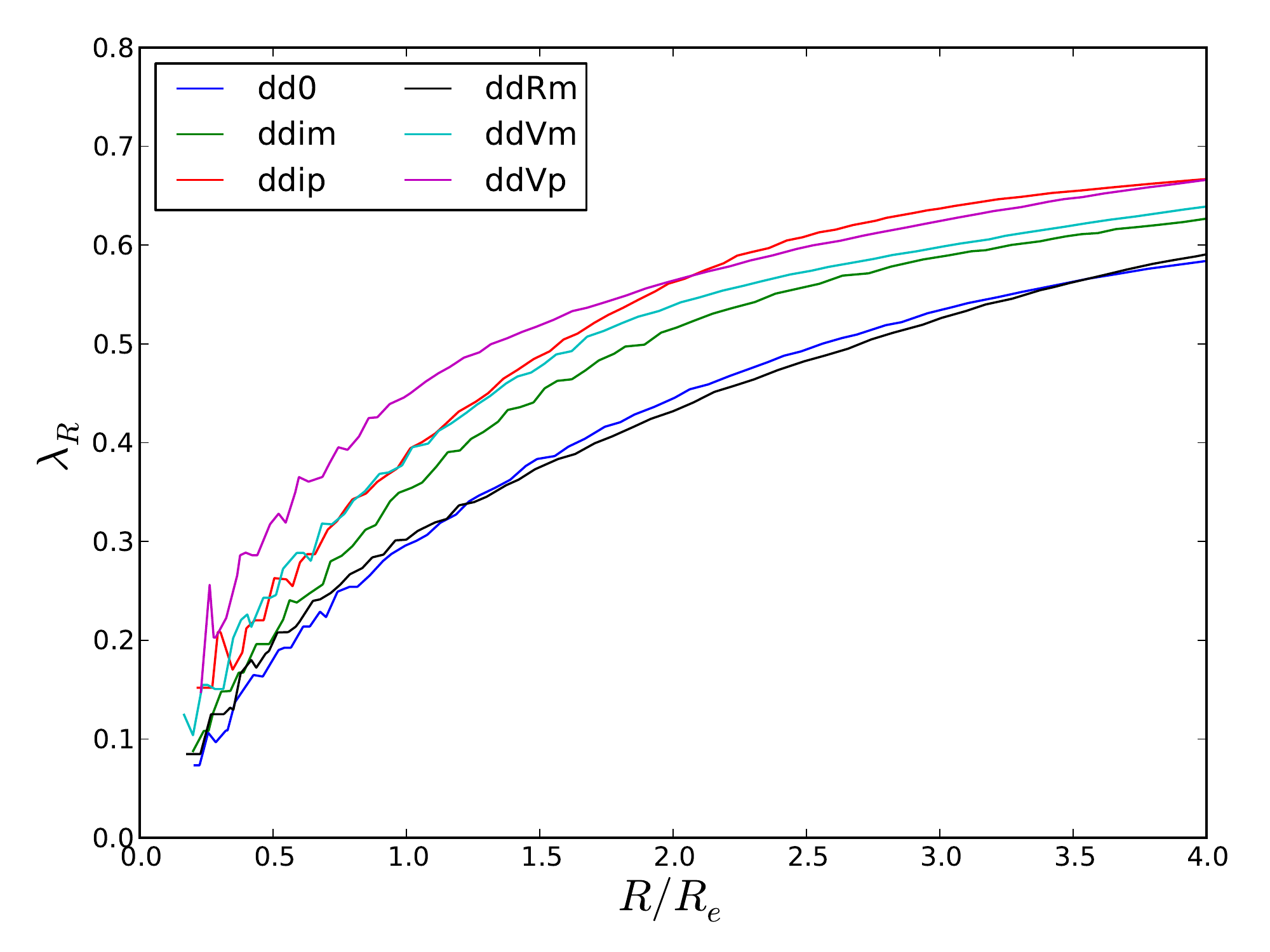}  \\
 & \\
  \hline
 & \\
  \includegraphics[width=1.05\columnwidth]{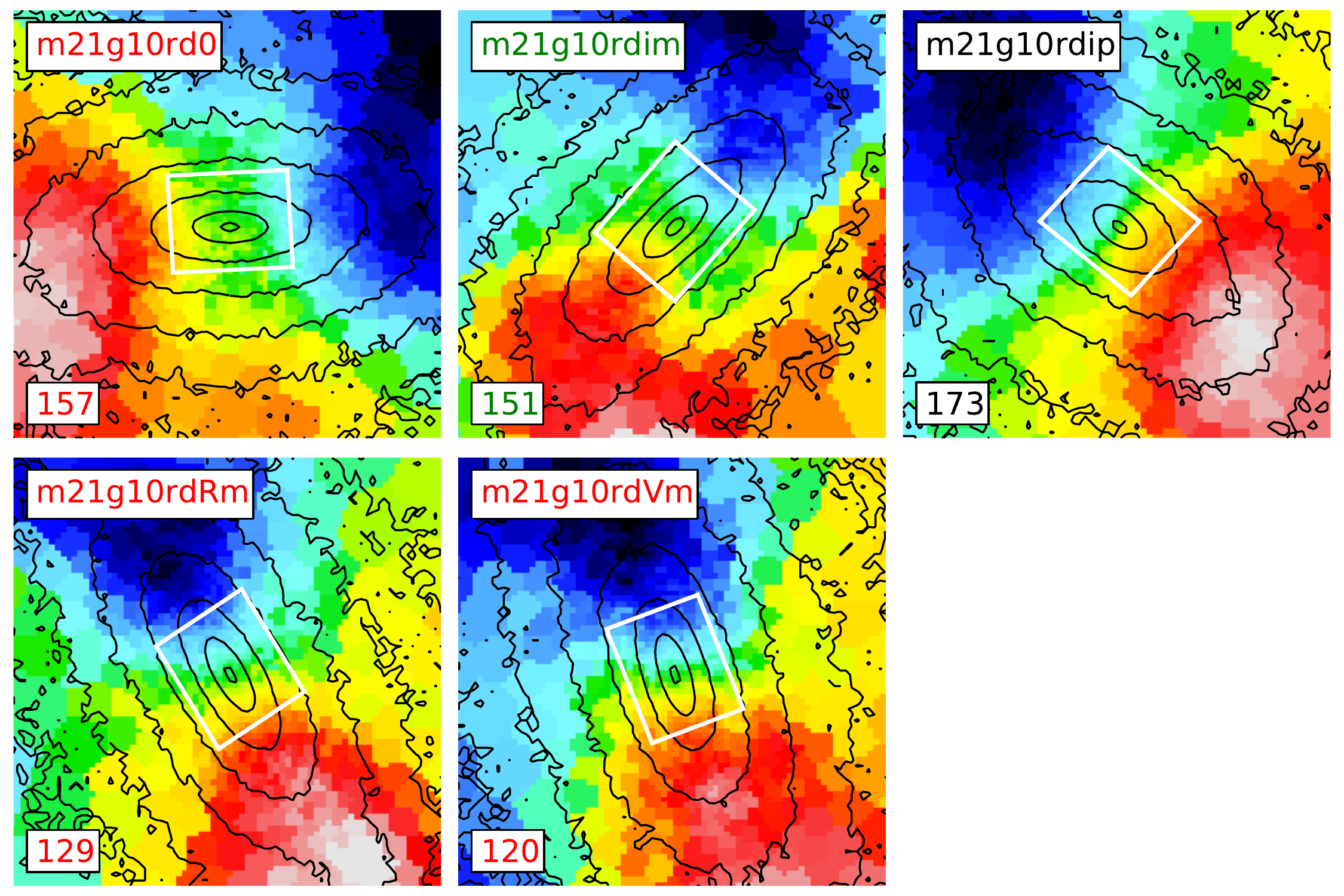} & \includegraphics[width=0.95\columnwidth]{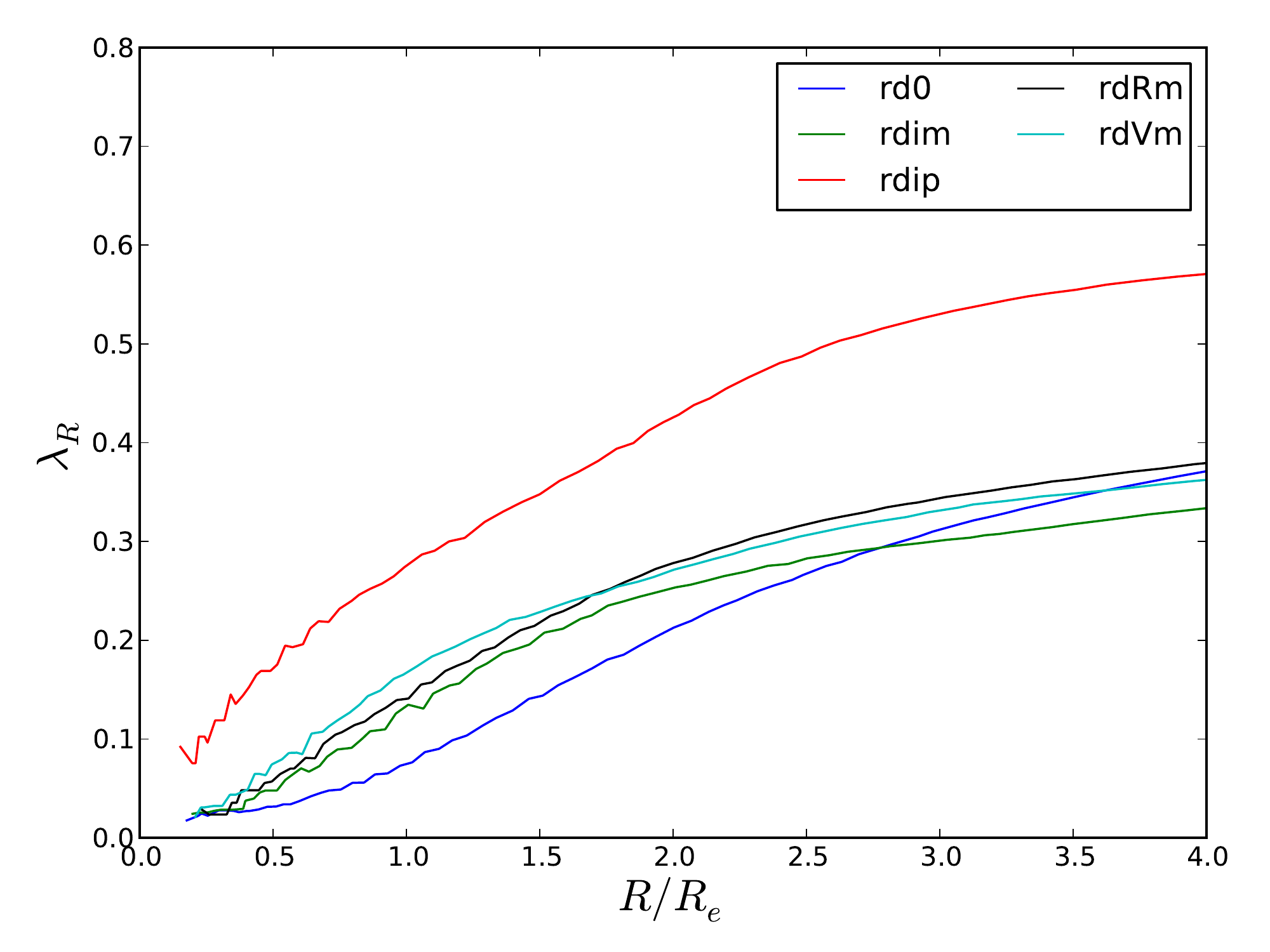}  \\
 & \\
  \hline
 & \\
  \includegraphics[width=1.05\columnwidth]{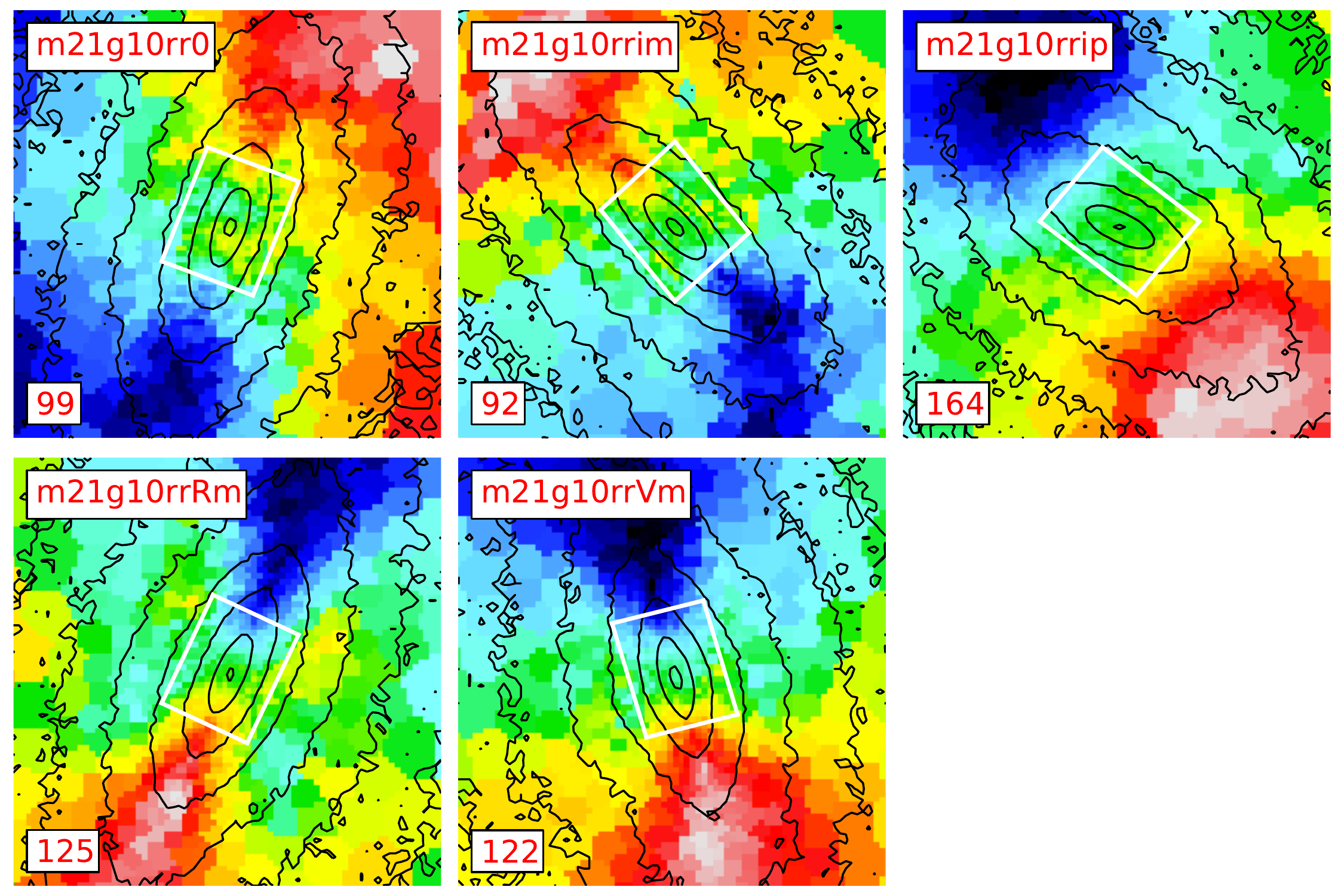} & \includegraphics[width=0.95\columnwidth]{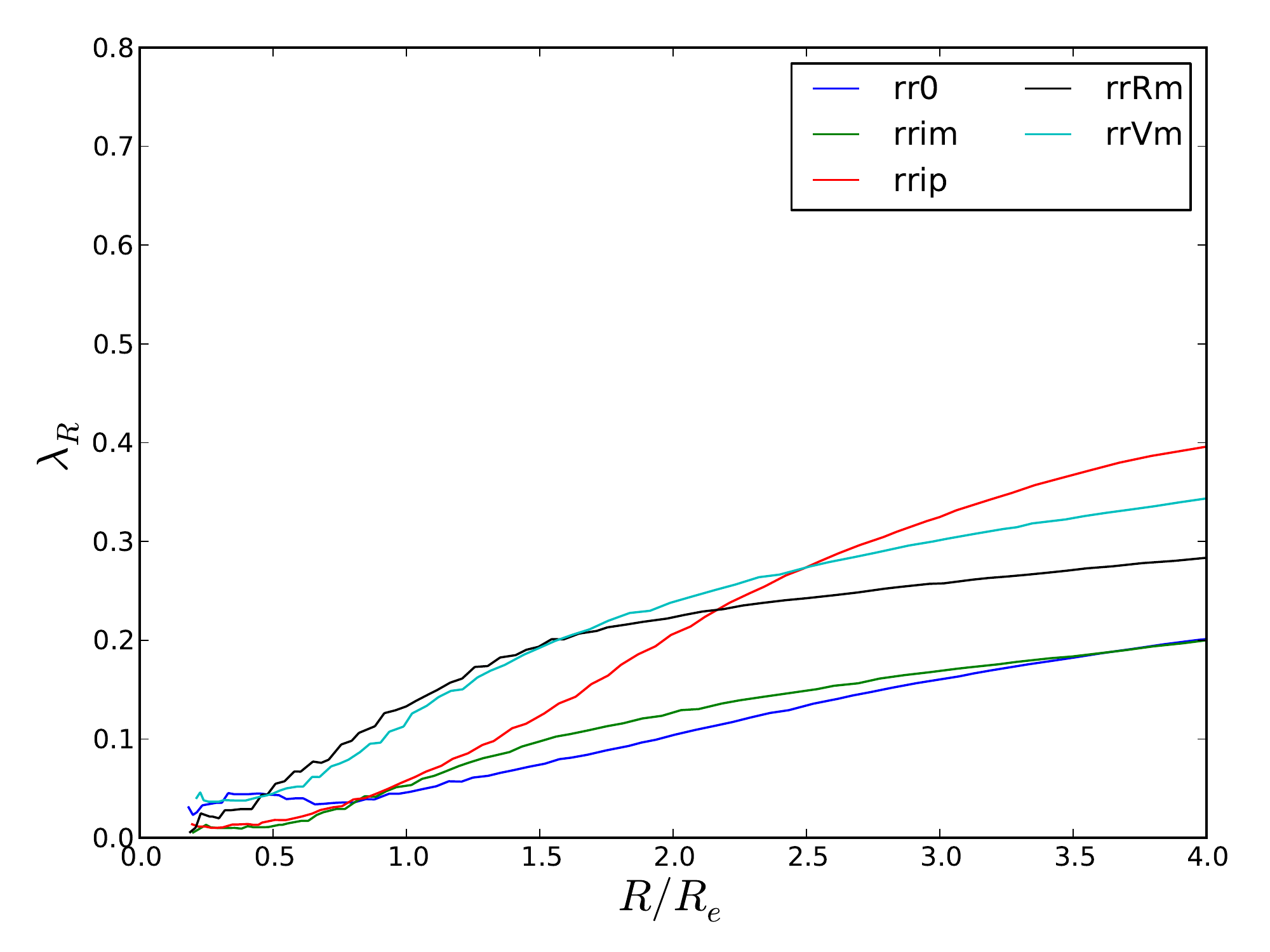}  \\
 & \\
  \hline
 \end{tabular}
\caption{Same as Fig.~\ref{fig:Vmerg11} for the binary mergers of mass ratio 2:1 with 10 per cent of gas.}
\label{fig:Vmerg2110}
\end{figure*}

\begin{figure*}
 \begin{tabular}{cc}
  \includegraphics[width=1.05\columnwidth]{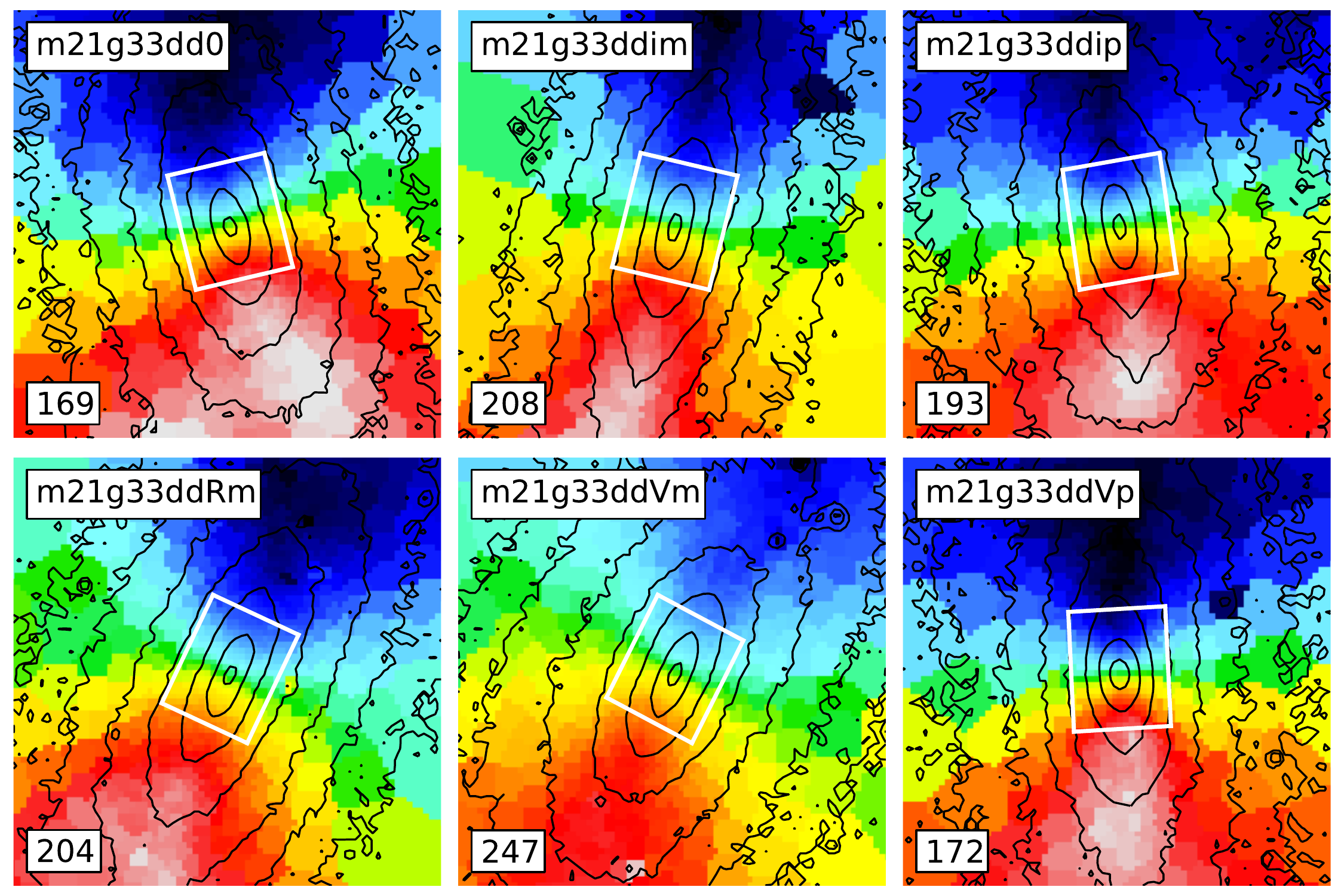} & \includegraphics[width=0.95\columnwidth]{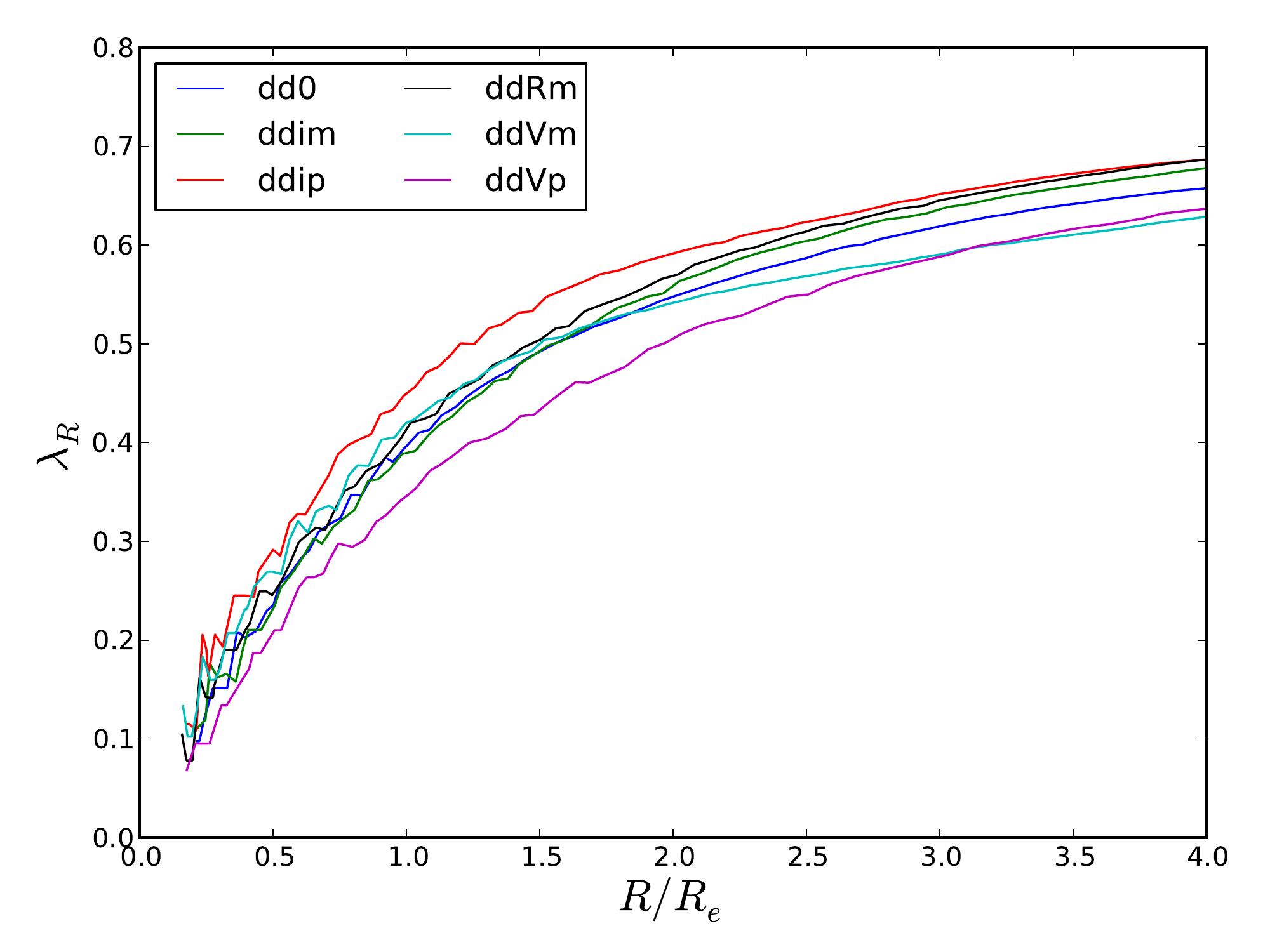}  \\
 & \\
  \hline
 & \\
  \includegraphics[width=1.05\columnwidth]{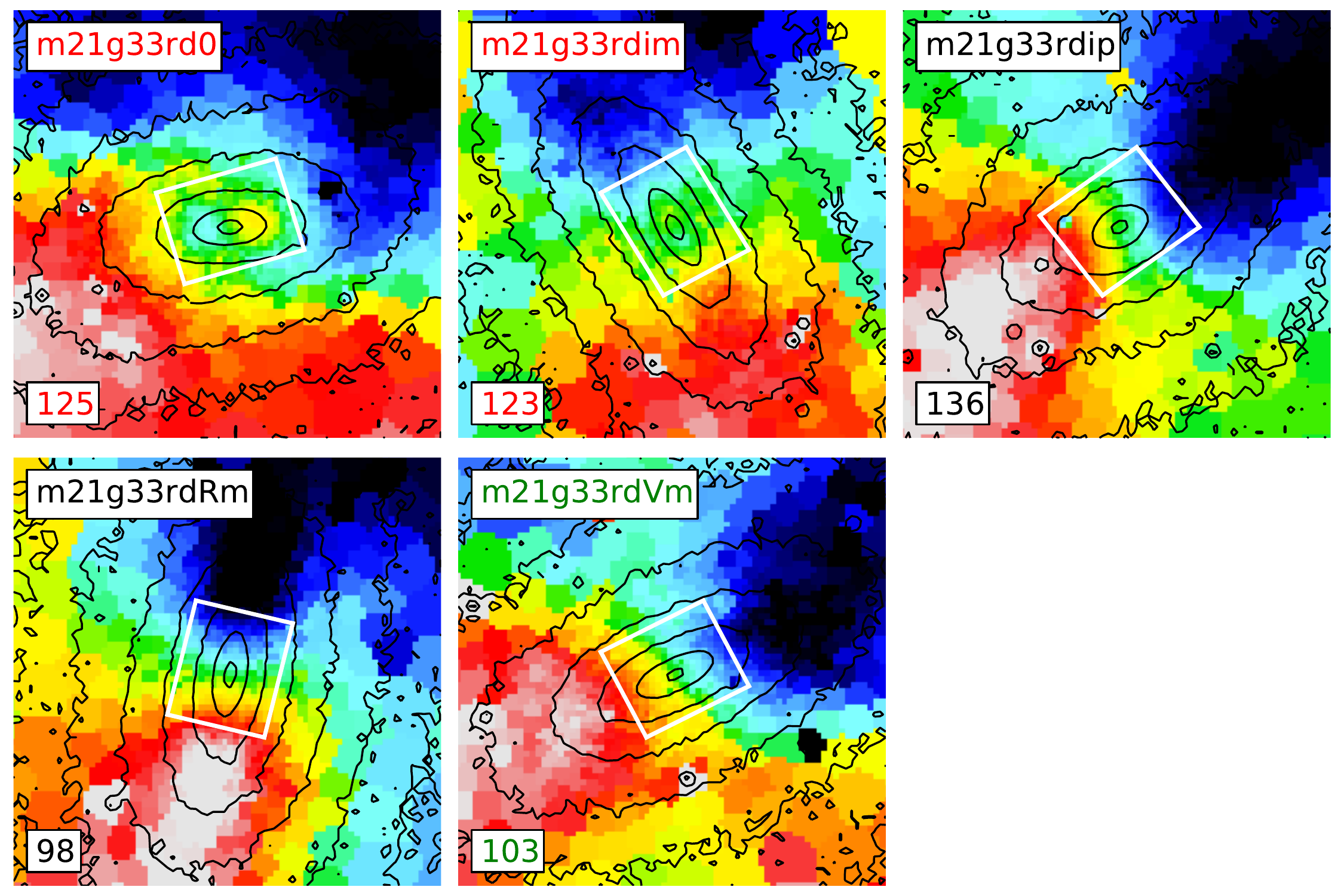} & \includegraphics[width=0.95\columnwidth]{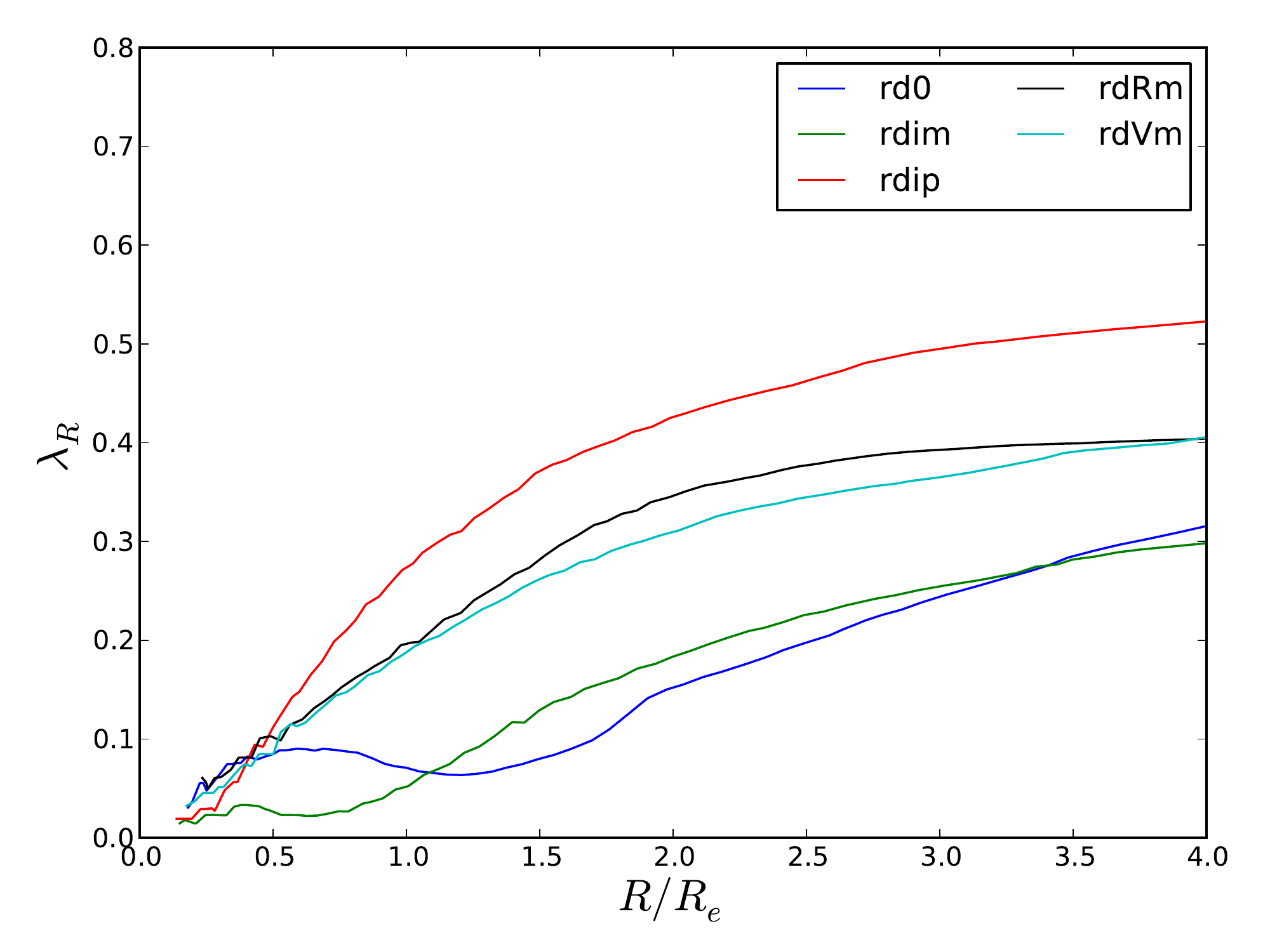}  \\
 & \\
  \hline
 & \\
  \includegraphics[width=1.05\columnwidth]{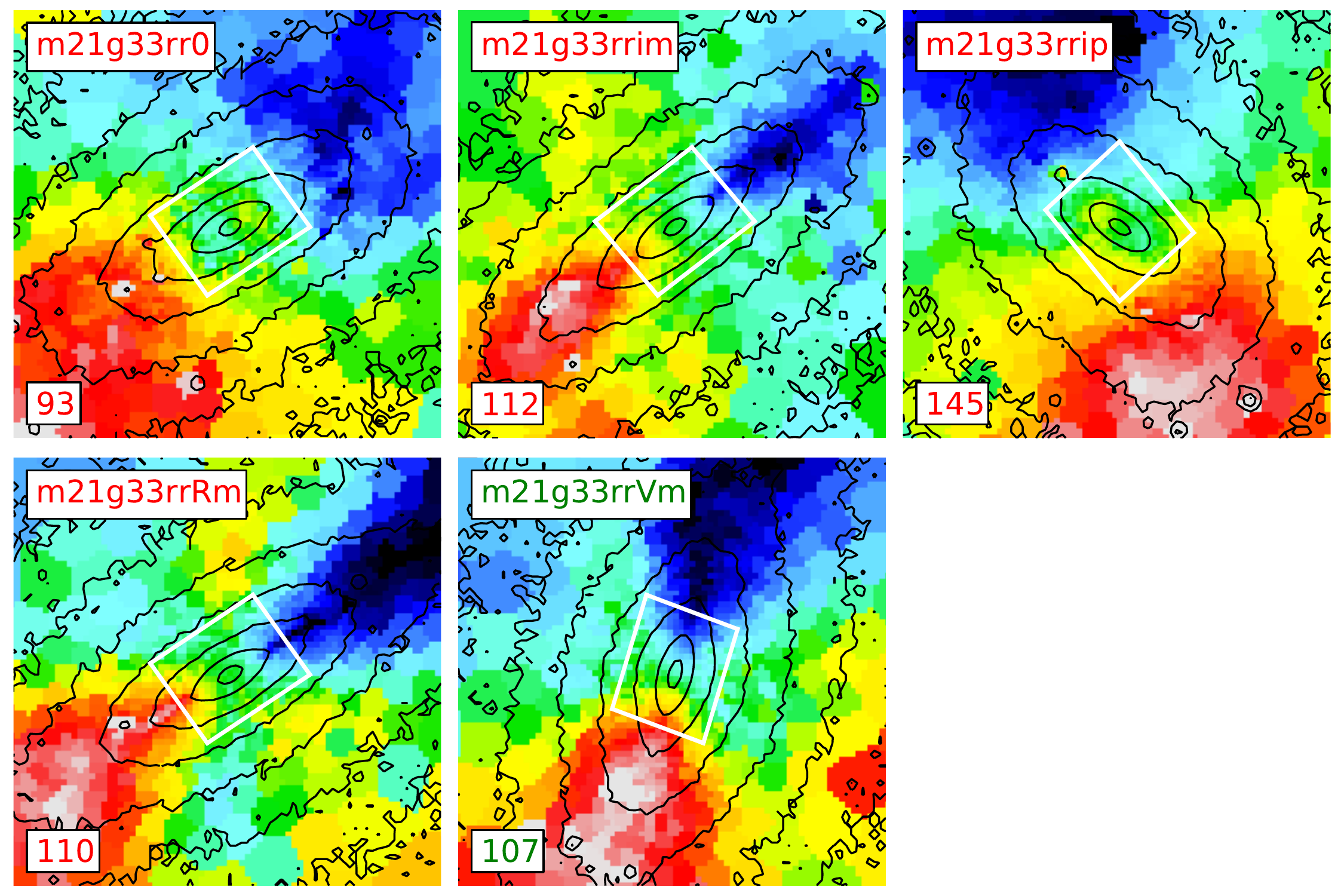} & \includegraphics[width=0.95\columnwidth]{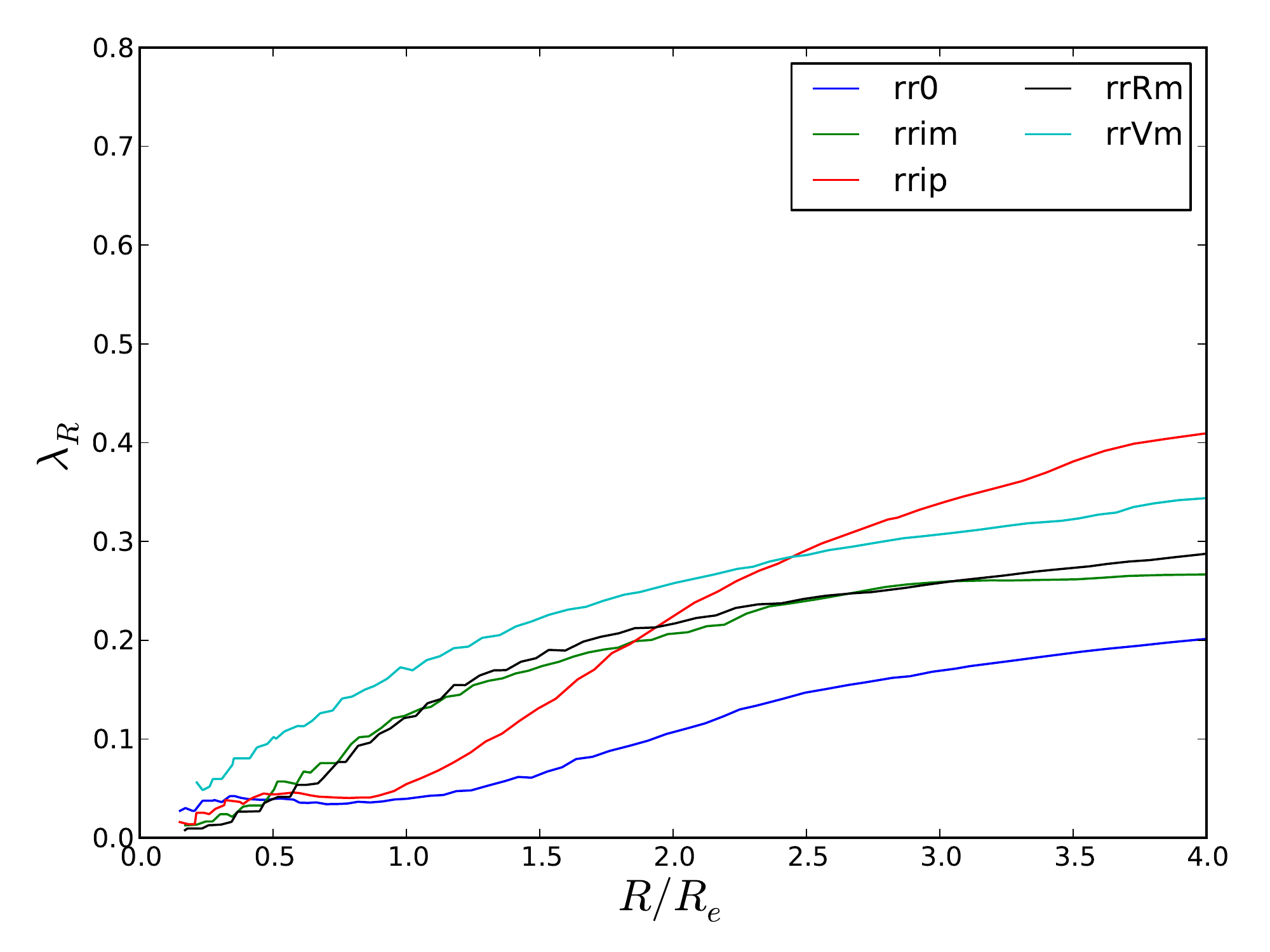}  \\
 & \\
  \hline
 \end{tabular}
\caption{Same as Fig.~\ref{fig:Vmerg11} for the binary mergers of mass ratio 2:1 with 33 per cent of gas.}
\label{fig:Vmerg2133}
\end{figure*}

\begin{figure*}
 \begin{tabular}{cc}
  \includegraphics[width=1.05\columnwidth]{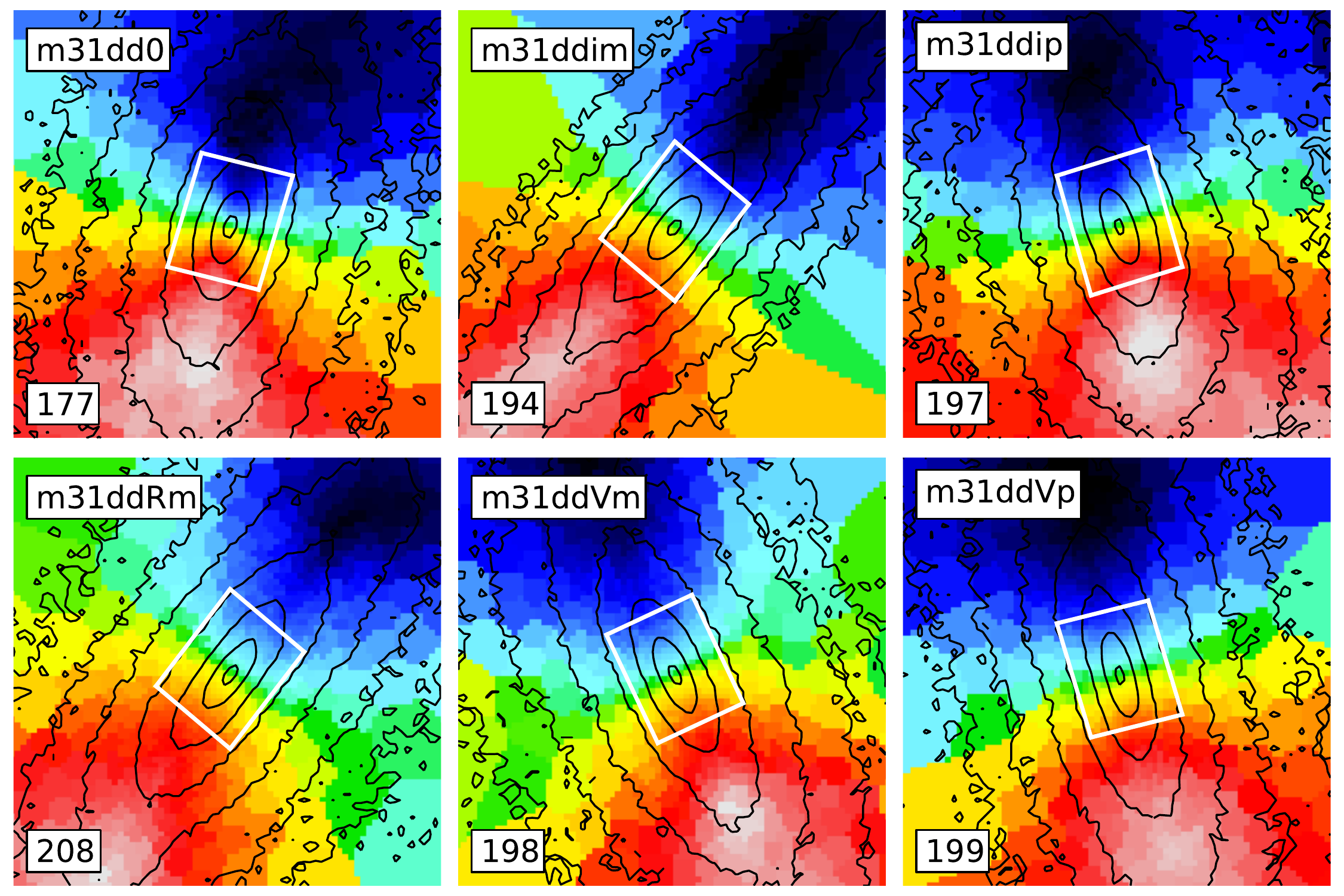} & \includegraphics[width=0.95\columnwidth]{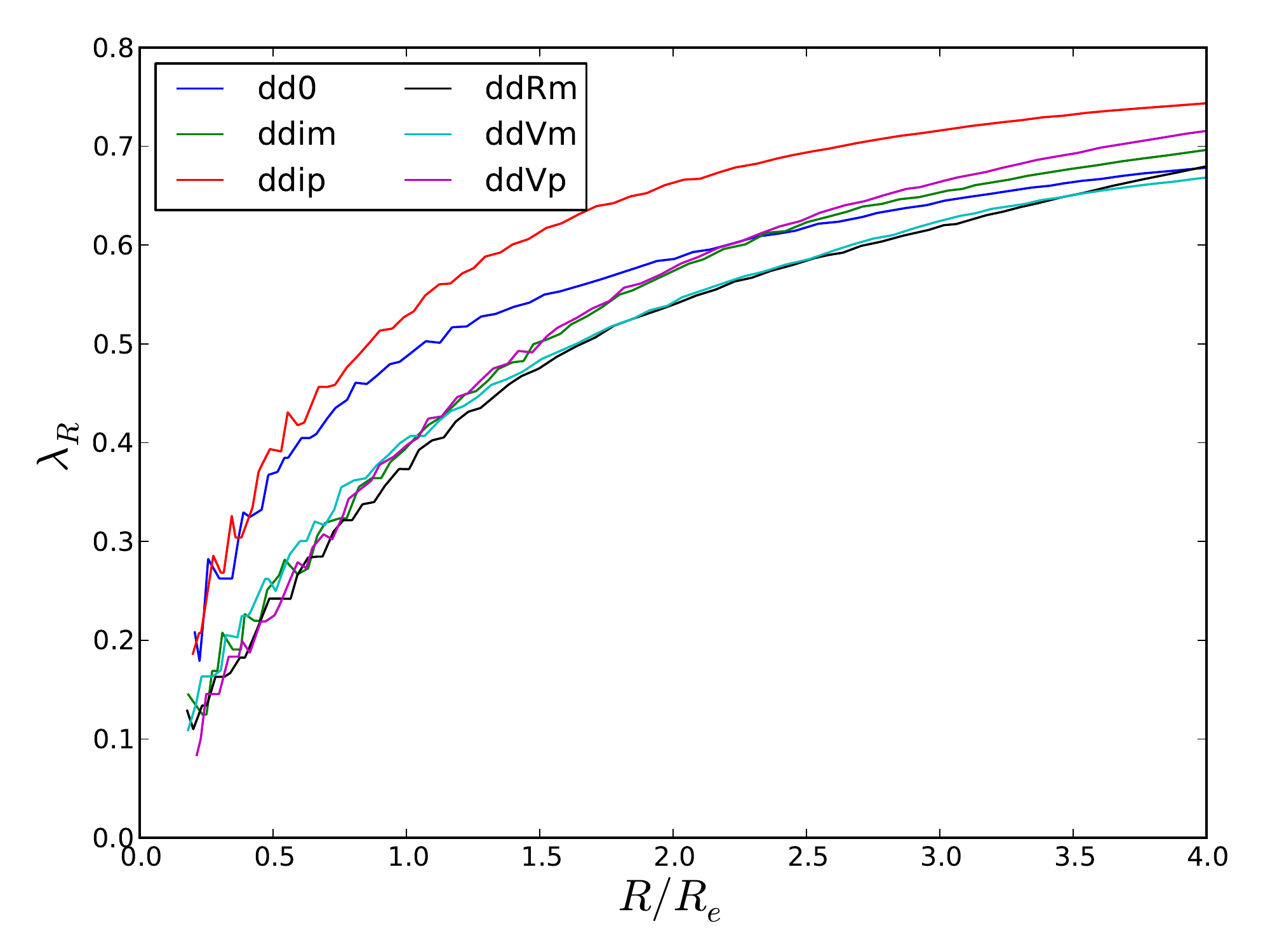}  \\
 & \\
  \hline
 & \\
  \includegraphics[width=1.05\columnwidth]{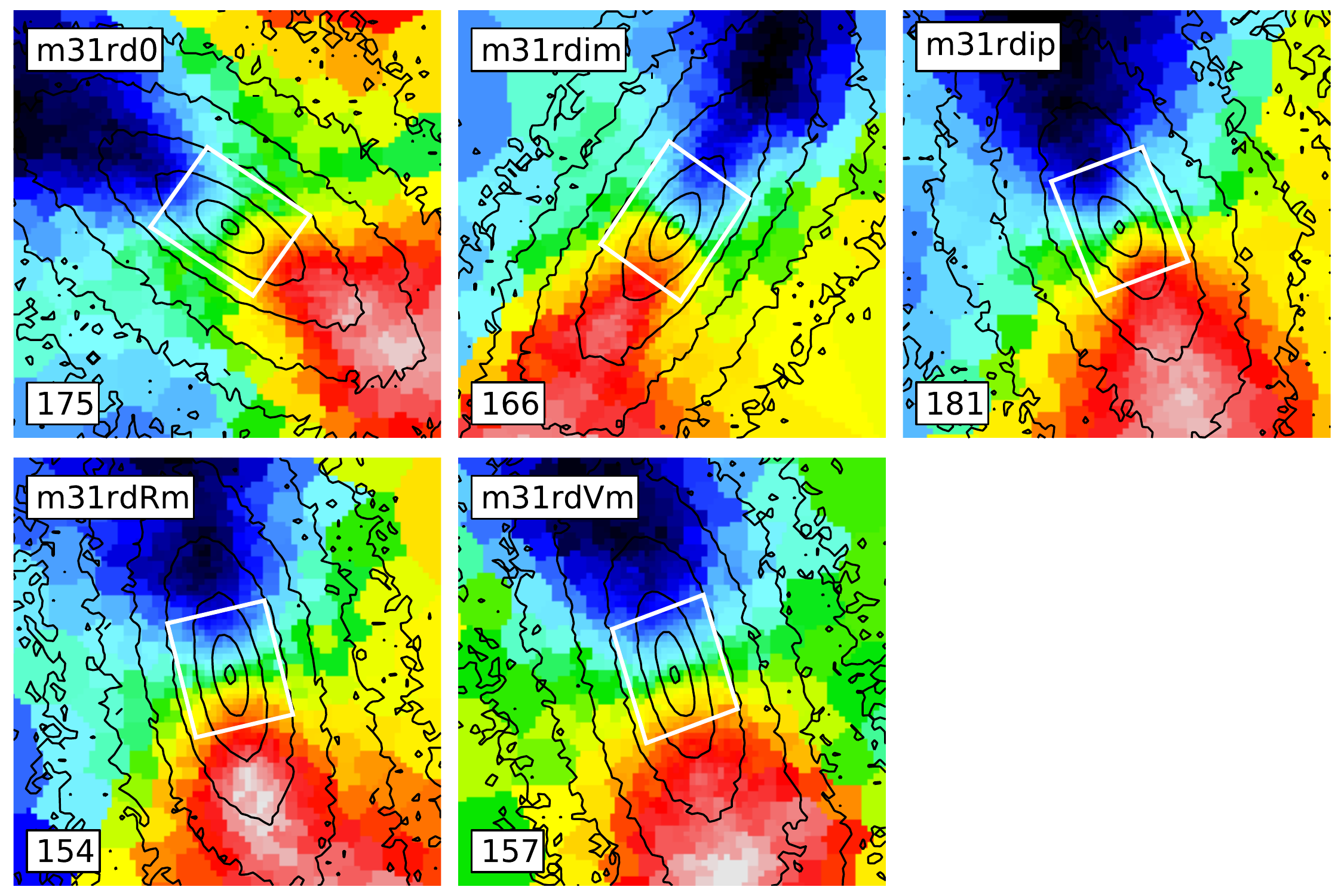} & \includegraphics[width=0.95\columnwidth]{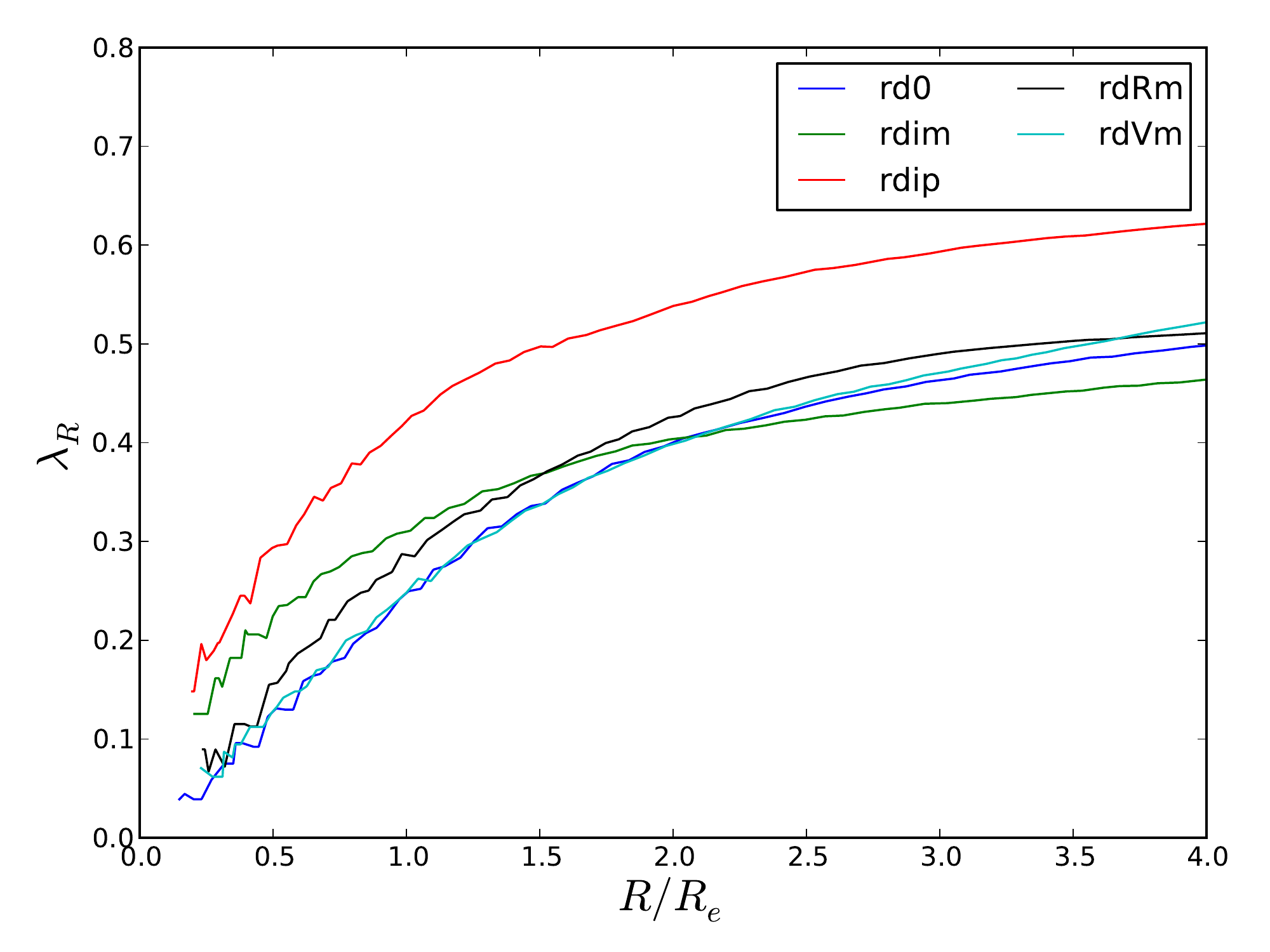}  \\
 & \\
  \hline
 & \\
 \includegraphics[width=1.05\columnwidth]{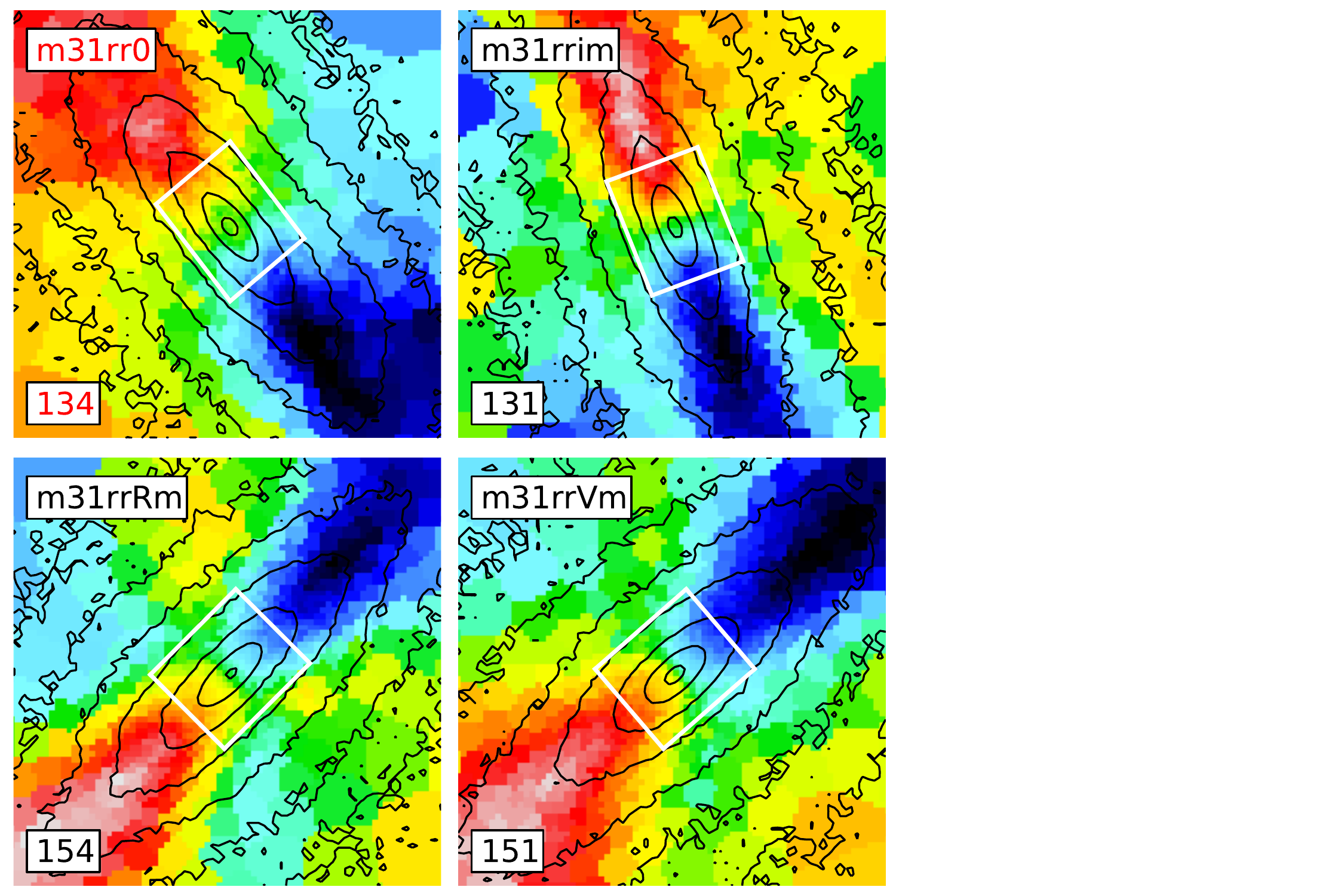} & \includegraphics[width=0.95\columnwidth]{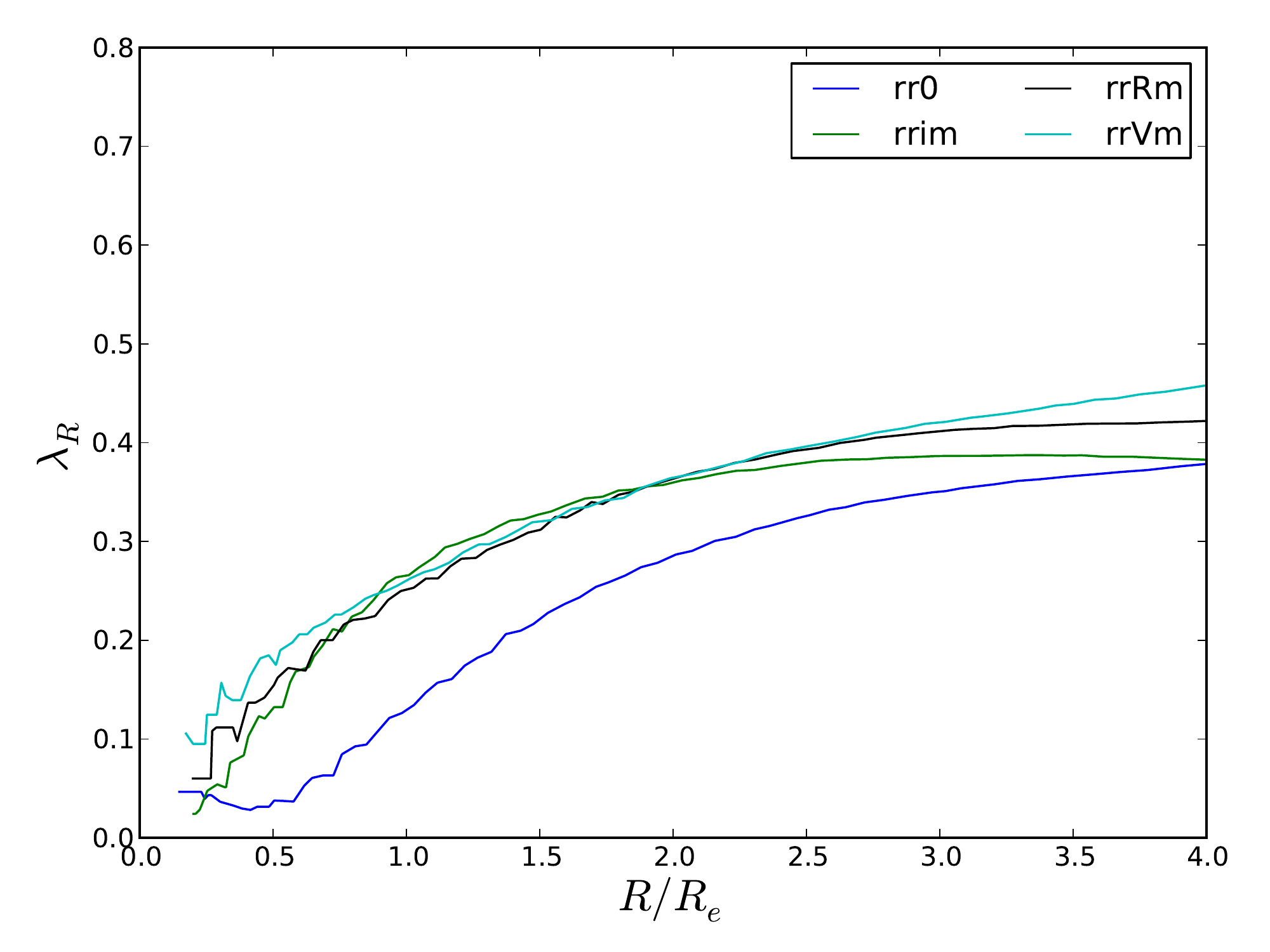}  \\
 & \\
 \hline
 \end{tabular}
\caption{Same as Fig.~\ref{fig:Vmerg11} for the binary mergers of mass ratio 3:1.}
\label{fig:Vmerg31}
\end{figure*}

\begin{figure*}
 \begin{tabular}{cc}
  \includegraphics[width=1.05\columnwidth]{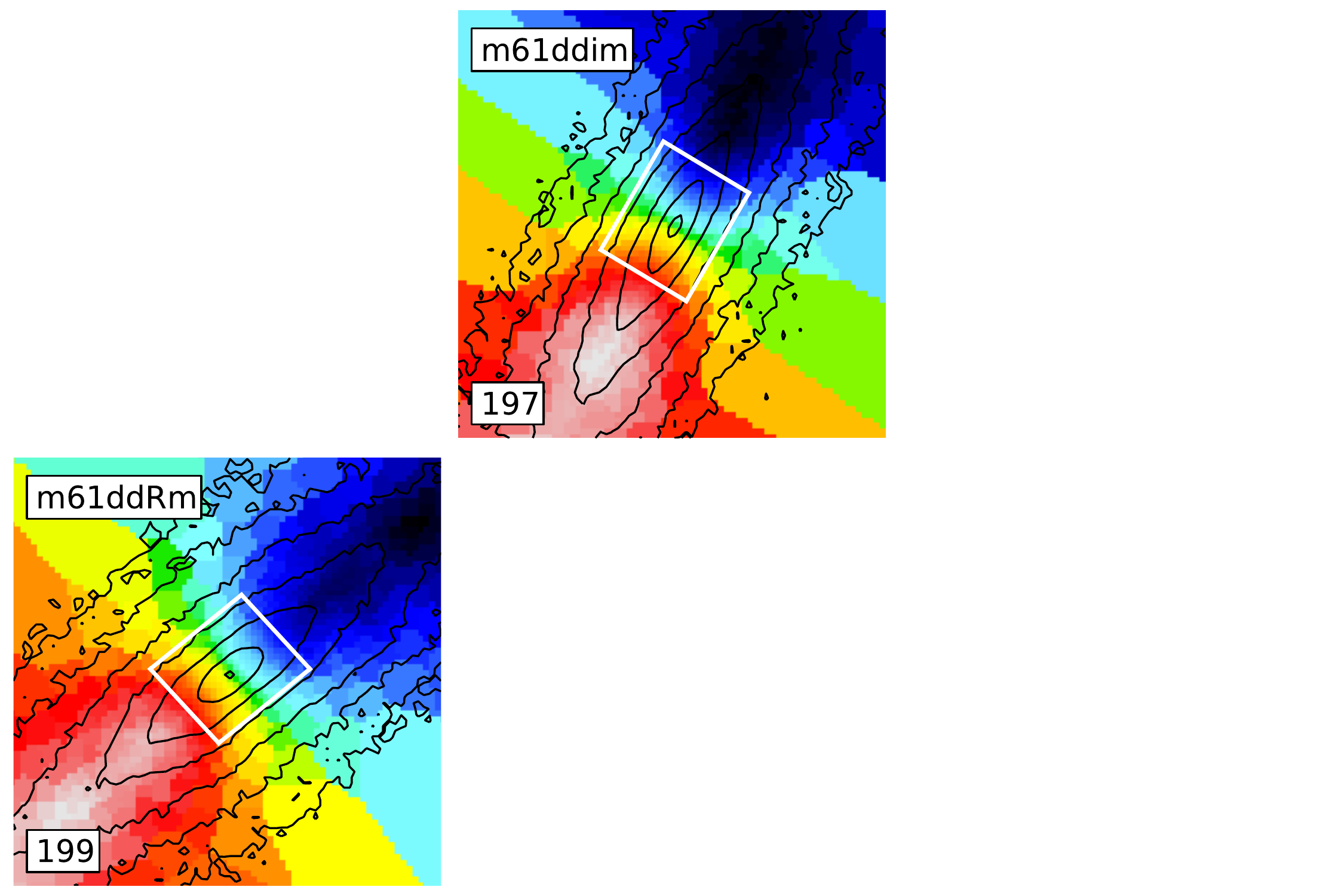} & \includegraphics[width=0.95\columnwidth]{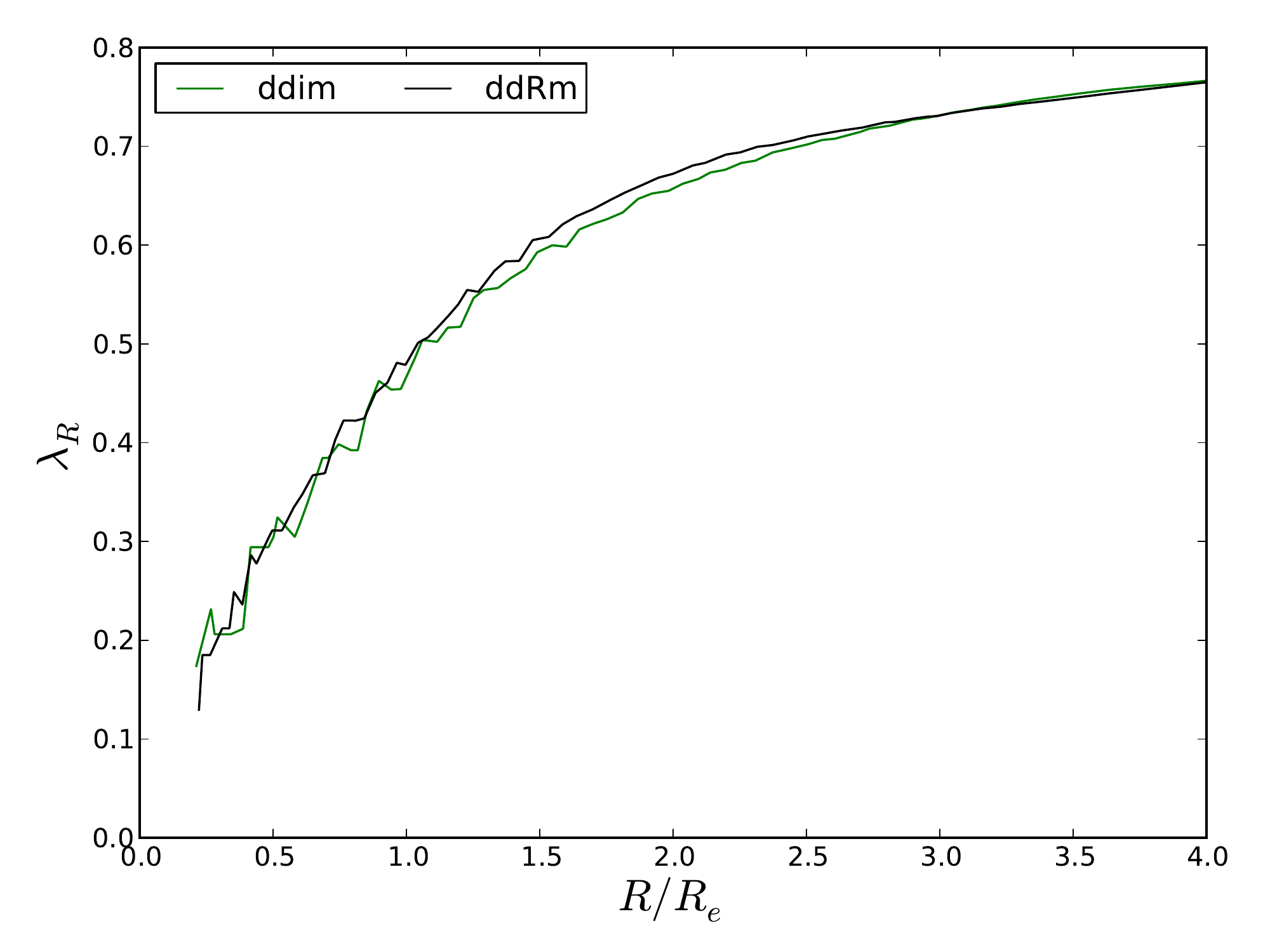}  \\
 & \\
  \hline
 & \\
  \includegraphics[width=1.05\columnwidth]{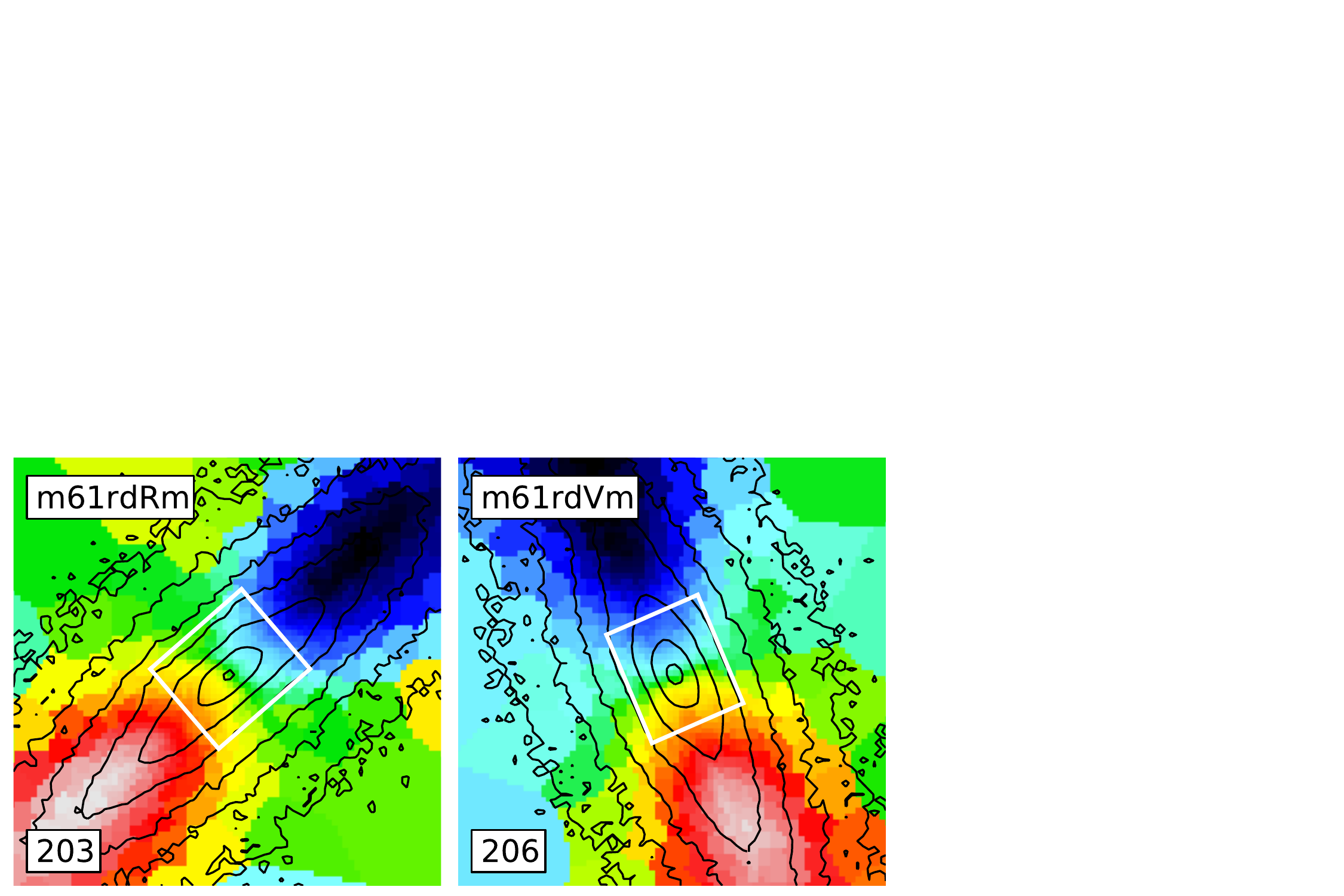} & \includegraphics[width=0.95\columnwidth]{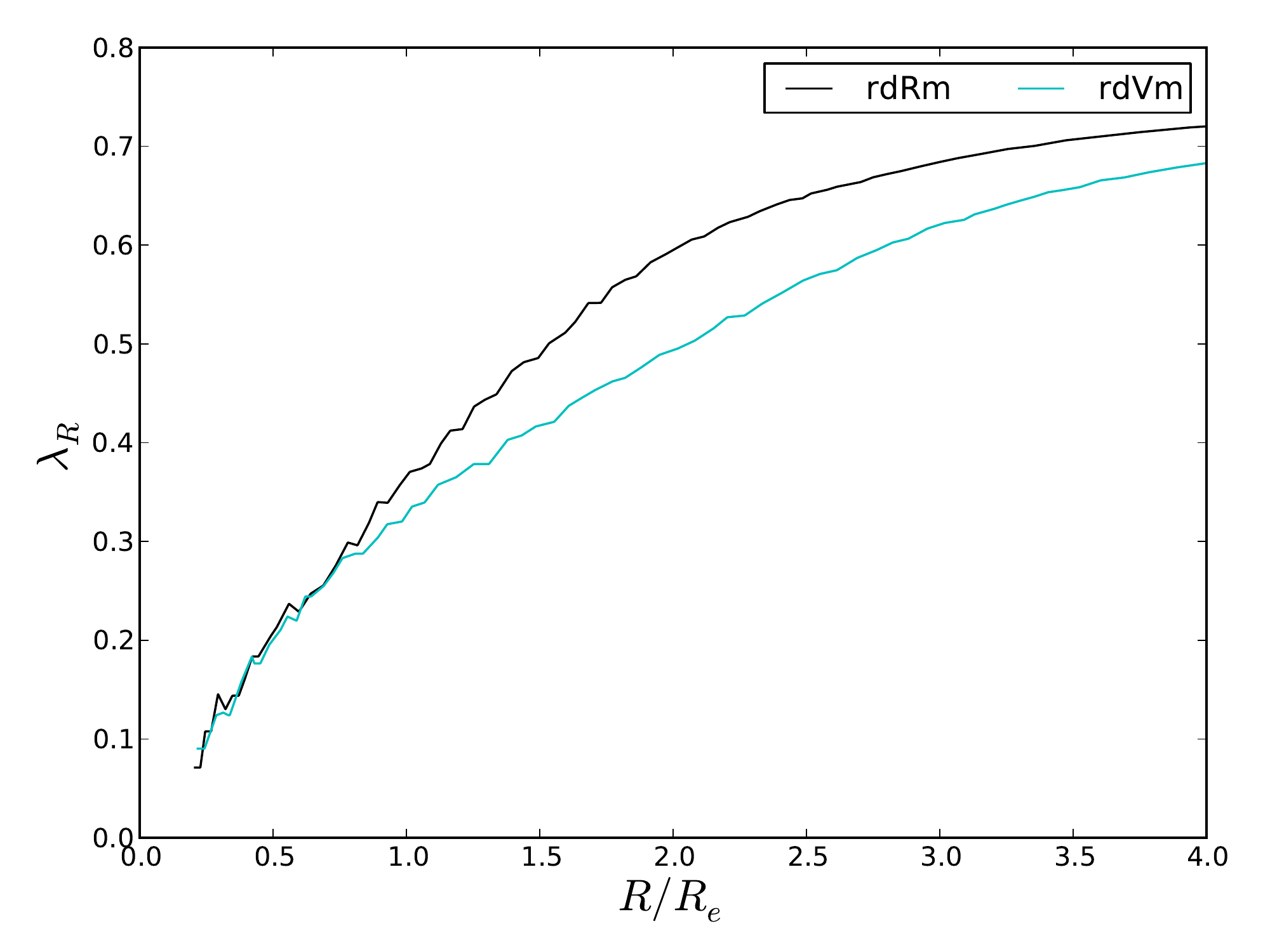}  \\
 & \\
  \hline
 & \\
  \includegraphics[width=1.05\columnwidth]{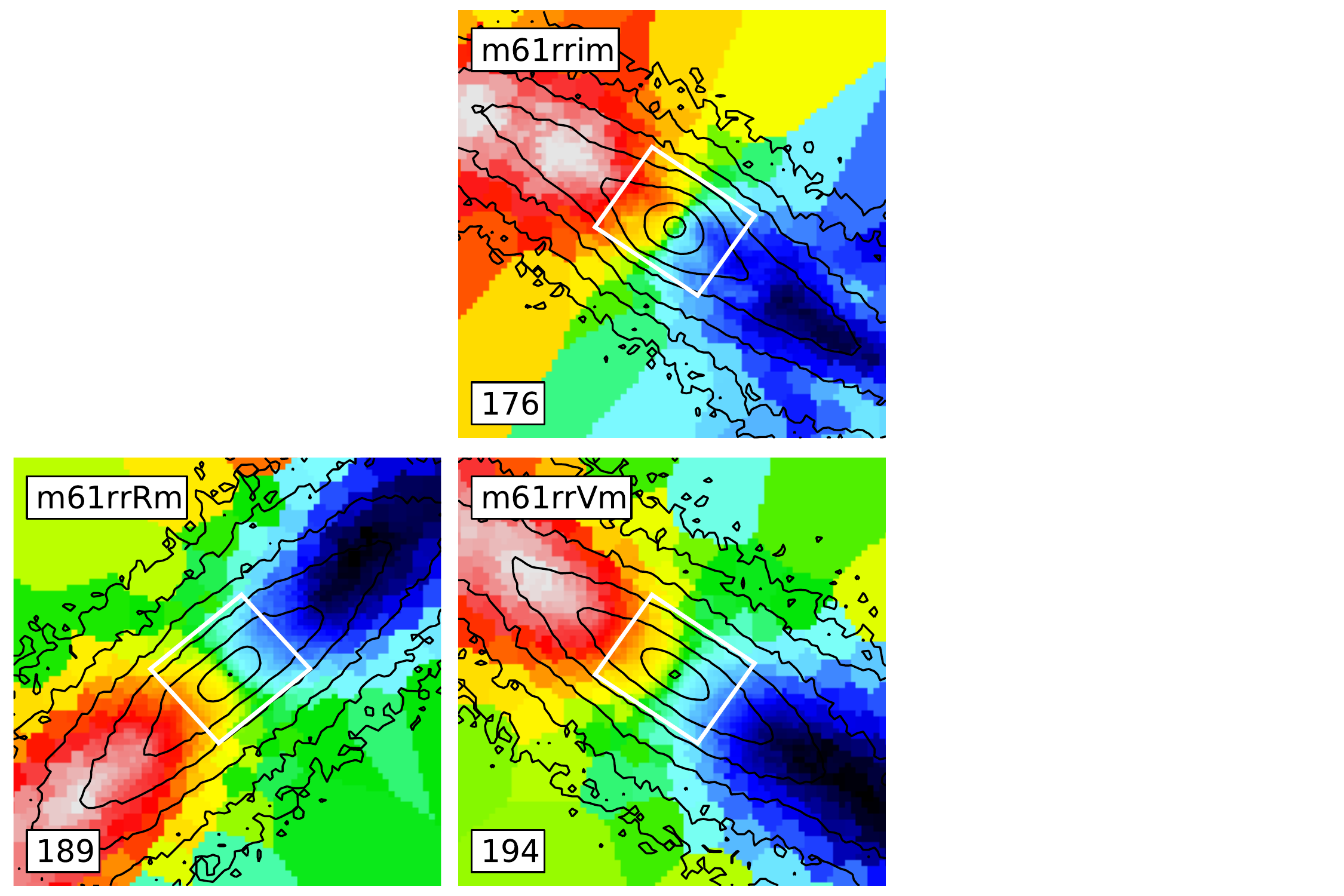} & \includegraphics[width=0.95\columnwidth]{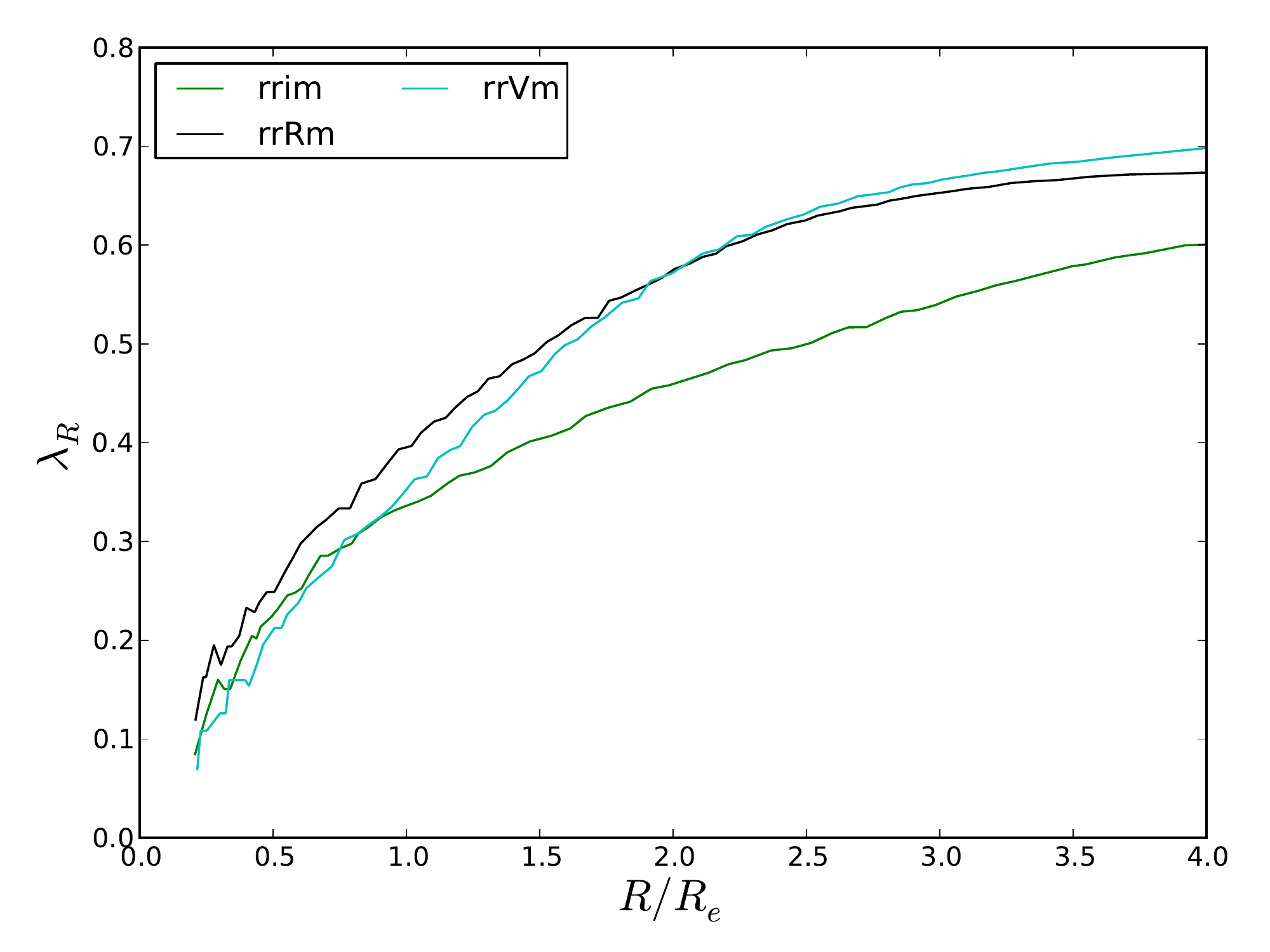}  \\
 & \\
  \hline
 \end{tabular}
\caption{Same as Fig.~\ref{fig:Vmerg11} for the binary mergers of mass ratio 6:1.}
\label{fig:Vmerg61}
\end{figure*}

\begin{figure*}
 \begin{tabular}{cc}
  \includegraphics[width=0.70\columnwidth]{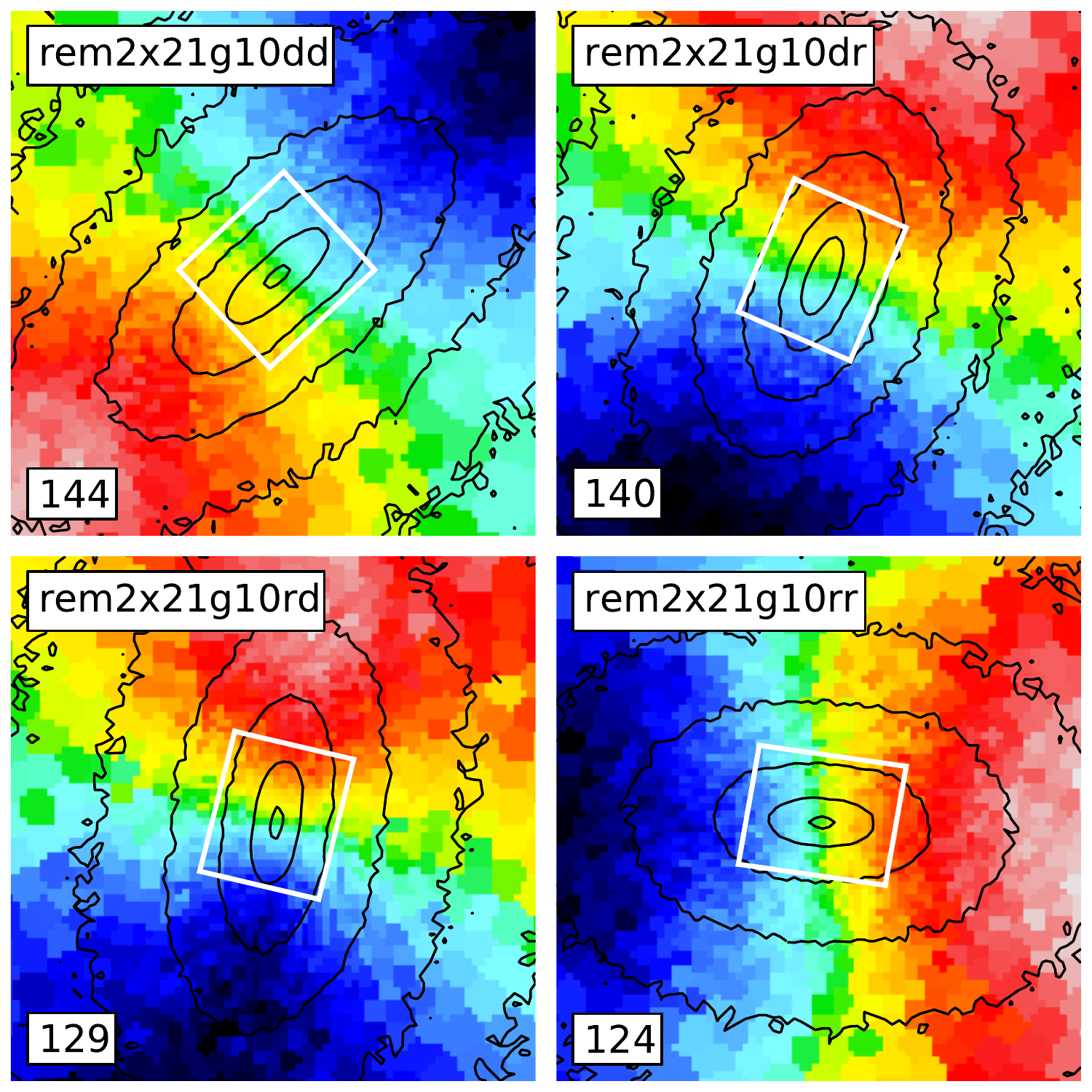} & \includegraphics[width=0.95\columnwidth]{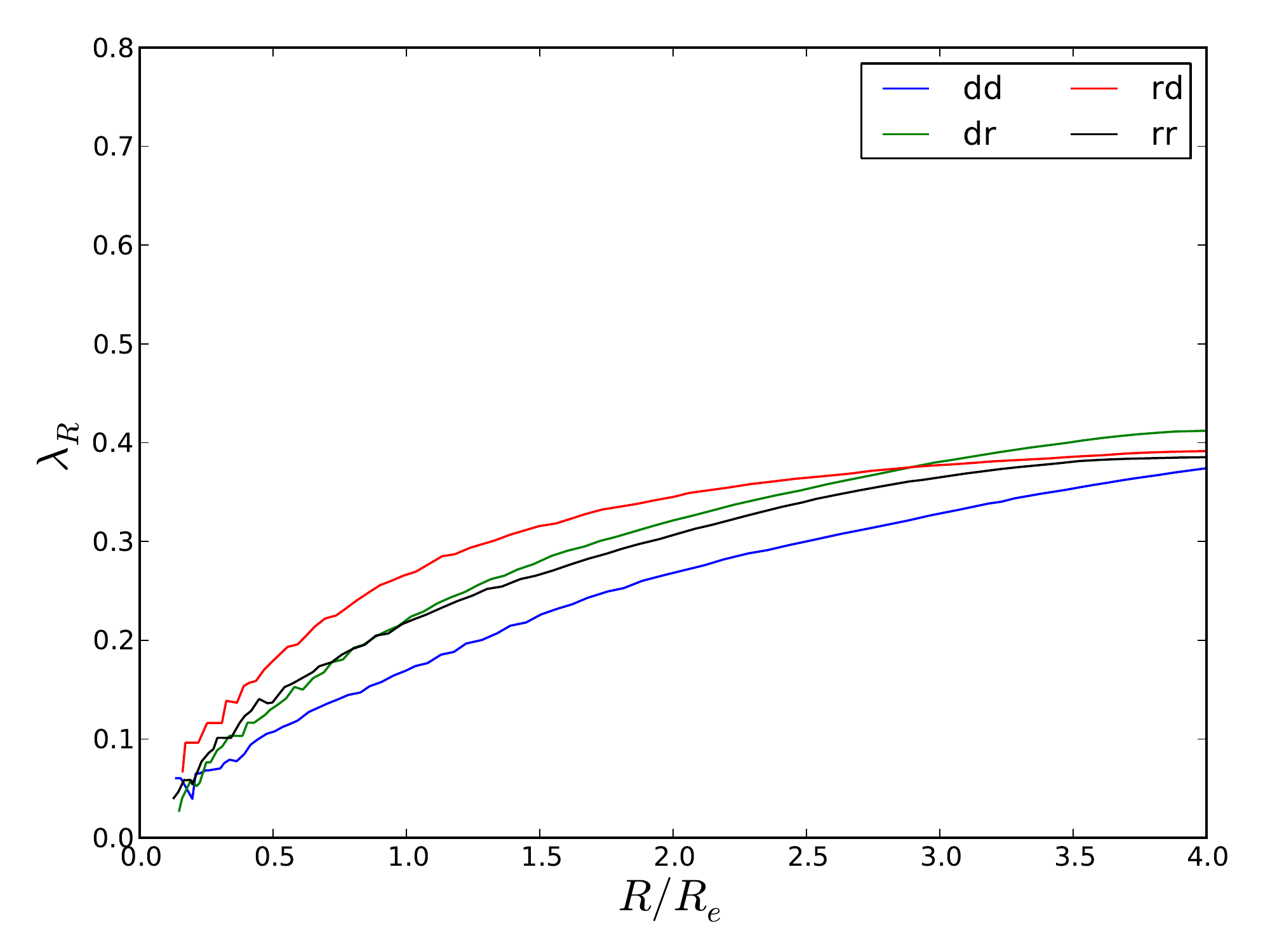}  \\
 \end{tabular}
\caption{Same as Fig.~\ref{fig:Vremergers} for remergers of two remnants of binary mergers of mass ratio 2:1 with 10 per cent of gas.}
\end{figure*}

\begin{figure*}
 \begin{tabular}{cc}
  \includegraphics[width=0.70\columnwidth]{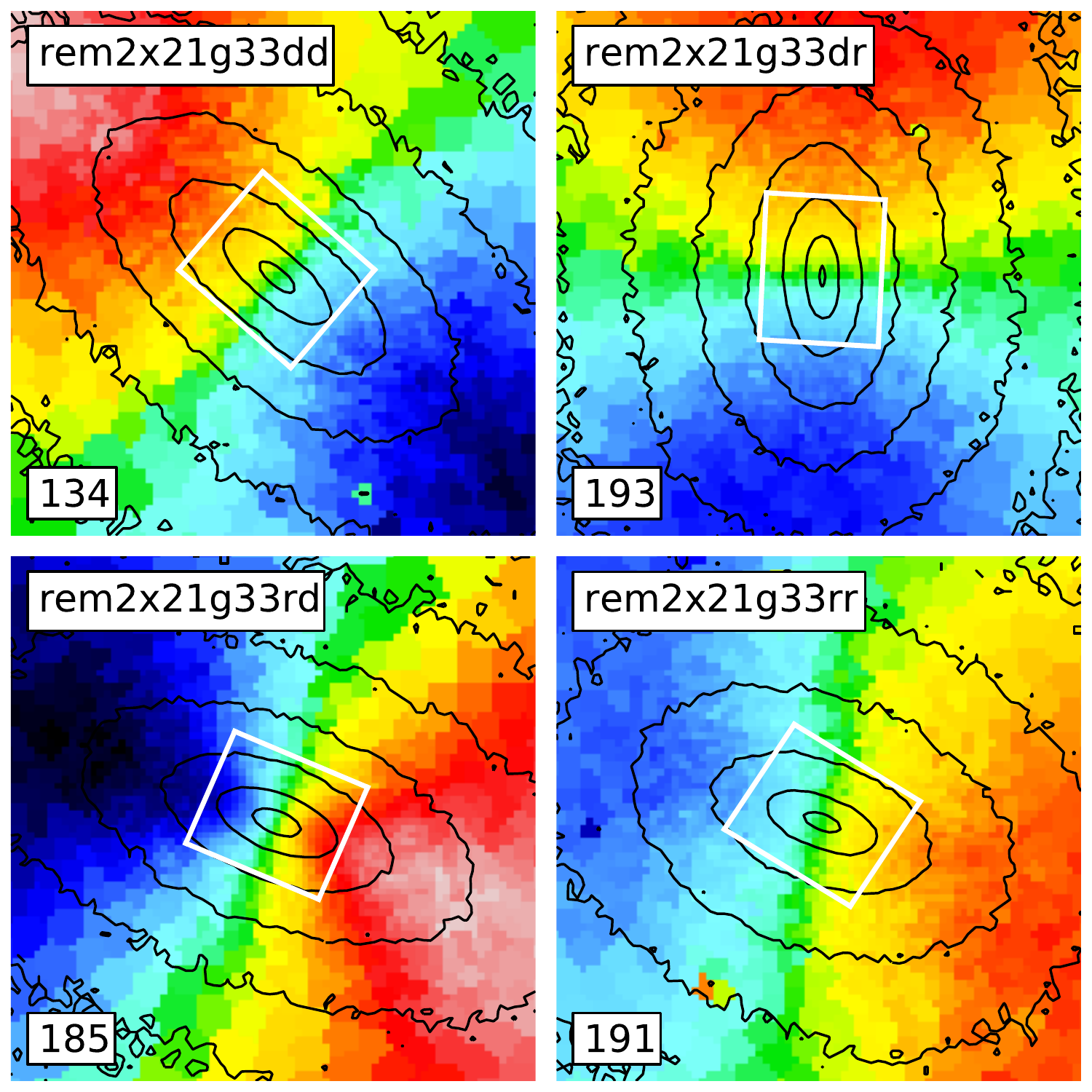} & \includegraphics[width=0.95\columnwidth]{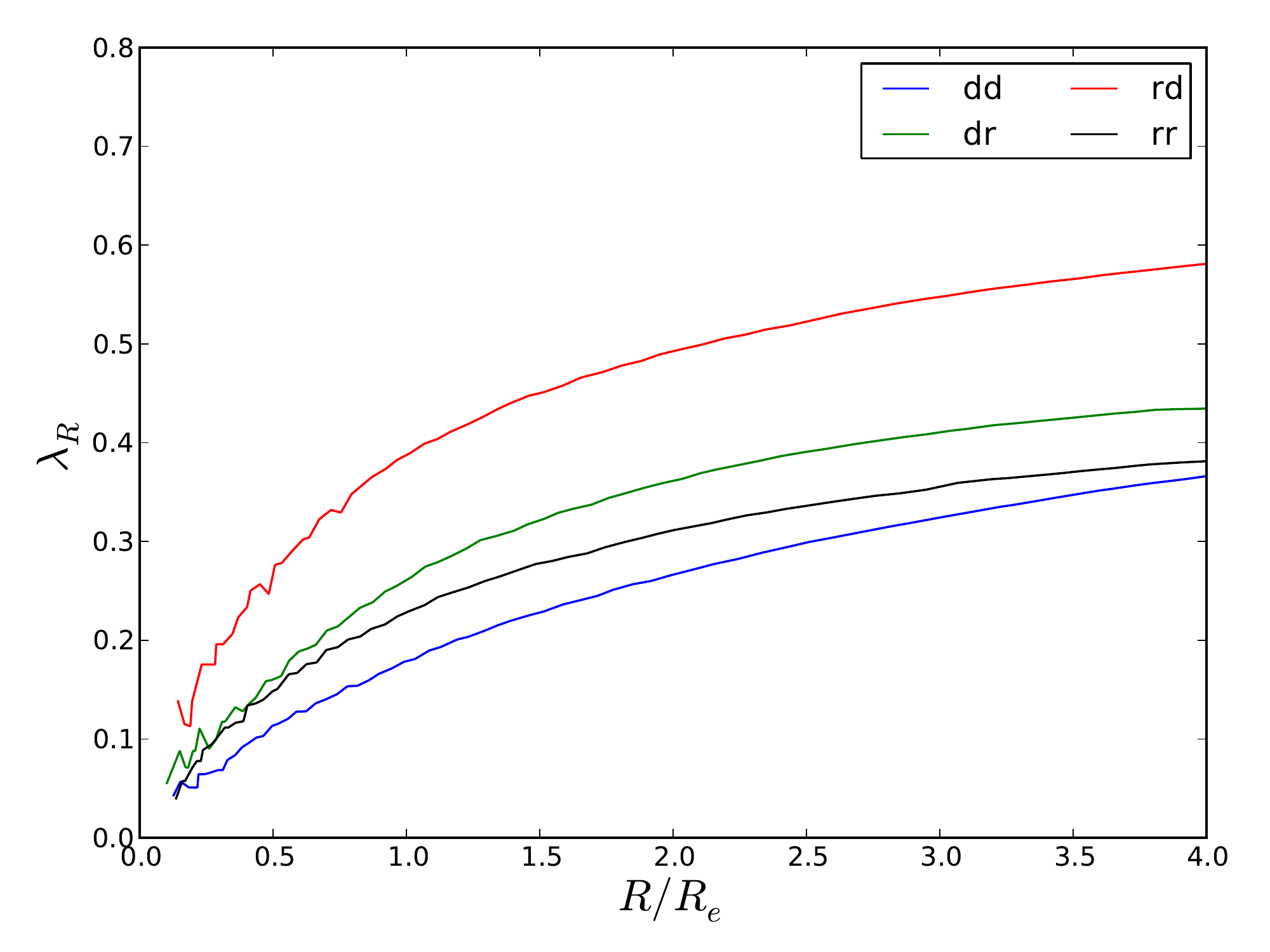}  \\
 \end{tabular}
\caption{Same as Fig.~\ref{fig:Vremergers} for remergers of two remnants of binary mergers of mass ratio 2:1 with 33 per cent of gas.}
\end{figure*}

\begin{figure*}
 \begin{tabular}{cc}
  \includegraphics[width=0.70\columnwidth]{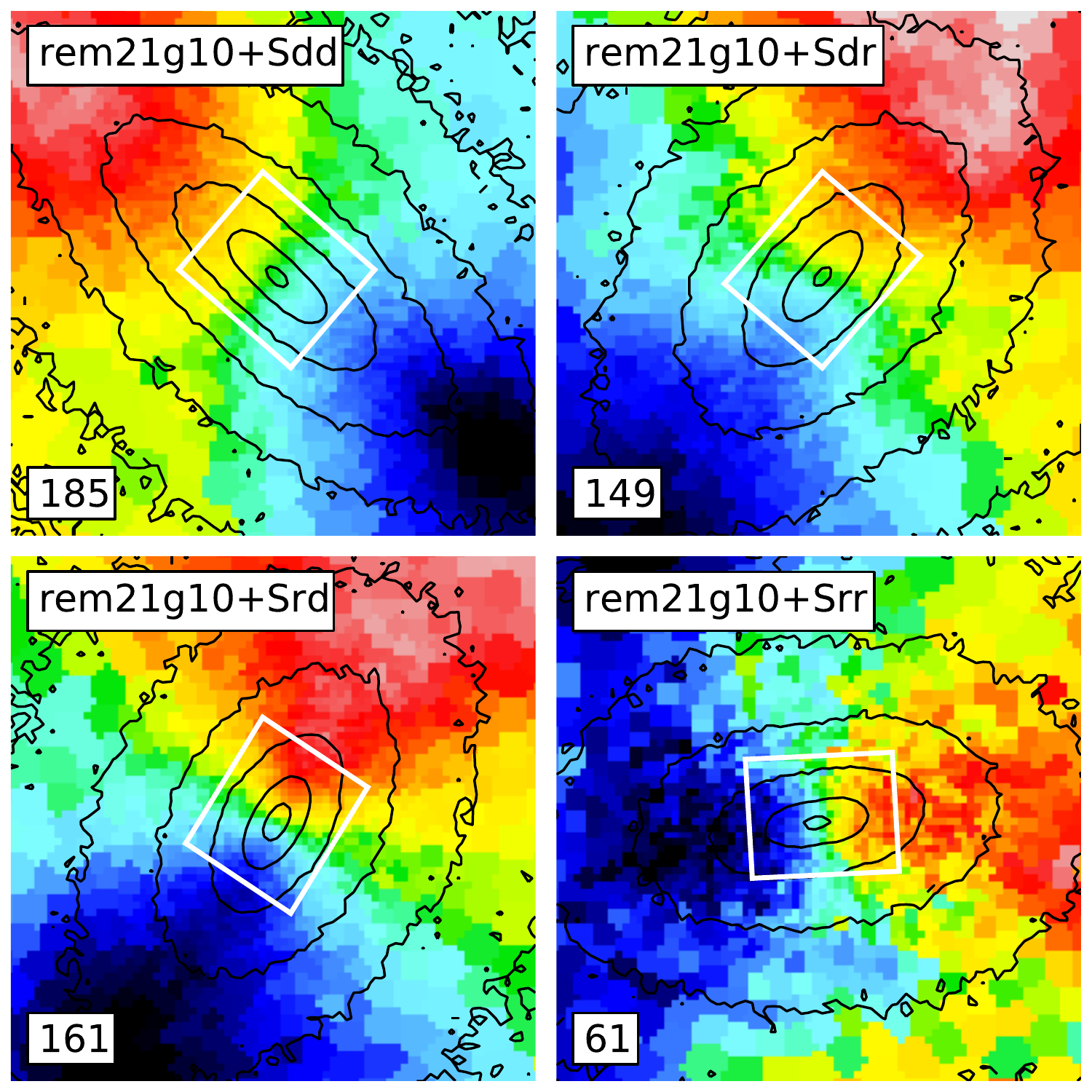} & \includegraphics[width=0.95\columnwidth]{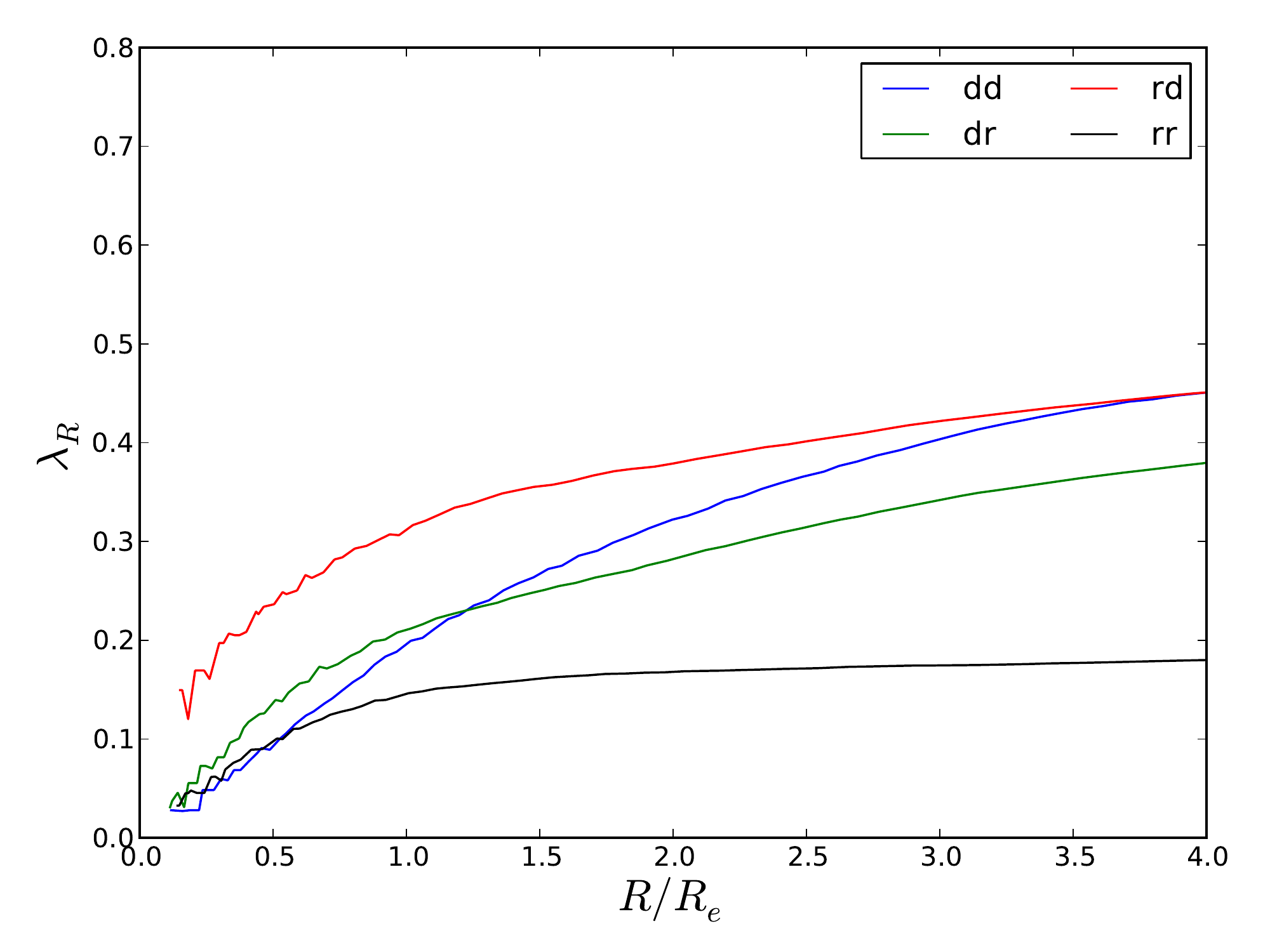}  \\
 \end{tabular}
 \caption{Same as Fig.~\ref{fig:Vremergers} for remergers between a remnant of binary mergers of mass ratio 2:1 with 10 per cent of gas and a spiral galaxy.}
 \label{fig:Vrem+s}
\end{figure*}

\label{lastpage}

\end{document}